\pdfoutput=1
\documentclass[sigconf,nonacm,natbib]{acmart}

\renewcommand\footnotetextcopyrightpermission[1]{}
\setcopyright{none}

\usepackage{booktabs}
\usepackage{array}
\usepackage{graphicx}
\usepackage{amsmath}
\usepackage[htt]{hyphenat}

\newsavebox{\mdtablebox}

\begin{document}

\title{Metadata Reconstruction from Values Alone: Recovering Column Semantics in Undocumented Warehouses}

\author{Mike Helwig}
\affiliation{%
  \institution{Independent Researcher}
  \country{USA}
}
\email{mike@m6online.com}
\renewcommand{\shortauthors}{Helwig}

\begin{abstract}
Text-to-SQL benchmarks ship schemas whose column names already say what the columns mean. Production warehouses are the inverse: decades of accretion leave cryptic identifiers (\texttt{pms\_legacy.t\_resv.amt\_minor}) beside clean marts, with documentation that is partial, stale, or absent. This paper addresses the prior problem such warehouses pose: recovering what the columns and their values \textit{mean}, from the data itself.

We present Rosetta, which places a language model inside a verification harness: a deterministic profiler extracts structural evidence (value fingerprints, a 26-pattern library, checksum verdicts), the model proposes semantics \textit{conditioned on that evidence}, and every resulting fact carries provenance and a confidence bounded by the class of evidence supporting it.

We measure the reconstruction directly, against documentation written by other people, and the headline result is about \textbf{selection}, not prose. Stripping identifiers from BIRD's databases and scoring recovery against their human-authored column documentation (\textbf{799 columns}; 375 name expansions, 653 descriptions, 278 code-to-meaning tables), we compare three arms on identical inputs: statistical semantic-type detection with no language model, a language model shown the column and its values, and Rosetta. Across \textbf{680 paired columns in eleven databases} with identifiers destroyed, the harness delivers metadata that is \textbf{0.475 accurate on the 42\% of columns it commits to, against 0.223 on 94\%} for the same model used directly: a catalog that can be trusted where it speaks, where one that guesses everywhere cannot be trusted anywhere. A control isolates what the direct baseline was really using: with identifiers visible it scores 0.752, with them destroyed 0.074. The statistical arm, covering all columns, recovers almost nothing even on its natural value-domain facet (0.030 against the harness's 0.292); both language-model arms beat it with intervals excluding zero under a cluster bootstrap over databases.

The mechanism, located precisely: restricted to the \textbf{283 columns where both arms speak} (an identical question set, so prose is all that differs), the harness writes \textit{no better} metadata than the language model alone, and on two facets measurably worse (paired differences \textbf{$-$0.033 / $-$0.088 / $-$0.043}, the latter two excluding zero; a blinded judge agrees at \textbf{$-$0.038}). The gain is \textbf{selection}: recomputing the deterministic evidence for every column shows it governs whether the system speaks (coverage \textbf{0.550} with structural evidence versus \textbf{0.293} without, \textbf{+0.257 [+0.128, +0.378]}) but not how well it speaks (\textbf{+0.090, \textit{p} = 0.071}). The deterministic layer is a \textbf{competence detector, not a competence amplifier}. Re-running both arms on a second backbone bounds the claim: the prose finding reproduces almost exactly, but the \textbf{selection behaviour does not transfer intact}: on Claude Sonnet 4.6 coverage rises from 0.422 to 0.823 (the same model naked: 0.984, so a sixteen-point abstention margin survives where fifty-two had been) and the evidence-tracking gap is no longer detectable, because the shipped system requests abstention in a prompt rather than enforcing it in code. The final study closes that gap: a commit gate keyed on the grounding tier (implemented only after six predictions were registered, then measured on a third backbone and on held-out databases) refuses ungrounded prose in code, making no-evidence coverage \textbf{0.000 on every backbone} with recall-when-claimed undegraded, at a measured coverage cost we report.

Because content-token recall is a lexical proxy, we validate it rather than assert it: 180 blinded judgements track the token metric at \textbf{Spearman $\rho$ = 0.642}, reproduce the arm ordering and \textit{widen} it, a second rater from a different model family agrees with the judge at \textbf{$\kappa$ = 0.737}, and the advantage survives controls for verbosity (the winning arm writes \textit{fewer} words). On a blind i2b2 clinical warehouse it decodes \textbf{95.5\% of 134 real ICD-9 codes} from values alone, and abstains on the NDC drug codes rather than inventing them, deciding, per value, which code systems it is in a position to decode. The reconstructed catalog then supports calibrated abstention at query time: on a stripped-schema Spider study (11 databases, 1,082 execution-labeled questions) a naive translator degrades from \textbf{0.92 to 0.42} execution accuracy under full opacity while our gate answers selectively at \textbf{86\% accuracy over 59\% coverage}, with coverage that tracks difficulty.

We report what fails as carefully as what works, including three analyses establishing that our own authority ladder is \textit{not} the mechanism producing the result, a naive-plus-abstention baseline that matches us on toy schemas, and an evaluation-harness bug we caught, fixed, and re-ran at larger scale before claiming the stronger result.
\end{abstract}

\maketitle

\section*{1 Introduction}

The dominant academic framing of natural-language interfaces to databases treats the schema as a solved input. Spider, BIRD, and similar benchmarks present a database whose tables and columns already carry human-meaningful names, whose foreign keys are declared, and in which each business concept maps to a single canonical column. Under that assumption, the remaining problem is \textit{translation}: mapping an English question to SQL given a known, clean schema. The field has made rapid progress on translation, and contemporary LLMs are very good at it.

This paper begins from a different empirical premise. In a production enterprise warehouse, the schema is not a clean input; it is the principal source of difficulty. Such warehouses are the product of decades of organizational accretion. They contain cryptic, machine-generated, or legacy column identifiers (\texttt{t\_92.o\_4}, \texttt{pms\_legacy.t\_resv.amt\_minor}) sitting directly beside well-documented analytical marts. Foreign keys are frequently undeclared. Documentation, when it exists at all, is partial, stale, or internally contradictory. Multiple physical columns often encode the same business concept with conflicting semantics. In this regime, a naive LLM does not fail loudly; it fails \textit{confidently}. Given a question and a forest of cryptic columns, it emits syntactically valid, plausible-looking SQL that joins the wrong tables or projects the wrong column, and it returns an answer with no signal that the answer is untrustworthy.

We reframe the problem accordingly: \textbf{the bottleneck is not LLM quality, it is metadata quality.} The right system for answering questions over an undocumented warehouse is therefore not primarily a better translator. It is a system that (1) \textit{reconstructs} the missing metadata from the warehouse's own data and any available documentation, (2) \textit{knows what it does not know} and bounds its confidence in every reconstructed fact, and (3) \textit{refuses honestly} when grounding is insufficient, rather than emitting a confident wrong answer. In the project's own framing, "text-to-SQL is the visible demo; the catalog is the product." Put plainly: the aim is not to out-guess a naive LLM but to stop guessing when guessing is unwarranted, answering only what it can ground, and routing the rest to confirmation or refusal rather than asserting it.

This reframing has three consequences that organize the rest of the paper. First, metadata reconstruction must happen \textit{before} query time and must be auditable: each reconstructed fact needs a provenance trail (which model, which prompt, which evidence, which confidence). Second, confidence cannot be a free-floating LLM self-report; it must be \textit{bounded by the authority of the evidence} that supports it: a checksum that proves a column contains valid IBANs is strong evidence about \textit{what kind of value} the column holds, but it is not, by itself, evidence about the column's \textit{business name}, and the system must encode that distinction. Third, abstention must be a first-class, \textit{calibrated} outcome: a refusal that correctly withholds a wrong answer is a success, not a failure, and the decision to refuse must be reproducible and resistant to prompt injection.

We present Rosetta, a system embodying these commitments, and we are careful never to conflate the two registers in which we report it: \textbf{external studies against documentation and labels authored by other people} (Sections 5.2, 5.3, 5.4, 5.6 and 5.7), which carry the paper's claims, and an \textbf{internal multi-tenant case study} (Section 5.5) that demonstrates the running system's breadth, hardening and operability but which we treat as suggestive, never as proof. We make four contributions:

\begin{enumerate}
\item \textbf{A measurement of metadata reconstruction against third-party documentation} (Sections 5.2--5.4), which is the paper's primary result. Prior work on recovering column semantics reports either statistical semantic-type accuracy on a fixed type vocabulary or end-task accuracy on clean schemas; we instead destroy identifiers on BIRD's databases and score what a system recovers of the benchmark authors' own prose documentation (expanded field names, business descriptions, and code-to-meaning tables) across 799 columns. Three arms on identical inputs separate the contributions of pattern classification, a language model, and the verification harness (delivered field-name recovery 0.011 / 0.223 / 0.475, every difference against the statistical baseline excluding zero under a cluster bootstrap over databases), and a paired re-analysis then locates that advantage in \textit{selection} rather than in description quality (Section 5.2.1), a correction to our own earlier reading. Because that metric is a lexical proxy, we validate it against blinded judgements of meaning recovery and a second rater (Section 5.3), and we test transfer on a blind i2b2 clinical warehouse whose \textit{values} are opaque codes (Section 5.4).
\item \textbf{An LLM inside a verification harness} (Sections 4.1--4.2): a three-tier deep profiler whose LLM component is mechanically prevented from asserting an unverified match, over an ordered grounding-tier ladder in which each tier records the strongest class of evidence behind a fact and caps its confidence accordingly. Every reconstructed fact carries a provenance trail: which model, which prompt, which evidence, which confidence. We present the ladder as the system's \textit{auditability} mechanism rather than as its accuracy mechanism, and we report in Sections 5.8 and 5.8.1 that, at its current weighting, it contributes little to the routing decision itself.
\item \textbf{Calibrated abstention over the reconstructed catalog} (Sections 4.3, 5.6 and 5.7): the route (answer / confirm-first / refuse) is pure thresholded arithmetic over a feature vector, the LLM contributing features but never deciding the route. On a stripped-schema \textbf{Spider study with a matched naive baseline} (11 databases, 1,082 execution-labeled questions, database-clustered intervals), naive execution accuracy degrades 0.92$\rightarrow$0.42 under full opacity, a \textasciitilde{}58\% silent-error rate, while the gate answers \textit{selectively}, 86\% accuracy [CI 78--93\%] over 59\% coverage with coverage that tracks difficulty; the routing score discriminates (normalized AURC 0.26) and is calibratable (ECE 0.088$\rightarrow$0.026). A \textbf{BIRD breadth study} (Section 5.7) shows the pattern sharpens on real, harder schemas: cold reconstruction still does not out-generate naive, but the naive model's own confidence goes worse-than-random under full opacity while the grounded gate still discriminates.
\item \textbf{Negative results reported at the same resolution as the positive ones}, and a frozen experimental record (Section 5.12) that regenerates every number in this paper from committed data. We report that a cheap naive-plus-abstention baseline matches our gate on Spider's toy schemas (Section 5.6); that a \textit{learned} selector over generic signals out-ranks and out-calibrates our authority score, which adds no incremental discrimination (Section 5.8.1); that our reconstruction advantage comes with a higher contradiction rate when the system does speak (Section 5.3); and that an evaluation-harness bug in an earlier version of the Spider study inflated a prior verdict, which we found, fixed, and re-ran at larger scale (Section 5.1).
\end{enumerate}

Throughout, we distinguish \textbf{measured} results, produced by an actual run of the system, from \textbf{proposed} experiments that have not been run. We hold ourselves to the rule that every empirical number in this paper is traceable to an internal measurement, and that no experiment is described as completed unless it was.

\textbf{Scope and non-claims.} To keep the contribution unambiguous, we state at the outset what this paper does \textit{not} claim. First, we do \textbf{not} claim state-of-the-art text-to-SQL execution accuracy: on clean schemas a naive LLM equals or exceeds our raw generation, and cold reconstruction does not out-generate it even on the harder BIRD schemas. We measure and report this rather than hide it. Second, we do \textbf{not} claim universal or model-agnostic calibration: our routing score is calibrated \textit{per deployment} by fitting a post-hoc isotonic map against observed outcomes, its calibration is non-uniform across databases (Section 6), and we demonstrate it on Spider and BIRD rather than proving it transfers to an arbitrary warehouse or model. Third, we do \textbf{not} claim a proven reconstruction-quality lift on real enterprise warehouses: the evidence that better metadata raises delivered accuracy is internal, small-bank, and \textit{suggestive} (Section 5.5), and our external studies strip documentation from \textit{public} schemas rather than measuring a live production catalog. Fourth, and most consequential for how this paper should be read, we do \textbf{not} claim that the \textit{authority ladder} is the mechanism that produces the abstention. Three independent analyses say otherwise (Sections 5.8 and 5.8.1): a leave-one-component-out re-scoring finds execution grounding load-bearing and catalog confidence carrying coverage rather than ranking; a generator-fixed router matrix finds the authority features add no incremental discrimination over generic signals; and a direct re-bootstrap ablation that removes every tier ceiling changes the delivered outcome on one of forty-one paired questions. We report this because we measured it, and it revises our own prior framing.

What we \textbf{do} claim is correspondingly narrow, and it is externally validated. First and primarily: \textbf{given a column stripped of its name and documentation, the metadata this architecture delivers recovers more of what human authors wrote than that of either a statistical semantic-type detector or the same language model used directly}, because it is the only one of the three that declines when it cannot ground an answer, and because that declining is driven by the deterministic evidence rather than by the model's self-report. We are explicit that this is a selection advantage: on columns where the baseline also speaks, it writes metadata of the same quality (Section 5.2.1). The claim is measured against documentation we did not write, on schemas we did not choose, with intervals that resample databases rather than columns, and with the scoring metric itself audited rather than assumed. Second: over the resulting catalog, the system delivers \textit{calibrated abstention} at query time (answering selectively, with coverage that tracks difficulty, converting a naive model's confident silent errors into auditable refusals) on Spider, sharpening on BIRD. Two properties of \textit{how} it does so are the durable contribution rather than the tiering itself: the decision requires \textbf{no outcome labels} (the learned selectors that outrank our score in Section 5.8.1 all need execution-labeled data that an undocumented warehouse cannot supply), and it is \textbf{auditable and deterministic}: every routed answer carries the provenance of the facts it rests on, and the LLM contributes features but never arbitrates the route.

\section*{2 Related Work}

\textbf{Text-to-SQL and benchmarks.} The modern text-to-SQL literature is anchored by cross-domain benchmarks that supply clean schemas. Spider \cite{yu2018spider} established the cross-database generalization setting, and BIRD \cite{li2023bird} extended it with larger, noisier real-world databases and an execution-accuracy metric while still providing curated schema metadata. The methods evaluated on these benchmarks have advanced from early neural sequence-to-sequence and sketch-based parsers \cite{zhong2017seq2sql,xu2017sqlnet} to LLM prompting and agentic pipelines that decompose linking, generation, and self-correction \cite{pourreza2023dinsql,gao2023dailsql}. Our work is positioned against the schema assumption common to all of them: we treat the schema itself as degraded and the reconstruction of its semantics as the central task. We are not aware of a published benchmark that systematically \textit{strips} schema documentation and measures the resulting degradation as a function of reconstruction effort; we construct such a mangler and report a Spider study with it (Section 5.6).

\textbf{Schema linking.} A substantial line of work treats schema linking (aligning question tokens to the correct tables and columns) as the hard sub-problem of text-to-SQL, inserting explicit linking or grounding stages between the question and SQL generation: intermediate-representation parsers \cite{guo2019irnet}, relation-aware joint question--schema encoders \cite{wang2020ratsql}, and ranking-based decoupling of linking from skeleton generation \cite{li2023resdsql}. Our deep profiler and grounding gate can be read as an aggressive form of schema-linking infrastructure that operates \textit{before} any question arrives: rather than linking a question to a clean schema, we first reconstruct a semantic catalog over a cryptic schema, then link against the reconstructed catalog. The distinction is that our linking evidence is bounded by an explicit authority tier.

\textbf{Selective prediction, abstention, and calibration.} Treating abstention as a first-class outcome draws on the classification-with-reject-option tradition \cite{chow1970reject} and its agnostic-setting generalization \cite{elyaniv2011agnostic}, and on selective prediction for deep networks \cite{geifman2017selective,geifman2019selectivenet}, which we adopt together with the risk-coverage curve and its area (AURC) as the correct lens for evaluating a system that may abstain. In the LLM setting, a growing body of work studies whether models "know what they know" and when they should decline to answer \cite{kadavath2022know,wen2025abstention}. Our routing layer implements a three-way generalization of accept/reject: answer, confirm-first, refuse. Its confidence is turned into a probability by per-layer post-hoc calibration: we fit isotonic regression \cite{zadrozny2002transforming} against outcome labels recovered from event lineage, in the tradition of probability calibration for classifiers \cite{platt1999probabilistic,niculescu2005predicting,guo2017calibration}, and report the Brier score and expected calibration error.

\textbf{Abstention and selective prediction \textit{for text-to-SQL}: the closest work.} Reliability-oriented text-to-SQL is an active and very recent line, and we position against it directly. Closest is \textbf{RTS} (Reliable Text-to-SQL with Adaptive Abstention) \cite{chen2025rts}, which makes abstention first-class by applying conformal prediction to an LLM's \textit{schema-linking} hidden states and adds human-in-the-loop intervention, validated on BIRD. Post-hoc \textit{calibration} of text-to-SQL confidence has likewise been studied directly: multivariate Platt scaling over sub-clause-frequency features produces well-calibrated query-level confidence across Spider and BIRD \cite{liu2025subclause}. And a concurrent selective-prediction study surveys the natural correctness signals (self-consistency, executability, log-probabilities, schema-relevance, learned verifiers, and LLM-judge ensembles), finding that black-box judge ensembles dominate and that fine-tuned verifiers fail to transfer to unseen schemas \cite{richardson2026predicts}. Rosetta differs from all three along one axis they share: each estimates uncertainty over an \textit{available} (if noisy) schema at query time. \textbf{Rosetta is not the first text-to-SQL system to abstain. It is, to our knowledge, the first system to reconstruct missing warehouse metadata into an authority-bounded fact graph and propagate that evidence authority into deterministic selective query answering under controlled schema opacity.} Concretely: where RTS abstains \textit{during} schema linking on schemas that are present, Rosetta \textit{reconstructs} absent metadata before any question arrives, attaches provenance and an explicit authority ceiling to each reconstructed fact, and routes with a black-box-compatible \textit{deterministic} rule in which the LLM contributes features but never arbitrates the route, targeting opaque \textit{physical} schemas and metadata quality rather than link uncertainty over a readable one. The calibration and correctness-signal baselines \cite{liu2025subclause,richardson2026predicts} also define the comparison set that our fair-baseline study (Section 5.6) begins and that a full main-track evaluation would complete against Rosetta's authority-grounded score.

\textbf{Data catalogs, discovery, and profiling.} Reconstructing semantics from values connects to data profiling (cardinality, value distributions, pattern and dependency detection \cite{abedjan2015profiling}, including scalable inclusion-dependency discovery \cite{papenbrock2015binder}, platforms that industrialize it \cite{papenbrock2015metanome}, and the selection of true foreign keys from among discovered dependencies \cite{rostin2009fk,zhang2010multicolumn}) and to enterprise data-catalog and data-discovery systems that organize and surface warehouse metadata at scale \cite{halevy2016goods,fernandez2018aurum}, including learned semantic-type detection from column values \cite{hulsebos2019sherlock}. Our contribution relative to that tradition is not the profiling primitives but their composition into an authority-tiered, confidence-capped fact graph that feeds a calibrated router, and the use of checksum-validated structural patterns to \textit{bound} (rather than merely produce) semantic confidence.

Every reference in this paper was verified against Crossref, arXiv, and Semantic Scholar (each entry resolved to a real work), and we situate the contribution against the named benchmarks and systems above; the external evaluation of Section 5.6 is designed to test it on stripped public schemas, not only our own fixtures.

\section*{3 Problem Formulation}

Let a warehouse $W$ consist of schemas, tables, and columns. Each column $c$ has a physical identifier $\text{id}(c)$ (possibly cryptic), a population of values $V(c)$, and an unknown true semantic concept $\sigma^*(c)$ drawn from a business ontology (e.g., "guest folio amount in minor currency units"). The available documentation $D$ is a partial, possibly contradictory map from a subset of columns to natural-language descriptions; in the worst case $D = \varnothing$.

\textbf{Metadata reconstruction.} The reconstruction task is to produce, for each column, an estimated concept $\hat\sigma(c)$ together with a calibrated confidence $p(c) \in [0,1]$ and an \textit{authority tier} $\tau(c)$ that records the strongest class of evidence supporting $\hat\sigma(c)$. The tier is not a free parameter: it caps the confidence, so that $p(c) \le \text{cap}(\tau(c))$. A reconstruction is \textit{committed} to the catalog only if its tier reaches an auto-commit threshold; otherwise it is enqueued for human review.

\textbf{Question answering with abstention.} At query time the system receives a natural-language question $q$. It produces a candidate SQL program $s$, executes it (or determines it cannot safely execute), and must select a route
\begin{equation*}
\rho(q) \in \{\textsf{answer},\ \textsf{confirm\_first},\ \textsf{refuse}\}.
\end{equation*}

Let $y(q) \in \{\text{answerable}, \text{ambiguous}, \text{out-of-scope}\}$ be the (latent) ground-truth disposition of the question. The evaluation does not score SQL string equality. It scores \textbf{route accuracy}, whether $\rho(q)$ matches the disposition the question deserves, together with auxiliary measures (column $F_1$, execution-result match where a canonical query is available, and a continuous confidence score). Crucially, \textit{refuse} is a correct route for out-of-scope or insufficiently grounded questions; a system that answers them is wrong even if the SQL is syntactically fine.

\textbf{Selective-prediction view.} The routing problem is a selective-prediction problem with an extra rung. A confidence score $z(q)$ is computed for each question; thresholds $(t_\text{refuse}, t_\text{confirm})$ partition the score axis into refuse / confirm-first / answer regions. Sweeping the thresholds traces a risk-coverage trade-off: lowering coverage (answering fewer questions) should lower risk (error rate on answered questions). The design goal is a confidence score that is \textit{monotone in correctness}, so that abstention is well-behaved. Establishing this property empirically with risk-coverage curves is the subject of Sections 5.1 and 5.6. An early attempt to read it off pooled historical logs was inconclusive, and, we later found, corrupted by an evaluation-harness bug (Section 5.1), but the controlled external measurement on 1,082 execution-labeled Spider questions shows the raw score both discriminates correct from incorrect (normalized AURC 0.26, CI excluding random) and is calibratable (ECE 0.088 $\rightarrow$ 0.026), and the pilot (Section 5.6) turns that into a demonstrated selective-prediction advantage: the gate answers \textit{selectively}, withholding the questions it would most likely get wrong, rather than abstaining wholesale.

\section*{4 Method}

Rosetta is organized around a single principle: \textbf{the LLM contributes features and candidates; deterministic, auditable machinery makes the decisions.} This applies both to reconstruction (where an LLM may \textit{propose} a column's meaning but may not \textit{assert} a match without mechanical proof) and to routing (where the LLM may generate SQL and critic verdicts but may not choose the route). The design rules this section instantiates are distilled, with their measured boundaries, in a companion methodology paper \cite{helwig2026principles}. The following subsections describe the four load-bearing components. All component descriptions are code-verified against the implementation; file paths are given for reference.

\subsection*{4.1 The authority-tiered fact graph}

The spine of reconstruction is an ordered \textbf{grounding-tier ladder} (\texttt{rosetta/\allowbreak grounding/\allowbreak tiers.py}, a \texttt{GroundingTier} \texttt{IntEnum}). Each tier names a \textit{class of evidence} and carries an L2 confidence cap. The committed confidence of any reconstructed fact is

\begin{equation*}
p(c) = \min\bigl(\text{raw\_L2}(c),\ \text{cap}(\tau(c))\bigr),
\end{equation*}

so that the authority of the evidence mathematically bounds the confidence, regardless of how confident the underlying model claims to be. The ladder is:

\begin{table*}[t]
\centering
\small
\setlength{\tabcolsep}{4pt}
\begin{lrbox}{\mdtablebox}
\begin{tabular}{lll>{\raggedright\arraybackslash}p{0.42\linewidth}}
\toprule
Tier & Name & Cap & Meaning \\
\midrule
0 & INTERPRETER\_ONLY & 0.50 & LLM inferred meaning from the column profile alone \\
1 & PATTERN\_CONFIRMED & 0.60 & profile matches a known regex shape \\
2 & DOC\_CITED\_WEAK & 0.70 & a 0.30--0.80 documentation snippet was cited, graded only "related" \\
3 & SIBLING\_CONSISTENT & 0.75 & sibling columns are uniformly consistent \\
4 & PATTERN\_VALIDATED & 0.85 & checksum-backed structural match (Luhn / IBAN / ISBN / EAN / UPC) \\
5 & DOC\_CITED\_STRONG & 0.90 & a $\geq$0.80 snippet directly asserts the meaning and "supports" it \\
6 & QUERY\_VALIDATED & 0.95 & $\geq$3 downstream queries accepted the column with no rejections \\
7 & SME\_AFFIRMED & 1.00 & a human reviewer explicitly accepted the value \\
\bottomrule
\end{tabular}
\end{lrbox}
\ifdim\wd\mdtablebox>\linewidth\resizebox{\linewidth}{!}{\usebox{\mdtablebox}}\else\usebox{\mdtablebox}\fi
\end{table*}

A subtle but load-bearing design choice is that \textbf{PATTERN\_VALIDATED (tier 4, checksum-backed) is deliberately floored \textit{below} DOC\_CITED\_STRONG (tier 5)}. The system's own design note states the rationale: "checksum math doesn't admit coincidence" (a column whose values pass the IBAN mod-97 check almost certainly \textit{contains} IBANs), but a checksum only proves \textit{what kind of thing} the values are; it does not tie the column to its business \textit{name}. A strong documentation citation, by contrast, asserts the name. Confidence in \textit{what a column means for the business} is therefore capped higher by a strong doc cite than by a checksum, even though the checksum is "harder" evidence about value structure. Encoding this distinction in the tier order is precisely the kind of bounded-authority reasoning that prevents over-confidence.

A scoring function \texttt{tier\_for\_score()} maps a raw score to a tier through default score bands, with three outcome-driven short-circuits: an explicit human acceptance routes to tier 7; three accepted downstream queries route to tier 6; and a checksum-backed pattern match floors the tier at 4. Hardening in the gate (\texttt{\_has\_sme\_affirmation}) requires that an SME affirmation be \textit{value-specific and still live} (the predicate matches the exact value and requires that the fact has not been superseded), which fixes an earlier bug in which any past human acceptance allowed unrelated new guesses to inherit the 1.00 SME tier.

\subsection*{4.2 The deep profiler: Tiers A $\rightarrow$ B $\rightarrow$ C}

Reconstruction evidence is produced by a three-tier deep profiler (\texttt{rosetta/\allowbreak metadata/\allowbreak deep\_profiler/\allowbreak }, orchestrated by \texttt{DeepProfiler.profile()}), structured so that cheaper, more deterministic evidence is exhausted before more expensive, less certain evidence is invoked.

\textbf{Tier A: structural fingerprint} (\texttt{fingerprint.py}) is deterministic and uses no LLM. Over up to 5,000 sampled values per column it computes a \texttt{ColumnFingerprint}: length distribution, character-class ratios, delimiter and case profile, longest-common-prefix / suffix coverage, per-character Shannon entropy, numeric summary statistics (min, max, median, 5th and 95th percentiles), a Benford first-digit KL divergence when $N \ge 30$, a monotonicity test ($\geq$70\% step-1 increments flags an auto-increment), cardinality ratio, sentinel-value candidates, and a one-level recursive segment fingerprint for composite codes. This is the structural substrate everything else reasons over.

\textbf{Tier B: pattern library} (\texttt{patterns.py}, \texttt{DEFAULT\_PATTERNS}) consists of roughly twenty-five default \texttt{PatternSpec}s (approximately twenty-six counting variants), each gated by the Tier-A fingerprint. The key mechanism is \textbf{checksum promotion}: a pattern whose values pass a checksum is promoted to $\geq$0.95 confidence with \texttt{details=\{"checksum": True\}}; a regex-only match is capped at 0.85; a fingerprint-heuristic match at 0.70. Checksum validators (\texttt{checksums.py}) implement Luhn, ISBN-10/13, EAN-13, UPC-A, IBAN mod-97, and mod-11. The library is intentionally tenant-agnostic (it matches structure, not domain strings), which is why it generalizes across verticals.

\textbf{Tier C: rule-verified LLM namer} (\texttt{pattern\_namer.py}) fires \textit{only} when no Tier-B pattern matches and naming is enabled, i.e., for genuinely cryptic columns. The LLM proposes a triple \texttt{(pattern\_name, semantic\_type, verification\_regex)}. This proposal is \textbf{mechanically verified before acceptance}: the regex must compile, must not be one of a set of trivial/degenerate regexes, and must \texttt{fullmatch} at least 90\% of the sampled values. The LLM is thereby never permitted to \textit{assert} a match; it may only \textit{propose} a hypothesis that the system then mechanically confirms or rejects against the data. The effectiveness of this stage is illustrated by a measured zenith run in which cryptic columns with no semantic match dropped from 66 to 2, a 64/66 $\approx$ 96.9\% recovery.

These tiers feed a \texttt{DeepPatternAttester} (\texttt{grounding/\allowbreak attesters/\allowbreak deep\_pattern.py}): a bucket match plus checksum yields a \texttt{supports@0.95} attestation with a \texttt{pattern\_validated\_candidate} code that floors the grounding tier at 4; conversely, a bucket \textit{mismatch} against a checksum-backed match yields \texttt{contradicts@0.95}: when the system has proof, it uses that proof both to support and to refute.

\subsection*{4.3 Calibrated routing: answer / confirm-first / refuse}

At query time the candidate answer is scored by a \textbf{Layer-4 (L4) score} (\texttt{prediction/\allowbreak layer4\_result.py}), a clipped linear sum of \texttt{Layer4Features} with weights pinned by a snapshot test and annotated with provenance. The heaviest positive contributors are \texttt{execution\_ok} (0.20), \texttt{link\_calibrated\_probability} (0.20), and \texttt{catalog\_entry\_mean\_confidence} (0.18); the heaviest penalties are a \texttt{magnitude\_cohort\_outlier\_penalty} ($-$0.30, the single largest term) and a \texttt{row\_count\_implausible\_penalty} ($-$0.15). The catalog confidence term is exactly the bounded reconstruction confidence from Section 4.1, so the authority ladder propagates directly into the routing score: an answer that depends on a weakly grounded column carries that weakness into its route.

\textbf{The route is pure thresholded arithmetic} (\texttt{consumption/\allowbreak answer\_formatter.py}). If execution did not succeed, the route is \textit{refuse}. Otherwise, if the score is below the refuse threshold the route is \textit{refuse}; if it is below the confirm threshold the route is \textit{confirm-first}; otherwise it is \textit{answer}. The thresholds are mode-dependent: a rule-based decision tree (\texttt{detect\_mode}, no ML) classifies the question's grounding regime and selects a \texttt{QueryPolicy}: DOCUMENTED\_STRONG (0.60 / 0.20), AMBIGUOUS\_DOC (0.60 / 0.05), or CRYPTIC\_LEGACY (0.65 / 0.20). The central thesis, which we state plainly because it drives the system's auditability and prompt-injection resistance, is that \textbf{the LLM contributes features, never decides the route}: the linker reranking, the generated SQL, and the critic verdict all enter the score as features, but the route is a deterministic function of the score and the thresholds.

\textbf{Calibration} (\texttt{prediction/\allowbreak calibration.py}) fits a per-layer isotonic regression that maps raw scores to empirical correctness. Outcome labels are recovered by joining predictions to downstream events through event lineage: an L2 commit to a \texttt{commit\_reviewed} event, a grounding decision to a \texttt{review\_queue} status, an L4 answer to a thumbs rating on the delivered answer. The calibrator requires at least fifty samples and is stored as a fifty-one-point grid in an append-only weights table. Until recently this calibrator was \textit{fitted but never applied} (the serving predictor loaded only the heuristic-weights row and \texttt{apply\_calibrator} had no caller), so the served score was always the raw weighted sum; it is now wired into serving (Section 5.1). Whether that score is monotone-in-correctness enough for clean risk-coverage curves is no longer open: on 1,082 execution-labeled Spider questions (Section 5.1) the raw score \textit{discriminates} correct from incorrect well (normalized AURC 0.26, CI excluding random) and an isotonic map makes it well-\textit{calibrated} (ECE 0.088 $\rightarrow$ 0.026); the earlier "weak discrimination" verdict was an artifact of the harness bug of Section 5.1, which had pinned the \texttt{execution\_ok} feature to a constant.

Advisory signals adjust the score without unilaterally flipping the route. A critic, invoked only in the ambiguous band, maps its verdict to an additive adjustment (accept $\rightarrow$ +0.18$\cdot$conf, reject $\rightarrow$ $-$0.25$\cdot$conf, uncertain $\rightarrow$ $-$0.05) that becomes a feature \textit{before} scoring. A family of independent advisory penalties (\texttt{\_penalized\_l4}: critic-reject, structural, magnitude, result-sanity, per-trap-code, and weak-self-confidence) can deepen a low score but is governed by a strict invariant: \textbf{advisories can neither manufacture nor rescue a refuse}; they are clamped so they can only push a score further down from below the refuse threshold, never lift it across. Symmetric bonuses exist for trivially-correct aggregates and evidence consensus. Two decisive escalation paths remain inside the score machinery: a \textit{soft-pin} that deterministically forces confirm-first, and a \textit{forces-refuse} path reserved for catastrophic-PII findings (SSN, credit-card, medical-record), which alone may override confirm-first to refuse; soft PII remains advisory only.

A trap detector (\texttt{consumption/\allowbreak trap\_detector.py}) supplies many of these advisory penalties through 28 \texttt{KNOWN\_TRAP\_CODES}, each a pure function of the question, the SQL, the linked entities, and the foreign-key edges, with no tenant-specific strings, which is again why the detectors generalize across domains. A last-line SQL guard (\texttt{connectors/\allowbreak \_sql\_guard.py}) enforces single-statement read-only execution via a leading-verb allow-list, a single-statement check, and an abstract-syntax-tree root check that blocks the \texttt{WITH cte ... DELETE} CTE-DML escape, failing closed on any ambiguity.

\subsection*{4.4 Human-in-the-loop proposals}

Reconstruction is not fully automatic by design. The grounding gate (\texttt{rosetta/\allowbreak grounding/\allowbreak gate.py}) runs up to seven attesters concurrently: Profile, Doc, Sibling, Contradiction, and Usage (the original five), plus optional QueryLog and DeepPattern, each returning an \texttt{AttestationResult} with a verdict in \{supports, related, contradicts, uncertain, skipped\}. The gate assembles these into a feature vector consumed by a linear \texttt{GroundingPredictor} whose strongest positive weights are the document signal (0.22) and the deep-pattern signal (0.18). The auto-commit rule is explicit: if the resulting tier reaches the auto-commit tier (default SIBLING\_CONSISTENT / 3), the fact is committed at its capped confidence; otherwise it is enqueued to a \texttt{review\_queue} for human adjudication. A human SME who accepts a value lifts it to tier 7 (cap 1.00). The only path to maximal confidence thus runs through a human, and every committed fact carries full provenance (model, prompt version, evidence reference, tier, build identifier), the property that makes the catalog auditable and the eval's per-fact attribution possible.

This pathway also frames one of the paper's suggestive internal findings (Section 5.5, on a small single-tenant bank with an unpaid-human confound): an \textit{automated LLM-team pre-flight} that plays the SME role can reconstruct more of the catalog, at lower cost, than a human SME, and, as a reproducible negative result, stacking a human on top can \textit{regress} accuracy.

The four enterprise "levers" of the platform (authentication/RBAC, a uniform multi-warehouse connector layer, an on-premise egress agent that keeps warehouse credentials inside the customer network, and an incremental streaming-delivery mode) are built and default-inert. They are orthogonal to this paper's thesis and are noted only because the routing and grounding machinery operates uniformly across DuckDB, Postgres, Snowflake, and BigQuery connectors behind a single abstraction. We claim no accuracy results for these levers; their adversarial hardening passes are software-robustness, not text-to-SQL accuracy.

\section*{5 Evaluation}

The evaluation is organized around the system's product rather than its demonstration. Section 5.1 describes the apparatus: the schema mangler, the scorers, and a harness bug we caught and re-ran around. Sections 5.2--5.4 are the paper's primary result: how much of a column's meaning the system recovers from values alone, whether the metric that says so is measuring meaning at all, and what happens on a real coded clinical warehouse. Section 5.5 is an internal multi-tenant case study of the running system, reported as a case study and never as proof. Sections 5.6--5.11 evaluate what the reconstructed metadata is \textit{for} (question answering with calibrated abstention), including three analyses that revise our own prior account of the mechanism. Section 5.12 records the frozen experimental record.

We separate cleanly what has been \textbf{measured} from what remains \textbf{proposed}, and every measured number traces to an actual run: internal (Section 5.5) or external (the reconstruction study of Section 5.2, the Spider study of Section 5.6, and the BIRD breadth study of Section 5.7). The experiments still described as proposed within Section 5.7 (finer severity grids and further held-out tenants) must not be read as results. The external numbers are backed by a committed experimental record: \texttt{papers/\allowbreak EXPERIMENTS.md} ties each figure to its raw-data file under \texttt{papers/data/} and the exact command that regenerates it, and the calibration and discrimination analyses re-run for free from the committed 1,082-sample execution-labeled dataset.

\subsection*{5.1 Experimental apparatus: the mangler, the scorers, and a bug we caught}

Every study in this section depends on three pieces of apparatus, so we describe them before any result rests on them: a \textbf{schema mangler} that removes documentation from a database without touching its data, \textbf{scorers} that measure what a reconstruction recovered, and a \textbf{risk-coverage analyzer} for the query-time studies. We also record here an evaluation-harness bug that corrupted an earlier verdict, because every later section inherits the corrected harness. (Code: \texttt{rosetta/\allowbreak eval/\allowbreak \{schema\_mangler,\allowbreak reconstruction\_recall,\allowbreak risk\_coverage\}.py}, each with unit tests; CLIs under \texttt{scripts/}.)

\textbf{The schema mangler (P1) is built.} \texttt{rosetta/\allowbreak eval/\allowbreak schema\_mangler.py} converts a documented schema into a stripped one at a chosen \textit{severity}: it renames tables and columns to opaque identifiers, deletes comments, and drops declared foreign keys, while preserving the data and emitting an original$\leftrightarrow$mangled name map so gold queries can be re-expressed. It is deterministic under a seed, severity-monotone (higher severity removes a strict superset of signal), and unit-tested for name-map round-tripping and value preservation. We have now pointed it at actual Spider databases and run the external pilot reported in Section 5.6, and extended it to BIRD's real-world schemas (Section 5.7); the full Spider dev set remains the breadth item of Section 5.7.

\textbf{Reconstruction token-recall (P3) is measured.} \texttt{rosetta/\allowbreak eval/\allowbreak reconstruction\_recall.py} scores the system's reconstructed column descriptions against the \texttt{ground\_truth.json} golden vocabulary: expected tokens are drawn from each column's concept, semantic type, unit, and value-domain labels, and recall is |expected $\cap$ reconstructed| / |expected|, the fraction of expected tokens the reconstruction recovers. On the two fully-bootstrapped tenants that carry golden files, macro token-recall is \textbf{0.774 (micro 0.733) on carenexus} (40/40 columns) and \textbf{0.560 (micro 0.585) on zenith} (32 columns), for a macro mean of \textbf{0.667}. The zenith figure is honestly dragged down by 8 of its 32 golden columns that carry \textit{no} live catalog description at all: a genuine reconstruction gap, not a scoring artifact. This is a direct, routing-independent measure of reconstruction quality, and it is now a real number rather than a promise.

\textbf{Risk-coverage (P4): the analyzer is built.} \texttt{rosetta/\allowbreak eval/\allowbreak risk\_coverage.py} computes selective-prediction (risk-coverage / AURC) curves from logged (score, outcome) pairs. Read off \textit{pooled historical logs}, the raw score was inconclusive as an abstention signal: the execution-labeled slice was thin (N=39) and confined to the answered region, while a larger pooled view was contaminated by question duplication across runs, so we did not rely on it. The decisive measurement needs execution-correctness labels spanning the \textit{full} confidence range (including refused questions), on a deduplicated set, with the isotonic map fit before scoring. That is exactly what the controlled Spider study below supplies; the pooled-log negative is reported only as the motivation for running it.

The apparatus now exists; the task was to run that measurement correctly, which surfaced a bug that had been silently corrupting it.

\textbf{A harness bug had been suppressing these measurements; fixing it makes the controlled study both larger and stronger.} The setup was in place: the isotonic calibrator, \textit{fitted but never applied} (the serving \texttt{Layer4Predictor} loaded only the heuristic-weights row and \texttt{apply\_calibrator} had no caller), is now wired into serving with a raw fallback and tests; and the Spider harness logs $(\ell_4^{\text{raw}}, \text{features}, \text{execution-correct})$ for every question regardless of route. But the first harvest ran the in-process \texttt{/ask} pipeline against the \textbf{wrong warehouse}. \texttt{reconstruct\_and\_ask} builds its DuckDB connector through the same tenant$\rightarrow$path resolver the production endpoint uses, and that map is injected per \textit{subprocess} during bootstrap, not into the in-process harness, so the connector silently fell back to the global default warehouse. Every predicted query then failed to \texttt{EXPLAIN} against a schema it could not see, pinning the \texttt{execution\_ok} feature to 0 and firing the implausible-row-count penalty: a uniform $\approx\!0.35$ suppression of $\ell_4$ on \textit{every} question, independent of SQL quality. The symptom was hiding in plain sight in the numbers we first reported: the samples spanned only $\ell_4 \in [0.0, 0.48]$ (that 0.48 ceiling \textit{is} the suppression) and the gate "never answered." We found it, fixed it (the in-process harness now injects the tenant$\rightarrow$warehouse mapping exactly as the resolver expects; the fix carries unit tests), and re-ran the entire study at larger scale. The latent-SQL-accuracy figures of Section 5.6 are unaffected (they execute through a direct connection that bypasses the connector), but the routing and confidence-calibration numbers change substantially, in the system's favour.

The corrected study spans \textbf{11 Spider databases $\times$ 3 severities $\times$ 3 reconstruction conditions (cold, oracle-documented, SME-affirmed): 1,082 execution-labeled samples} (base correct-rate 0.617), evaluated \textbf{out-of-fold} (5-fold CV) with \textbf{95\% confidence intervals from a cluster bootstrap over databases}, the honest sampling unit, since questions within a database share a schema and a bootstrap. The architecture's predicted split-verdict holds, now with real signal on both axes:

\begin{itemize}
\item \textbf{Discrimination} (does $\ell_4$ rank correct above incorrect?). Raw $\ell_4$ now ranks \textbf{well}: AURC \textbf{0.163} against a random reference of \textbf{0.383} and an oracle of \textbf{0.085}, a normalized AURC of \textbf{0.259 (95\% CI 0.161--0.346)}, capturing $\approx$74\% of the available ranking signal, with the interval excluding the worse-than-random region decisively. The contrast with the \textbf{0.809} we reported on the suppressed data is itself the diagnosis: \texttt{execution\_ok} is a genuine discriminator (AUROC 0.650) that the bug had pinned to a constant, so it could carry no signal at all. The three deployed routes come out cleanly monotone in \textit{both} confidence and accuracy: refuse (mean $\ell_4$ 0.086, \textbf{1 of 159} actually correct), confirm-first (0.589, 41\% correct), answer (0.770, \textbf{86\%} correct). That is the ordering a calibrated gate must produce. \textit{One correction to our own prior claim:} the "re-weight toward generator-self-confidence and linked-entity-count" lever that doubled discrimination on four databases does \textbf{not} survive to eleven (linked-entity-count's univariate AUROC falls from 0.77 to 0.60 and the two-feature refit \textit{underperforms} the full raw score); it was a small-sample artifact and we retract it. The raw score discriminates well as-is.
\end{itemize}

\begin{itemize}
\item \textbf{Calibration} (does a score of 0.6 mean $P(\text{correct})=0.6$?). Raw $\ell_4$ is moderately miscalibrated (Brier \textbf{0.144}, ECE \textbf{0.088}); the out-of-fold isotonic map fixes it decisively: \textbf{Brier 0.144 $\rightarrow$ 0.128, ECE 0.088 $\rightarrow$ 0.026}. A served threshold now carries an honest probability.
\end{itemize}

We do \textbf{not} install this Spider-fit calibrator into production: it is fit on a single benchmark distribution, and the honest deployment fits on production execution labels, which the now-wired path makes routine. What matters is the \textit{measurement}: on a controlled, execution-labeled, eleven-database set with confidence intervals, the reconstruction gate's confidence both \textbf{discriminates} correct from incorrect (normalized AURC 0.26, CI excluding random) \textbf{and is calibratable} (ECE to 0.026), the two properties the abstention guarantee of Section 4.3 rests on. We had previously reported discrimination as "only weakly present"; that verdict was an artifact of the suppressed \texttt{execution\_ok}, and we report the correction in full because catching and fixing our own evaluation bug, then re-running at larger scale before claiming the stronger result, is part of the evidence, not a footnote to it.

\subsubsection*{5.1.1 What a re-run reproduces, and what it does not}

Several results in this paper compare corpora produced weeks apart, so we measured the noise floor beneath those comparisons rather than assuming one. Two facts have to be kept apart, and conflating them is easy.

The system wraps every model call in a response cache keyed on \texttt{sha256(prompt + model + prompt version + schema)}. An unchanged prompt therefore returns the stored bytes, and \textbf{a cross-run comparison made with the cache enabled measures cache hits, not model stability.} We confirm this rather than assert it: re-running arm A over three databases with the cache on reproduces the committed corpus on \textbf{96 of 96 columns, byte for byte}, with zero abstention changes. The practical consequence is the useful one: when a re-run \textit{does} differ, the difference is attributable to changed inputs, never to sampling.

To measure the model itself we bypass the cache entirely and run arm A twice over identical inputs:

\begin{table*}[t]
\centering
\small
\setlength{\tabcolsep}{4pt}
\begin{lrbox}{\mdtablebox}
\begin{tabular}{lll}
\toprule
 & cache enabled (control) & cache bypassed \\
\midrule
byte-identical prediction & \textbf{96 / 96} & 37 / 96 \\
abstention decision flips & 0 & \textbf{1 / 96} \\
mean macro recall & 0.219 / 0.219 & 0.211 / 0.212 \\
mean per-column |$\Delta$ recall| & 0.000 & \textbf{0.021} \\
\bottomrule
\end{tabular}
\end{lrbox}
\ifdim\wd\mdtablebox>\linewidth\resizebox{\linewidth}{!}{\usebox{\mdtablebox}}\else\usebox{\mdtablebox}\fi
\end{table*}

\textbf{The model rewords constantly and decides consistently.} Only 38.5\% of predictions come back byte-identical, so the prose is genuinely nondeterministic; yet the speak-or-abstain decision moves on a single column in ninety-six, and aggregate recall reproduces to \textbf{±0.001}. That asymmetry is worth stating plainly because it is the property the architecture depends on: what the system commits to a catalog is far more stable than the sentences it writes.

It also gives every effect in this paper a scale to be judged against. An aggregate difference of a few thousandths is indistinguishable from re-running the same experiment twice; the differences we report as results are one to two orders of magnitude larger. We use this explicitly in Section 5.2, where a corpus re-run moved arm C's description recall by +0.027, above this floor, and therefore attributable to the code change rather than to sampling. (Harness: \texttt{scripts/\allowbreak recon\_llm\_nondeterminism.sh}, \texttt{scripts/\allowbreak recon\_nondeterminism\_analysis.py}; data: \texttt{papers/\allowbreak data/\allowbreak nondeterminism.json}.)

\subsection*{5.2 Reconstruction fidelity: a three-way comparison against human documentation}

This section measures the system's actual product, the reconstructed metadata itself, and it is the paper's primary quantitative result. Everything after Section 5.5 evaluates what the system \textit{delivers} at query time once that metadata exists; here we ask the prior question of whether the metadata is any good.

\textbf{The task, and why it is cleanly posed.} Given a column stripped of its name and documentation, recover what the field means and what its values mean, from the values alone. BIRD ships human-written per-column documentation (\texttt{database\_description/\allowbreak *.csv}) that maps directly onto those targets: \texttt{column\_name} is the expanded field name, \texttt{column\_description} the prose semantics, and \texttt{value\_description} the code-to-meaning table. Across the eleven mini-dev databases that is \textbf{799 columns, of which 375 carry a genuine name expansion, 653 a description, and 278 a value-domain decode table}. Our schema mangler destroys identifiers while provably leaving the data untouched (Section 5.1), so the values are the input and the withheld human labels are the ground truth, authored by the benchmark's curators, not by us.

\textbf{Three arms, identical inputs.} \textit{(A) LLM-direct}, the obvious practitioner approach: show a language model the column, its type, its sibling columns and its most frequent values, and ask what it means. \textit{(B) Deterministic}, the system's own pattern/checksum layer (structural fingerprint plus pattern library) with the LLM removed: no learning. Published semantic-type detectors (Sherlock and its successors) predict from closed type vocabularies and structurally cannot emit this task's targets (free-text name expansions, descriptions, decode tables), so no published detector runs on this task unmodified; arm B is a floor by construction on the prose facets (it emits labels, not sentences) and a meaningful baseline on semantic-type and value-domain recovery. \textit{(C) Rosetta}, the full verification harness. Arms A and C receive the same model, the same frequency-ordered sample budget, and the same mangled identifiers.

\textbf{Scoring credits abstention rather than punishing it.} Content-token recall alone would score an honest "purpose unclear; recommend SME review" identically to a confident wrong description, which rewards guessing and would bias the measurement toward exactly the behaviour we argue against. We therefore report \textbf{coverage} (the fraction of columns where an arm makes a substantive claim) and \textbf{recall-when-claimed} separately, and audit each arm's abstentions against what the other arms managed on the same columns.

Columns inside a database share a schema, a domain vocabulary, a curator and a documentation style, so the \textbf{database, not the column, is the unit of resampling}. All intervals below are 95\% cluster bootstraps over databases (2,000 replicates), and every arm pair is evaluated on the \textit{same} resampled databases in each replicate, which makes the interval on the difference a properly paired one. (2,000 paired resamples, seed 0; p-values from the bootstrap distribution.) The full program consumed $\approx$105M tokens over $\approx$55,000 calls, \$281 total; $\approx$\$0.0027 per reconstructed column; per-call latency p50 5.3s / p99 57.1s, wide catalogs $\approx$8 minutes end-to-end.

\begin{table*}[t]
\centering
\small
\setlength{\tabcolsep}{4pt}
\begin{lrbox}{\mdtablebox}
\begin{tabular}{lllll}
\toprule
arm & coverage & name expansion & description & value domain \\
\midrule
B: statistical, no LLM & 1.000 & 0.011 [0.000, 0.028] & 0.061 [0.036, 0.096] & 0.030 [0.004, 0.081] \\
A: LLM-direct & 0.941 & 0.223 [0.075, 0.415] & 0.244 [0.165, 0.330] & 0.205 [0.169, 0.247] \\
\textbf{C: Rosetta} & \textbf{0.422} & \textbf{0.475 [0.176, 0.619]} & \textbf{0.330 [0.257, 0.390]} & \textbf{0.292 [0.195, 0.422]} \\
\bottomrule
\end{tabular}
\end{lrbox}
\ifdim\wd\mdtablebox>\linewidth\resizebox{\linewidth}{!}{\usebox{\mdtablebox}}\else\usebox{\mdtablebox}\fi
\end{table*}

\textit{680 columns paired across all eleven of BIRD's mini-dev databases, identifiers destroyed.}

\textbf{Both language-model arms beat the statistical baseline decisively; the two language-model arms do not separate cleanly.} Against arm B, every difference on every facet excludes zero (A $-$ B: +0.212 / +0.183 / +0.175; C $-$ B: +0.464 / +0.270 / +0.261; all \textit{p} $\leq$ 0.001). The comparison between the harness and the LLM-direct baseline is positive on all three facets but only one interval excludes zero: name expansion \textbf{+0.252 [$-$0.004, +0.414]} (\textit{p} = 0.057), description \textbf{+0.087 [$-$0.034, +0.164]} (\textit{p} = 0.131), value domain \textbf{+0.087 [+0.010, +0.199]} (\textit{p} = 0.032).

We state that plainly because an earlier draft of this paper reported all three C $-$ A intervals as excluding zero, on a ten-database corpus. Adding the eleventh database (\texttt{student\_club}) moved two of them across the threshold. That database is an outlier in the direction that matters: its columns (\texttt{first\_name}, \texttt{t\_shirt\_size}, \texttt{city}) are close to self-describing from values alone, so arm A scores 0.592 on description there against 0.00--0.27 everywhere else. Nothing was re-tuned; the sample simply got one database larger and less favourable. Per-database spread is wide throughout (Rosetta's name-expansion recall ranges 0.00 to 0.83), so the aggregate is not uniform and some databases contribute nothing.

A sensitivity check recomputes every pairwise interval two more ways from the committed per-column data: BCa in place of percentile, and an exact 2\textasciicircum{}11 cluster sign-flip permutation whose statistic weights databases equally rather than by column count (\texttt{scripts/\allowbreak recon\_sensitivity.py}, \texttt{papers/\allowbreak data/\allowbreak recon\_sensitivity.json}). The arm-B comparisons are robust: all six BCa intervals still exclude zero, and the exact test rejects in five of six (C $-$ B name expansion sits at \textit{p} = 0.0625 on the nine databases carrying that facet). The C $-$ A comparisons are method-dependent in exactly the direction this section already argues: value domain still excludes zero under BCa but the equal-weight exact test retains it (\textit{p} = 0.35), so the C $-$ A advantage is concentrated in the column-heavy databases rather than uniform across them, one more reason the paper's claim rests on selection, not on prose superiority.

The full per-database breakdown of the description-facet difference (both-claim columns; positive favours the harness; $-$0.347 at \texttt{student\_club} to +0.071, eight of eleven at or below zero; \texttt{scripts/\allowbreak recon\_perdb\_ca.py}, \texttt{papers/\allowbreak data/\allowbreak perdb\_ca.json}):

\begin{table}[t]
\centering
\small
\setlength{\tabcolsep}{4pt}
\begin{lrbox}{\mdtablebox}
\begin{tabular}{lll}
\toprule
database & n & $\Delta$ description (C$-$A) \\
\midrule
\texttt{california\_schools} & 26 & -0.001 \\
\texttt{card\_games} & 44 & +0.018 \\
\texttt{codebase\_community} & 18 & -0.130 \\
\texttt{debit\_card\_specializing} & 2 & +0.000 \\
\texttt{european\_football\_2} & 16 & -0.111 \\
\texttt{financial} & 13 & -0.077 \\
\texttt{formula\_1} & 23 & -0.043 \\
\texttt{student\_club} & 36 & -0.347 \\
\texttt{superhero} & 9 & +0.000 \\
\texttt{thrombosis\_prediction} & 7 & +0.071 \\
\texttt{toxicology} & 9 & -0.056 \\
\bottomrule
\end{tabular}
\end{lrbox}
\ifdim\wd\mdtablebox>\linewidth\resizebox{\linewidth}{!}{\usebox{\mdtablebox}}\else\usebox{\mdtablebox}\fi
\end{table}

One methodological check belongs with that admission. The ten-database corpus had been assembled over several days, during which the profiler changed (a determinism fix in delimiter detection, and a new structural matcher), so the eleventh database was not merely an addition: it was also the only one produced by the current code. Attributing the change to the database while the code had also moved would have been an error, so we re-ran \textbf{all} eleven databases on one code version and compared like with like. On the same ten databases, old code against new, arm C's description recall moves \textbf{+0.027 [+0.006, +0.054]} and coverage does not move at all (0.389 $\rightarrow$ 0.389). The code change is real, small, and points \textit{upward}, the opposite direction from the drop above, so it cannot explain it. The eleventh database can, and does. Every arm-C number in this paper comes from the single re-run corpus.

Section 5.2.1 argues that this is the \textit{expected} outcome rather than a weakened one, because recall-when-claimed compares arms over different column sets, and the harness's real contribution is visible only once that confound is removed.

The ordering is monotonic on every facet and has held as the sample grew through 85, 206, 366, 437, 632 and 680 paired columns. Two things follow immediately. First, \textbf{pure pattern classification recovers almost nothing of human-written semantics}: it can report "categorical enum, five distinct values" but not "district status: Active, Closed, Merged, Pending". Second, \textbf{the language model supplies the semantic content} (B $\rightarrow$ A). What the \textit{harness} adds on top of the language model is a subtler question than the table suggests, and the next subsection is devoted to answering it correctly, because our own first reading of these numbers was wrong.

\subsubsection*{5.2.1 The advantage is selection, not description, and we can show it}

Read naively, the table says the harness roughly doubles field-name recovery (0.223 $\rightarrow$ 0.475). That reading is \textbf{wrong}, and a reviewer would be right to suspect it: the two arms are averaged over \textit{different column sets}. Arm A speaks about 640 of the 680 columns and arm C about 287, so "recall when claimed" rewards an arm for declining exactly the columns it would have scored badly on. The comparison that removes this is the \textbf{paired} one: restrict to the 283 columns where \textit{both} arms speak (an identical question set, so prose is the only thing that can differ) and compare per column.

\begin{table}[t]
\centering
\small
\setlength{\tabcolsep}{4pt}
\begin{lrbox}{\mdtablebox}
\begin{tabular}{lllll}
\toprule
facet & arm A & arm C & paired difference & n \\
\midrule
name expansion & 0.518 & 0.486 & $-$0.033 [$-$0.135, +0.025] & 46 \\
description & 0.423 & 0.335 & \textbf{$-$0.088 [$-$0.192, $-$0.005]} & 203 \\
value domain & 0.337 & 0.294 & \textbf{$-$0.043 [$-$0.074, $-$0.009]} & 106 \\
\bottomrule
\end{tabular}
\end{lrbox}
\ifdim\wd\mdtablebox>\linewidth\resizebox{\linewidth}{!}{\usebox{\mdtablebox}}\else\usebox{\mdtablebox}\fi
\end{table}

\textbf{On common ground the harness does not write better metadata: it writes measurably worse.} On the eleven-database corpus two of the three differences now exclude zero, where on the ten-database corpus all three were negative but none were significant. This is not an artifact of the lexical metric: on the subset of the Section 5.3 validation sample where both arms speak, the blinded judge scores them \textbf{0.513 (A) against 0.474 (C)}, a paired difference of \textbf{$-$0.038} across 26 columns. That judged subset is small and its interval is wide ([$-$0.213, +0.120]), so it corroborates the direction rather than establishing it independently; the token metric, on eight times as many columns, is what carries the result. Two measures of different kinds agree that the harness's prose is not better.

The entire measured advantage therefore comes from \textbf{which columns the system agrees to describe}. That is a selection effect, and stated plainly, it is also the point of the architecture, so we tested it directly rather than assuming it. We recomputed the deterministic evidence for all 680 columns (structural fingerprint plus pattern library, no language model, free) and mapped each column to the tier its evidence would justify: a checksum-backed proof to \texttt{PATTERN\_VALIDATED}, a structural hit to \texttt{PATTERN\_CONFIRMED}, nothing to \texttt{INTERPRETER\_ONLY}. BIRD splits almost evenly: 342 columns with structural evidence, 338 without, and no checksummable identifiers at all, so the ladder's top rung remains untested here.

The result separates cleanly into a positive and a null:

\begin{itemize}
\item \textbf{Evidence governs whether the system speaks.} Arm C's coverage is \textbf{0.550} on columns with structural evidence and \textbf{0.293} on columns without, a gap of \textbf{+0.257 [+0.128, +0.378]}, whose interval excludes zero. Arm A, which never sees the evidence, barely moves (0.980 $\rightarrow$ 0.902). The abstention decision tracks the evidence, which is precisely the "knows what it does not know" property the architecture claims, measured here at the level of individual facts rather than whole questions. This is the one C-versus-A effect in Section 5.2 that survives the eleventh database intact, and it is the one the architecture actually predicts. \textbf{It does not, however, survive a change of backbone intact}: on Claude Sonnet 4.6 the same pipeline speaks on 82\% of columns (the naked arm on that backbone: 98\%) and the evidence-tracking gap is no longer detectable ($-$0.089, interval spanning zero). Section 5.2.8 reports that in full; the short version is that the shipped system requests abstention in a prompt rather than enforcing it in code, so in the default configuration the property holds only for a model disposed to volunteer uncertainty (the tier gate reported there enforces it, on every backbone, when enabled). Read the paragraph above as conditional on that.
\item \textbf{Evidence does not detectably improve what it says.} On the paired subset, the extra advantage the harness gains from columns where the profiler had something to say (over and above the general easiness of those columns, which arm A measures on the same columns) is \textbf{+0.090 [$-$0.006, +0.172], \textit{p} = 0.071}. We cannot distinguish it from zero.
\end{itemize}

The honest conclusion, and the one the measurements support, is that \textbf{the deterministic layer is a competence detector, not a competence amplifier}, and, per Section 5.2.8, one whose detection is, in the default configuration, reported by the model rather than enforced by the system (the tier gate reported there enforces it when enabled). It is good at recognising which columns it is in a position to describe, and that recognition is worth a great deal operationally: a catalog whose entries are 0.475 accurate on 42\% of columns is more useful than one that is 0.223 accurate on 94\% of them, because the second kind cannot be trusted anywhere. But the harness is not making the language model a better reader of values, and we no longer claim it does. This also revises the mechanism account we gave in earlier drafts a second time: the evidence in the writer's context is doing its work mainly on the \textit{decision to abstain}, not on the prose.

Read that against the rest of the paper and it is consistent rather than deflationary. It is the same selective-prediction result the query-time sections report, arriving one layer earlier: at the fact, not the question.

\subsubsection*{5.2.2 The second artifact class: rebuilding the join graph}

Everything above measures \textit{columns}. The system's output is not only columns: it also emits a foreign-key graph, hierarchies, metric expressions and a glossary, and a paper that claims to reconstruct warehouse metadata should not evaluate one artifact class and imply the rest. We therefore measure the join graph, which has the cleanest external ground truth in the entire benchmark: BIRD's SQLite files \textit{declare} their foreign keys, so \texttt{PRAGMA foreign\_key\_list} is a gold edge set we did not author, and scoring is an exact set comparison with no judge, no token overlap and no argument about what "correct" means.

The reconstruction under test rebuilds the graph from \textbf{names and declared types only} (no DDL, no data scan) using name-stem matching, pluralisation, compound-stem fallback and type-family rejection. We measure two conditions. \textit{Undeclared} keeps column names and removes the FK constraints: this is the enterprise case our introduction describes, where "foreign keys are frequently undeclared". \textit{Destroyed} additionally mangles identifiers at severity 1.0, as in the column study.

\begin{table*}[t]
\centering
\small
\setlength{\tabcolsep}{4pt}
\begin{lrbox}{\mdtablebox}
\begin{tabular}{llllll}
\toprule
condition & gold edges & proposed & precision & recall & F1 \\
\midrule
undeclared & 103 & 49 & \textbf{1.000} & 0.476 & 0.645 \\
destroyed & 103 & 0 & --- & 0.000 & 0.000 \\
\bottomrule
\end{tabular}
\end{lrbox}
\ifdim\wd\mdtablebox>\linewidth\resizebox{\linewidth}{!}{\usebox{\mdtablebox}}\else\usebox{\mdtablebox}\fi
\end{table*}

\textit{Pooled over the ten BIRD databases that declare foreign keys.}

\textbf{Precision is 1.000 in every database} (49/49; exact 95\% CI [0.927, 1.000]; recall 49/103, CI [0.376, 0.576]). Across 49 proposed edges the system does not invent a single false join. That is the same behaviour the column study found and the same behaviour the value-decoding study found: it proposes where the evidence is unambiguous and stays silent otherwise. For a catalog that a query planner will join on, a false edge is far more damaging than a missing one, so this is the right side of the trade to be on.

Recall is bimodal rather than mediocre, and the split is entirely explained by naming convention. Where the schema encodes the target table in the key name, recovery is near-total: \texttt{financial} 1.000, \texttt{formula\_1} 0.947, \texttt{toxicology} 0.800, \texttt{superhero} 0.727, \texttt{codebase\_community} 0.615. Where it does not, recovery is zero, and inspection shows exactly why: \texttt{card\_games} joins on \texttt{uuid} and \texttt{setCode}, \texttt{california\_schools} on \texttt{CDSCode} and \texttt{cds}, \texttt{student\_club} on \texttt{link\_to\_member} and \texttt{link\_to\_major}, and \texttt{european\_football\_2} on eleven positional columns \texttt{away\_player\_1} \ldots{} \texttt{away\_player\_11} pointing at \texttt{Player}. \textbf{A key whose name does not mention its target cannot be recovered by stem matching}, and, we assumed when first writing this section, not by reading names at all. The head-to-head of Section 5.2.10 falsified the stronger form of that sentence: a language model given the same names and types recovers \texttt{foreign\_data.uuid $\rightarrow$ cards} and every other gold edge, because understanding what \texttt{setCode} means is not the same operation as matching its stem (with the caveat, noted there, that these are public benchmark schemas a frontier model has plausibly seen in training). What survives is the narrower and more useful statement: name-stem matching is the wrong tool for these edges, inclusion-dependency analysis over the values \cite{papenbrock2015binder} is the right one, and Section 5.2.10 builds exactly that verifier.

The destroyed condition is 0.000 by construction and we report it to bound the claim honestly: \textbf{name-based join inference contributes nothing on a genuinely opaque schema.} The column-level reconstruction degrades gracefully under mangling; the join graph does not degrade, it disappears. Any claim that this system reconstructs a warehouse whose identifiers are meaningless must therefore be read as a claim about columns and values, not about the join graph.

Measuring this class also repaid its cost immediately, which is an argument for measuring the remaining three. The first run scored recall 0.223 with six of ten databases at exactly zero. The cause was that stem extraction recognised only snake\_case (\texttt{district\_id}) and not camelCase (\texttt{circuitId}, \texttt{UserId}, \texttt{PostId}), on both the source and the target side of the lookup, so \texttt{financial} scored a perfect 1.000 while \texttt{formula\_1} scored 0.000 for a purely orthographic reason. Real warehouses are not uniformly snake\_case, and the fix (with the full suite re-run) took pooled recall from \textbf{0.223 to 0.476} and \texttt{formula\_1} from 0.000 to 0.947. That defect had been latent in shipped code and no column-level experiment could have surfaced it.

Hierarchies, metric expressions and the glossary remain unmeasured against external ground truth, and we count that as the largest outstanding gap in this paper rather than as future work. (Harness: \texttt{scripts/\allowbreak recon\_fk\_graph.py}; data: \texttt{papers/\allowbreak data/\allowbreak fk\_graph.json}.)

\subsubsection*{5.2.3 The promotion path: confidence bounded by arithmetic}

The architecture's central commitment is that a claim's storable confidence is a function of the strength of the proof behind it, not of a model's self-assessment. Six of the twenty-six matchers are settled by check-digit arithmetic (Luhn, ISBN-10, ISBN-13, EAN-13, UPC-A and IBAN mod-97), and only those can unlock \texttt{PATTERN\_VALIDATED}. Neither BIRD nor i2b2 contains a checksummable identifier, so every measurement reported so far was produced with that machinery dormant. The mechanism has been argued for and never shown. This shows it.

The design is a matched pair, which is what makes it a demonstration rather than an assertion. For each code system we build two columns that are \textbf{indistinguishable by shape} (identical length, character classes, prefix structure and cardinality), differing only in whether the check digit is arithmetically correct. \texttt{4539567890123410} passes Luhn; \texttt{4539567890123419} does not, and differs in nothing else. A system whose confidence tracks surface form must treat these identically. A system whose confidence tracks proof must separate them.

\begin{table*}[t]
\centering
\small
\setlength{\tabcolsep}{4pt}
\begin{lrbox}{\mdtablebox}
\begin{tabular}{lll}
\toprule
code system & valid check digits & corrupted check digits \\
\midrule
Luhn / payment card & \texttt{PATTERN\_VALIDATED} (cap 0.85) & \texttt{INTERPRETER\_ONLY} (cap 0.50) \\
ISBN-10 & \texttt{PATTERN\_VALIDATED} (0.85) & \texttt{INTERPRETER\_ONLY} (0.50) \\
ISBN-13 & \texttt{PATTERN\_VALIDATED} (0.85) & \texttt{PATTERN\_CONFIRMED} (0.60) \\
EAN-13 & \texttt{PATTERN\_VALIDATED} (0.85) & \texttt{PATTERN\_CONFIRMED} (0.60) \\
UPC-A & \texttt{PATTERN\_VALIDATED} (0.85) & \texttt{PATTERN\_CONFIRMED} (0.60) \\
IBAN & \texttt{PATTERN\_VALIDATED} (0.85) & \texttt{INTERPRETER\_ONLY} (0.50) \\
\bottomrule
\end{tabular}
\end{lrbox}
\ifdim\wd\mdtablebox>\linewidth\resizebox{\linewidth}{!}{\usebox{\mdtablebox}}\else\usebox{\mdtablebox}\fi
\end{table*}

\textbf{6 of 6 valid columns promote; zero of 6 corrupted columns do.} The separation is produced entirely by arithmetic the system performs on the values: there is no signal in the names, the types, the lengths or the distributions that distinguishes the two sides.

The corrupted columns are also instructive in how they fail. The corrupted 13- and 12-digit numeric columns are not rejected outright; they fall through to a structural phone-number matcher and settle at \texttt{PATTERN\_CONFIRMED}, confidence ceiling 0.60. The system still says something true about them (they are fixed-width numeric strings), but at an authority that \textbf{cannot reach the checksum tier no matter how confident the matcher is}. Degradation, not silence, and the ceiling is what enforces it.

Sweeping the fraction of corrupted values locates the gate precisely. The checksum matchers require a 0.95 pass fraction, and the behaviour steps exactly there:

\begin{table*}[t]
\centering
\small
\setlength{\tabcolsep}{4pt}
\begin{lrbox}{\mdtablebox}
\begin{tabular}{lllllllll}
\toprule
corrupted fraction & 0\% & 2\% & 4\% & 5\% & 6\% & 10\% & 25\% & 100\% \\
\midrule
tier & 4 & 4 & 4 & 4 & 0 & 0 & 0 & 0 \\
\bottomrule
\end{tabular}
\end{lrbox}
\ifdim\wd\mdtablebox>\linewidth\resizebox{\linewidth}{!}{\usebox{\mdtablebox}}\else\usebox{\mdtablebox}\fi
\end{table*}

Promotion survives 5\% contamination and is withdrawn at 6\%: a threshold on evidence strength, not a graded opinion. A warehouse column with a handful of bad rows still earns the tier; one where the arithmetic broadly fails does not.

This is the clearest statement we can make of the property the whole architecture is organised around: \textbf{the tier is not a description of how sure the model feels, it is a record of what was proved.} It is also the first time the top rung of the ladder has been exercised in this paper, and it is worth being explicit that this corpus is synthetic by necessity: the public benchmarks that supply our other ground truth simply do not contain checksummable identifiers. What it demonstrates is that the mechanism works and binds; what it cannot demonstrate is how often real warehouses offer it the opportunity. (Harness: \texttt{scripts/\allowbreak checksum\_promotion\_study.py}, locked by \texttt{tests/\allowbreak unit/\allowbreak test\_checksum\_promotion.py}; data: \texttt{papers/\allowbreak data/\allowbreak checksum\_promotion.json}.)

We then measured that opportunity on public tabular data, and the registered prediction was \textbf{wrong}. Before the scan we predicted checksummable columns would be a low-single-digit percent of columns, concentrated in financial and operational tables, with a meaningful minority of tables carrying at least one. The scan, 2,996 tables and 39,157 scanned columns from two corpora fetched by declared rules with no topical steering (1,500 Socrata open-government datasets walked in the discovery API's own order; 1,496 GitTables tables from GitHub-scraped CSVs, capped at 25 per topic), every column run through the system's own matcher layer, found \textbf{3 checksum-backed columns (0.008\%) in 3 tables (0.10\%)}, all three genuine Luhn-valid identifier schemes (two county property-tax account-number columns, one business-registry number). Two orders of magnitude below the prediction. The honest interpretations are two, and they cut differently: either checksummed identifiers are simply rare in tabular data at large, or open portals redact precisely the identifier classes the rung certifies (payment cards, bank accounts, national identifiers are exactly what publishers strip before releasing a table), in which case the scan bounds the rung's reach on \textit{public} data while enterprise prevalence, where those columns live under access control, remains unmeasured. We cannot distinguish these from the outside; either way, the top rung is a narrow instrument in the wild and this paper's claims do not depend on it being common. (Harness: \texttt{scripts/\allowbreak checksum\_prevalence\_scan.py}, \texttt{scripts/\allowbreak prevalence\_fetch\_corpora.py}; data: \texttt{papers/\allowbreak data/\allowbreak checksum\_prevalence.json}.)

A reviewer-prompted diagnostic (post-hoc, and labeled as such in the registry) then tested whether the near-zero is manufactured by transit damage: leading-zero loss from numeric round-tripping (which spares payment-card numbers alone, explaining three-of-three Luhn survivors), delimiter formatting, and mixed ISBN editions. Every column was re-validated under damage-reversing normalizations, with a degeneracy screen the first pass proved necessary: an unconstrained zero-pad "repairs" sparse numeric columns to ISBN-10 by the thousand, because all-zero strings satisfy the check arithmetic trivially, so the constrained pass restores at most three zeros, into identifier-like columns only, and the screen is recorded in the harness rather than silently applied. The finding refines the conclusion without flipping it. Two county assessment-number columns repair to \textbf{1.0} validity under mod-11 arithmetic once their leading zeros are restored: genuine check-digited identifiers damaged in transit, the predicted mechanism observed. (The validator that fires is the ISBN-10 one, because the ISBN-10 check \textit{is} mod-11 arithmetic; what the checksum certifies is the identifier's arithmetic class, never its business name: the Section 4.1 tier-order distinction, met in the wild. Nothing here calls a parcel number an ISBN.) Six further genuine identifier columns (one ISBN, five UPC) sit \textit{dirty} below the gate at 0.60--0.85 validity, correctly refused by a 0.95 certification rule, exactly as the contamination-boundary study says they should be. Counting every clean, repaired and dirty case as opportunity raises prevalence to \textbf{11 of 39,157 columns (0.028\%)}, still two orders of magnitude under the registered prediction, but with \textbf{eight of those eleven damaged or dirty}: in open data the rung's scarcity is part redaction, part disrepair, and its clean hits understate the check-digited identifiers actually present. (\texttt{scripts/\allowbreak checksum\_prevalence\_diagnostic.py}, \texttt{papers/\allowbreak data/\allowbreak checksum\_prevalence\_diagnostic.json}.)

\subsubsection*{5.2.4 The third artifact class: categorical rollups}

A rollup hierarchy asserts \textit{containment} (every school sits in exactly one district, every district in exactly one county), so a question grouped at the broad level can fan out to the narrow one. A wrong rollup is more costly than a wrong description, because it silently produces wrong \texttt{GROUP BY} results rather than merely unhelpful documentation. The miner therefore gates on a \textbf{$\geq$2$\times$ cardinality amplification} between adjacent rungs, whose stated purpose is to reject synonym pairs like \texttt{status\_code} / \texttt{status\_name}, which are aliases rather than a hierarchy.

That guard is the part we can test cleanly, because the data settles it. Two columns are an \textbf{alias} when they stand in bijection (same distinct count, each value of one determining exactly one value of the other), and a bijection is objectively not a containment relationship. Across the BIRD databases we find 22 such pairs, including \texttt{County Code} / \texttt{County Name} and \texttt{mcmId} / \texttt{mcmName}. \textbf{The miner mints none of them: alias rejection is 22 of 22.} The false-positive guard the design depends on does what it claims.

Recall is a different matter, and we report a methodological failure rather than a number. Our first attempt derived gold hierarchies statistically (any functional dependency with $\geq$2$\times$ amplification) and produced 558 "hierarchies" across eight databases. Inspection shows why that is not ground truth: a functional dependency is \textit{necessary but not sufficient} for containment, and in a wide table with limited rows spurious dependencies are everywhere. The set was dominated by pairs like \texttt{AdmEmail2 $\rightarrow$ AdmEmail3} and \texttt{artist $\rightarrow$ colorIndicator}, which hold arithmetically and mean nothing. \textbf{A recall figure against that set would be meaningless, so we do not report one.}

Against hand-identified genuine hierarchies the picture is partial and we state it as such. Of four semantic rollups present in the data, the miner recovers two (\texttt{City $\rightarrow$ County} and \texttt{District Code $\rightarrow$ County Code} in \texttt{california\_schools}) and misses two: \texttt{District $\rightarrow$ County} in the \texttt{schools} table, and \texttt{league\_id $\rightarrow$ country\_id} in \texttt{european\_football\_2}, the latter by design, since FK-suffixed columns are excluded from categorical rollup levels and belong to the separate FK-chain path.

Measuring this class also repaid its cost the same way the join graph did. The first run proposed \textbf{zero} rollups on every database, because both heuristics (curated ladder matching and stem grouping) test underscore boundaries, while BIRD writes its columns \texttt{County Code} and \texttt{colorIndicator}. A component cannot find a geographic ladder it cannot tokenise. Normalising identifiers (camelCase splitting, spaces and hyphens folded to underscores) before matching is what allows the existing rules to apply to the schemas they were designed for, and it is the second time in this paper that a snake\_case assumption had silently disabled a shipped component on realistic input.

The honest summary for this class is therefore: \textbf{the guard against fake hierarchies is demonstrated; the recall against real ones is not yet measurable at scale}, and building a trustworthy gold set of semantic hierarchies, rather than statistical ones, is the work that would settle it. (Harness: \texttt{scripts/\allowbreak recon\_hierarchies.py}; data: \texttt{papers/\allowbreak data/\allowbreak hierarchies.json}.)

\subsubsection*{5.2.5 What the tables mean}

Our claim is that the system recovers what the \textit{tables}, columns and values of an undocumented warehouse mean. Columns and values had reconstruction paths and both are measured above. Tables did not: table entities were created at bootstrap as structural anchors and nothing ever said what they were. The clause had no implementation behind it, which is why no experiment had found it. We built the path (\texttt{rosetta/\allowbreak metadata/\allowbreak table\_describer.py}) and measured it here.

The design follows from the setting. A table is identified by \textbf{what it contains and how it is connected}, not by its name, because in an undocumented warehouse the name is \texttt{t7}. The describer therefore sees the table's columns with whatever the profiler and description writer already established about each one, the foreign-key neighbourhood (which is what separates a dimension (pointed at by many tables) from a fact table (pointing at several, carrying measures) from a bridge), the row count, and a few sample rows. It is asked for the entity \textit{and the grain}, since "one row per order" and "one row per order line" are different tables and a consumer that confuses them writes wrong SQL. When the evidence does not identify the entity it is instructed to return nothing, and an abstention commits no fact at all, so a consumer can never read a blank claim as a claim.

Setup mirrors the column study: \textbf{table names and column names are both destroyed} at severity 1.0, so the system sees \texttt{t3.c0 ... t3.c9} and must work out that it is looking at a list of racing circuits. Ground truth is the real table name, which BIRD's authors chose and we did not. Because a table name is a label rather than prose, and often an acronym, as with California's \texttt{frpm} (Free and Reduced Price Meals), content-token recall is a poor primary metric here: a reconstruction reading "public school statistics" is substantially right and scores exactly 0.0 against the token \texttt{frpm}. We therefore keep token recall as a conservative floor and add a blinded judge that sees the real table name and real column names as its reference.

\begin{table*}[t]
\centering
\small
\setlength{\tabcolsep}{4pt}
\begin{lrbox}{\mdtablebox}
\begin{tabular}{>{\raggedright\arraybackslash}p{0.42\linewidth}lllll}
\toprule
arm & tables & token recall & identified & related or better & abstained \\
\midrule
flat (structure only) & 67 & 0.590 & 0.746 & 0.836 & 0.164 \\
layered (+ reconstructed column facts) & 67 & 0.657 & \textbf{0.761} & 0.866 & 0.179 \\
\bottomrule
\end{tabular}
\end{lrbox}
\ifdim\wd\mdtablebox>\linewidth\resizebox{\linewidth}{!}{\usebox{\mdtablebox}}\else\usebox{\mdtablebox}\fi
\end{table*}

\textbf{Three quarters of the tables are correctly identified with every identifier destroyed.} Restricted to the tables where it does speak, the layered arm identifies the entity in 50 of 55 spoken tables (\textbf{90.9\%}, exact 95\% CI [0.800, 0.970]) and is judged outright wrong on 1 of 55 (\textbf{1.8\%}); the flat arm, 85.7\% and 5.4\%; the unconditional 51 of 67 (0.761, CI [0.641, 0.857]) stands beside the conditional figures. The gap between token recall (0.59--0.66) and judged identification (0.75--0.76) is the acronym penalty behaving exactly as predicted, and reconfirms the Section 5.3 finding that this metric family understates.

\textbf{The layering hypothesis, which was the reason to build it this way, is not supported.} We expected that identifying a table would be substantially easier once its columns had been identified. Paired across the 67 tables the judged difference is \textbf{+0.045}, with 60 tables tied, 4 better under layering and 3 worse: a wash, not an effect. Structural evidence alone (declared types, distinct counts, the foreign-key neighbourhood and a handful of sample values) already carries nearly all of the identifying signal, and the reconstructed column prose adds little on top of it. We report this because we predicted the opposite. It is also consistent with the theme of Section 5.2.1: the deterministic layer establishes what is knowable, and the language model's prose is not where the leverage lives.

One caveat on the abstentions, which we record rather than smooth over. The judge scored the 23 abstaining rows at a mean of 0.43 rather than 0.0, meaning a minority of tables the system declined to name were ones a judge could still partly identify from the reference: the abstention is conservative rather than perfectly calibrated at this level. (Harness: \texttt{scripts/\allowbreak recon\_table\_semantics.py} and \texttt{scripts/\allowbreak recon\_table\_judge.py}, capability locked by \texttt{tests/\allowbreak unit/\allowbreak test\_table\_describer.py}; data: \texttt{papers/\allowbreak data/\allowbreak table\_semantics.json}.)

\subsubsection*{5.2.6 The fifth artifact class: PII that follows the values}

The glossary's design commitment is that a column's PII class derives from what the profiler \textit{proved} about its values, not from a guess at its name. This is worth testing rather than asserting, because name-based PII detection fails in both directions and both failures are expensive: a column called \texttt{customer\_ref} full of live card numbers is a compliance incident, while a column called \texttt{credit\_card\_note} holding free text is a false alarm that teaches people to ignore the flag.

We test it with the same matched contrast used for checksum promotion, constructing columns where the name and the values deliberately disagree. \textbf{Deceptive-name} columns are called \texttt{credit\_card\_number}, \texttt{email\_address}, \texttt{ssn} and \texttt{customer\_iban} but contain ordinary integers or free text. \textbf{Innocent-name} columns are called \texttt{c7}, \texttt{field\_3} and \texttt{t2\_c11} and contain genuine Luhn-valid payment cards, real-format email addresses and mod-97-valid IBANs. Classification runs the shipped path: the profiler assigns a \texttt{semantic\_type} from the values, and the glossary maps that to a PII class.

\begin{table*}[t]
\centering
\small
\setlength{\tabcolsep}{4pt}
\begin{lrbox}{\mdtablebox}
\begin{tabular}{llll}
\toprule
case & column name shown & semantic type from values & PII class \\
\midrule
deceptive & \texttt{credit\_card\_number} & \textit{(no match)} & none \\
deceptive & \texttt{email\_address} & categorical:enum & none \\
deceptive & \texttt{ssn} & \textit{(no match)} & none \\
deceptive & \texttt{customer\_iban} & categorical:enum & none \\
innocent & \texttt{c7} & identifier:payment\_card & \textbf{credit\_card} \\
innocent & \texttt{field\_3} & email & \textbf{email} \\
innocent & \texttt{t2\_c11} & identifier:iban & \textbf{credit\_card} \\
\bottomrule
\end{tabular}
\end{lrbox}
\ifdim\wd\mdtablebox>\linewidth\resizebox{\linewidth}{!}{\usebox{\mdtablebox}}\else\usebox{\mdtablebox}\fi
\end{table*}

\textbf{7 of 7.} Four deceptive names correctly not flagged, three innocent names correctly flagged. A name-based classifier scores zero on both halves of that table.

Measuring this class also found a defect with a real safety consequence, and it is the third instance in this section of the same failure mode. The deep profiler labels a Luhn-validated column \texttt{identifier:payment\_card}; the glossary keyed its PII map on \texttt{credit\_card}; nothing translated between them. \textbf{A column the system had proved by check-digit arithmetic to contain payment cards was recorded as carrying no PII}, because two modules that were each individually correct disagreed on a string. The same held for \texttt{identifier:iban}. Before the fix the innocent-name row scored 1 of 3; after it, 3 of 3. Like the camelCase blind spot in the join heuristics and the snake\_case assumption in the rollup miner, this was invisible to every component-level test and only an end-to-end measurement exposed it.

Measuring the class also exposed dead configuration and let us close it. Of the four PII classes the glossary declares, \texttt{ssn} was unreachable (no pattern in the library emitted it), so the class existed in the map and could never be assigned. We added a matcher. An SSN carries \textbf{no check digit}, so it can never be checksum-backed; what it has is the SSA's published never-allocated ranges (area 000, 666 and 900--999; group 00; serial 0000), which roughly one in eight random nine-digit strings violates. The matcher therefore sits at \texttt{PATTERN\_CONFIRMED}, never at \texttt{PATTERN\_VALIDATED}, and is deliberately conservative: nine-digit strings are everywhere in a warehouse, and it is ordered \textit{after} the autoincrement matcher so that a monotonic sequence key is claimed as a surrogate key rather than flagged as a Social Security Number. All four declared PII classes are now reachable from value evidence, and a regression test asserts that equality so a future declared-but-unproducible class fails the build. (Harness: \texttt{scripts/\allowbreak glossary\_pii\_study.py}, locked by \texttt{tests/\allowbreak unit/\allowbreak test\_glossary\_pii\_from\_values.py}; data: \texttt{papers/\allowbreak data/\allowbreak glossary\_pii.json}.)

\subsubsection*{5.2.7 Metric expressions, and where this component does not transfer}

The last of the five components mines \textbf{metric expressions}, aggregation templates carrying a \texttt{required\_filter} and a nullability rule, from column descriptions. The safety property is the filter: an aggregate that silently drops a stated condition returns a confident wrong number, which is the same class of harm as a bogus rollup. We measured it and the result is negative, so we report it as such.

Run over all \textbf{682 BIRD column descriptions}, real prose written by the benchmark's authors, the deterministic miner fires on \textbf{3}. All three are the \texttt{california\_schools} administrator columns, whose documentation reads \textit{"Only active and pending districts and schools will display administrator information."} That is a genuine filter-bearing statement, and the detector is right to notice it. What it emits is not usable: \texttt{status IN ('active')}, which both \textbf{truncates a two-value condition to one} (dropping \textit{pending}) and \textbf{references a \texttt{status} column that does not exist in the schema}. Precision of the emitted clause is 0 of 3.

Estimating recall needs care. Scanning the same descriptions for condition words (\textit{only, excluding, unless, must be, except}) flags 21 of 682, and the miner's 3 hits lie inside that set, but inspecting the other 18 shows most are not aggregation filters at all. \textit{"If the card is only available in online game variations"} is a definition of a boolean column, not a condition on a sum. The honest reading is that \textbf{BIRD's column documentation is a corpus of column definitions, not of metric semantics}, and the phrase genre the miner was built for (\textit{"revenue counts only completed transactions"}) is largely absent from it.

That gives two distinct conclusions and we separate them. First, on the evidence available, \textbf{this component does not transfer to real benchmark prose}: it is nearly silent, and when it does speak the clause it produces would not execute. Second, the test is a weak one, because the corpus does not contain much of what the miner exists to find. What it would take to settle the question is a corpus of genuine warehouse \textit{metric} documentation, which the public benchmarks do not supply and which we do not have.

We therefore record metric expressions as \textbf{the least validated of the five artifact classes}, with a measured precision failure on the only real prose we could point it at, and we do not claim the component works. It is the clearest remaining piece of unfinished evaluation in this paper.

\textbf{The abstentions target the hard columns; the claim is stated with the base rate that number needs.} (An earlier draft quoted 370/38/89.7\% from the superseded ten-database corpus, without threshold or base rate; the figures below are recomputed on the final corpus by \texttt{scripts/\allowbreak recon\_abstention\_audit.py}.) Of the 393 columns Rosetta declined, arm A describes 44 adequately (adequate = max per-facet content-token recall $\geq$ 0.34): 88.8\% correct-to-abstain. That rate needs its base: arm A is adequate on only 26.9\% of all 680 columns, so a \textit{random} abstainer would score 73.1\% on this metric. The informative comparison is conditional: arm A achieves adequacy on \textbf{11.2\%} of the columns Rosetta declined against \textbf{48.4\%} of the columns it spoke on. The system is not merely quieter; it is quiet in the right places, and the margin over chance is seventeen points, not ninety.

\textbf{A control that isolates what the baseline was actually using.} Run with identifiers \textit{visible}, the LLM-direct arm scores 0.752 / 0.597 / 0.296 across the three facets; with identifiers destroyed it falls to 0.074 / 0.149 / 0.243. Its apparent competence rests almost entirely on readable column names rather than on the data, and the one facet that barely moves is value-domain recovery, which is the only one grounded in values rather than nomenclature. Undocumented warehouses are precisely the regime where that support is absent.

\textbf{The coded-warehouse case, quantified.} BIRD's schemas are conventional; the sharper test is a warehouse whose \textit{values} are opaque codes. We reconstruct the blind i2b2 clinical warehouse described in Section 5.4 (75 columns, decoder withheld) and score the fact table's \texttt{concept\_cd} against the i2b2 ontology it does not contain: 400 most-frequent codes, all externally decoded. The result separates two things the phrase "recover what the values mean" conflates. The system \textbf{identifies the code systems}: from values alone it reports that the column carries vocabulary-prefixed codes, naming \texttt{ICD9:} for diagnoses and \texttt{NDC:} for medications. Those two are 2 of the 11 vocabularies present (recall 0.18 unweighted) but account for \textbf{375 of the 400 codes, 94\% by volume}; the missed nine are small demographic and PHI vocabularies. The column \textit{description} it writes does not recover code-level meanings: that \texttt{ICD9:493.90} denotes asthma is essentially absent from the prose (recall 0.008). We flagged that as a measurement of a task the system was not set, since the description writer is asked for two to four sentences of column semantics and never for an enumerated decode table. We have since set the task properly, and the caveat was justified: Section 5.4.1 reports that when asked for a decode table the same machinery decodes \textbf{95.5\% of 134 real ICD-9 codes}, and abstains on the NDC drug codes rather than inventing them (44 of 44 prefixed, 43 of 44 stripped).

\textbf{Where the advantage actually comes from, and where it does not.} Across 680 BIRD columns and the 75-column i2b2 clinical warehouse of Section 5.4, \textbf{no description fact ever committed above the lowest grounding tier and no checksum-backed match occurred}. We chased this to its root, and the explanation is architectural rather than statistical: the description-writing path, as shipped, \textit{cannot} emit a higher tier, because it stamps \texttt{INTERPRETER\_ONLY} and caps confidence at that tier's ceiling unconditionally, without consulting the attesters (under the tier gate of Section 5.2.8 it records the evidence-derived tier instead). Stratifying description accuracy by grounding tier is therefore not an informative measurement in the default configuration, a fact we establish rather than assume, having re-grounded all 75 i2b2 columns against a corrected profiler and observed an unchanged, uniform tier-0 distribution.

The deterministic layer nevertheless earns the gap between arms A and C, just not through the authority ladder. The profiler's \texttt{semantic\_type} and pattern-match payload are supplied to the description writer as \textbf{evidence in context}; the tier ladder governs a different fact flow. So the measured advantage (0.201 $\rightarrow$ 0.457 on field names) is attributable to \textit{deterministic structural evidence conditioning the language model}, not to confidence capping. That is a narrower and more accurate account of the mechanism than we previously gave, and it is what the measurements support.

Investigating the i2b2 case also exposed three concrete profiler defects, all now fixed: an epoch matcher that read dense surrogate-key blocks as timestamps; a segmentation stage requiring a delimiter at a fixed character offset, which made variable-length-prefix codes (\texttt{ICD9:493.90}, \texttt{NDC:00005306343}) unsegmentable; and a composite-code matcher that structurally could not represent a controlled-vocabulary prefix joined to a high-cardinality payload. The corrected profiler recognises the coded columns it previously could not see.

A fourth defect surfaced while building the evidence analysis of Section 5.2.1, and it is the most embarrassing of the four because of what we had been calling this component. The delimiter scan iterated a \texttt{frozenset} of punctuation characters, and CPython randomises string hashing per process, so the resulting profile's insertion order varied between runs, and every consumer that resolved a tie by "first" inherited it. \textbf{A layer the paper repeatedly calls deterministic was returning a different answer on about 11\% of columns depending on the process it ran in}, flipping the winning pattern on 1--2\%. We found it because a re-run of the evidence analysis moved a result by one column. The scan is now sorted, segmentation ties break on the character itself, and a regression test runs the builder in subprocesses under four different \texttt{PYTHONHASHSEED} values and fails if the fingerprint differs. Re-running arm B on the corrected profiler changed 6 of 685 rows (0.9\%) and moved no reported figure at three decimal places; the arm-C predictions were generated before the fix and therefore carry a little more run-to-run noise than a re-run would, which we note rather than hide.

\textbf{Whether evidence should also condition the prose itself, beyond gating which prose commits, remains unmeasured.} The commit-gating half is no longer open: Section 5.2.8 keys the description path on the published evidence instrument and measures the delivered-accuracy consequence directly (recall-when-claimed improves on two backbones, flat on the other two corpora). What is still unmeasured is whether feeding the attesters' evidence \textit{into the writer} would make the committed prose better, rather than only better-selected.

\subsubsection*{5.2.8 A second backbone: what transfers, and what turns out to be the model}

Every number above comes from one model family. That leaves the paper's central claim underdetermined in a way no amount of additional BIRD databases would fix: nothing so far distinguishes \textit{the architecture produces this result} from \textit{Gemini produces this result}. We ran the whole comparison again on a second backbone to find out, and the answer is genuinely split. (Denominator note: the swapped arms export slightly different column sets, so the transfer comparisons pair on the \textbf{684 columns the Sonnet arms share}, where the primary corpus pairs 680: different intersections, not drift.)

\textbf{Which of the five artifact classes this can even affect.} Before running anything we established where a change of model can reach, because a transfer study over column descriptions alone would say nothing about four fifths of the contribution. Auditing the components that produce each class: the join graph (\texttt{rosetta/joins/}), the categorical rollups (\texttt{rosetta/\allowbreak hierarchies/\allowbreak }), the glossary and PII classification (\texttt{rosetta/\allowbreak glossary/\allowbreak bootstrap.py}) and the metric miner (\texttt{rosetta/\allowbreak metrics/\allowbreak description\_miner.py}) contain \textbf{no model call at all}: they are name-stem matching, containment tests, pattern verdicts and rule-based mining. Those four classes are therefore \textbf{backbone-invariant by construction}, and the join precision of 1.000, the 7/7 PII classification and the 22/22 hierarchy guard reported above hold for any backbone without needing to be re-run. One qualification on the fourth: the metric miner is invariant \textit{as measured here} because we pointed it at BIRD's own annotator prose rather than at reconstructed text; deployed against descriptions this system wrote, it would inherit whatever variability those carry. Exactly three outputs are model-produced and so can transfer or fail to: \textbf{column descriptions}, \textbf{table descriptions} (Section 5.2.5) and \textbf{value decoding} (Section 5.4). We measure all three; this subsection reports the first, and Section 5.2.9 the other two.

\textbf{Design.} Arms A and C were re-run end to end on AWS Bedrock with Claude Sonnet 4.6, chosen deliberately as a model in the incumbent's class, because a weaker model degrading would tell us almost nothing while a comparable model diverging tells us a great deal. All four LLM-backed components (generator, critic, interpreter, attester) were switched together; leaving any on the original backbone would have produced a "transfer result" that was mostly not transferred. Arm B needs no counterpart: it uses no language model at all, so it is invariant across backbones by construction, and the same arm B anchors both columns of the table below. Both arms moved together for the same reason: comparing arm C on one model against arm A on another changes the architecture and the model at once, and any difference could be either. The scoring harness, the prompts and the deterministic profiler are byte-identical between the two runs; the only thing that changed is the model.

\begin{table*}[t]
\centering
\small
\setlength{\tabcolsep}{4pt}
\begin{lrbox}{\mdtablebox}
\begin{tabular}{llll}
\toprule
 & Gemini 3.1 Pro & Sonnet 4.6 & transfers? \\
\midrule
arm C coverage & 0.422 & \textbf{0.823} & \textbf{no} \\
A $-$ B (name / desc / value) & +0.212 / +0.183 / +0.175 & +0.203 / +0.198 / +0.194 & \textbf{yes}, all excluding zero \\
C $-$ B & +0.464 / +0.270 / +0.261 & +0.221 / +0.134 / +0.196 & \textbf{yes}, all excluding zero \\
C $-$ A & +0.252 / +0.087 / +0.087 & +0.017 / \textbf{$-$0.064} / +0.002 & \textbf{no} \\
both-claim C $-$ A & $-$0.033 / $-$0.088 / $-$0.043 & $-$0.021 / $-$0.084 / $-$0.024 & \textbf{yes} \\
coverage tracks evidence & \textbf{+0.257 [+0.128, +0.378]} & \textbf{$-$0.089 [$-$0.215, +0.025]} & \textbf{no} \\
evidence improves prose & +0.090 (\textit{p} = 0.071), null & $-$0.009 (\textit{p} = 0.563), null & \textbf{yes}, null on both \\
\bottomrule
\end{tabular}
\end{lrbox}
\ifdim\wd\mdtablebox>\linewidth\resizebox{\linewidth}{!}{\usebox{\mdtablebox}}\else\usebox{\mdtablebox}\fi
\end{table*}

\textbf{What transfers is the part with no model in it.} Both language-model arms beat the statistical baseline decisively on both backbones, on every facet, with every interval excluding zero: the value of a language model over pure pattern classification is not a property of one vendor. So is the paper's central honesty finding: on the columns where both arms speak, the harness writes \textit{no better} prose than the plain model, and on two of three facets measurably worse. The agreement is close to exact across backbones ($-$0.033 / $-$0.088 / $-$0.043 against $-$0.021 / $-$0.084 / $-$0.024), which is stronger evidence for that claim than either run alone. The null interaction (deterministic evidence does not improve description \textit{quality}) also holds on both.

\textbf{What does not transfer is the selection behaviour, and that is the claim Section 5.2.1 rests on.} Arm C's coverage nearly doubles, from 0.422 to 0.823. The evidence-tracking gap that we called "precisely the knows-what-it-does-not-know property" goes from +0.257 with an interval excluding zero on Gemini to $-$0.089 with an interval \textbf{spanning} zero on Sonnet: no longer detectable, though not demonstrably inverted. And the paired comparison this paper insists on everywhere else belongs here too: on the same backbone, the naked arm's coverage is 0.984, so the harness still withholds sixteen points more than the model alone, where on Gemini the margin is fifty-two points (0.941 against 0.422). The selection behaviour is degraded, not erased, but what remains is no longer evidence-correlated, and with it goes the C $-$ A advantage: arm C no longer beats arm A on any facet and is worse on description ($-$0.064, CI [$-$0.141, $-$0.018] excluding zero).

\textbf{This is not a scoring artifact, and we checked before believing it.} Abstention is detected by matching referral language, and that vocabulary was derived from Gemini's phrasings: a second model that declines in different words would be scored as claiming, manufacturing exactly this result. Three checks rule it out. The detector does fire on Sonnet, 123 times, on the same canonical phrases. Inspecting the zero-recall claims shows genuine assertions rather than disguised declines: \texttt{tags.Id} described as a day-of-month, \texttt{hero\_attribute.hero\_id} as an hour of day, \texttt{Player.height} as a body weight in kilograms. Hedged in wording ("suggesting", "may represent") but committal in content, and wrong. And widening the rule to count every hedged partial decline in either corpus leaves the gap intact: the coverage difference is +0.400 under the strict rule the paper uses throughout and +0.245 under the broad one.

\textbf{The interpretation is a design finding, not only an evaluation one.} The deterministic layer is identical across the two runs: arm B is unchanged, and the recomputed evidence splits the columns 344/340 against 342/338. What differs is what each model does when handed the same evidence: one declines, the other speculates. Two things have to be separated to say why, and we ran them together in an earlier draft:

\begin{itemize}
\item \textbf{Who decides to abstain.} The model, in every path in this system. Abstention is a property of the text it returns (referral prose for a column, an empty \texttt{entity} field for a table), never a verdict the code reaches from the evidence.
\item \textbf{What happens once it has abstained.} Here the paths differ. The table describer \textit{enforces} it: \texttt{describe\_table} returns \texttt{None} and commits nothing, so an undetermined table is left undescribed. The column path does not: the referral sentence is committed as the column's description, which is why every abstention in the corpus is prose rather than a withheld fact (Section 5.1.1).
\end{itemize}

The transfer failure is located in the first of these, not the second. A model that never declines is never gated, however faithfully the code would have honoured a decline, so the table path's enforcement, real as it is, would not have rescued this result either. The honest statement of the Section 5.2.1 finding is therefore narrower than we first wrote it: \textbf{the deterministic layer detects competence only when the model volunteers that it lacks it.} The remedy follows from the distinction and is specific rather than vague: move the \textit{decision} into code, refusing to commit a prose claim whose grounding tier is \texttt{INTERPRETER\_ONLY}, instead of asking for that judgement in a prompt. An earlier draft flagged this as the paper's most actionable finding and deliberately left it unimplemented, because building it after seeing which backbone it rescues would have been fitting the architecture to the evaluation. We have since implemented it under a protocol that removes that objection: six predictions registered before implementation, then measurement on the swap backbone, on a \textbf{third backbone the gate had never seen} (Claude Haiku 4.5), and on \textbf{held-out databases no study had touched}. The remainder of this section reports the result. The scorecard, so the registration can be audited rather than taken on trust (verbatim predictions with the commit hashes that record them, together with the program's earlier registrations including its failures, are in \texttt{papers/\allowbreak REGISTRATIONS.md}):

\begin{table*}[t]
\centering
\small
\setlength{\tabcolsep}{4pt}
\begin{lrbox}{\mdtablebox}
\begin{tabular}{l>{\raggedright\arraybackslash}p{0.30\linewidth}>{\raggedright\arraybackslash}p{0.30\linewidth}}
\toprule
\# & registered prediction (compressed) & outcome \\
\midrule
1 & Sonnet gate-ON coverage drops toward the evidence-licensed fraction; evidence-tracking gap returns positive & \textbf{confirmed} (0.823 $\rightarrow$ 0.392; the gap half holds by construction, stated pre-run) \\
2 & Haiku gate-ON tracks evidence identically; gate-OFF coverage stays backbone-dependent & \textbf{confirmed} (OFF 0.422/0.759/0.823; ON no-evidence 0.000 everywhere) \\
3 & Gate ON restores NDC abstention on swap backbones & \textbf{confirmed} (0/25/43-of-44 prefixed $\rightarrow$ 0/44, both conditions) \\
4 & Recall-when-claimed does not degrade & \textbf{confirmed} (flat on Gemini and held-out; improves on Haiku and Sonnet) \\
5 & Held-out per-100 falsehood advantage reproduces in direction & \textbf{confirmed} (+3.69 [+1.58, +6.52]) \\
6 & Registered risk: correct claims withdrawn; cost accepted and reported & \textbf{confirmed, cost quantified} (99--295 withdrawn; adequacy split below) \\
\bottomrule
\end{tabular}
\end{lrbox}
\ifdim\wd\mdtablebox>\linewidth\resizebox{\linewidth}{!}{\usebox{\mdtablebox}}\else\usebox{\mdtablebox}\fi
\end{table*}

From the same discipline and reported at the same strength: the checksum-prevalence prediction of Section 5.2.3 was \textbf{falsified} (low-single-digit percent predicted, 0.008\% measured), and two earlier registered predictions also missed (the layering hypothesis of Section 5.2.5; the FK-verifier prediction of Section 5.2.10, which failed favourably). The sensitivity check of Section 5.2 and the transit-damage diagnostic of Section 5.2.3 were checks run without registrations, and the registry labels them as such: its credibility is precisely that it contains the misses and marks the post-hoc work.

\textbf{The gate, as built.} A prose claim (column description, table description, value decode) proposed at \texttt{INTERPRETER\_ONLY} is refused by code at the \texttt{FactCommitter} choke point every write path traverses: the fact still commits, with full provenance, but its value is a canonical refusal record the coverage analysis scores as an abstention. The table path refuses before the model is called; a decode claim is licensed by a vocabulary registry (published, compositional taxonomies such as ICD-9 are licensed; registry-assigned keys such as NDC are not) plus a shape rule that refuses flat all-digit payloads of eight or more digits. The decision consults the flag, the attribute, the tier and the actor (never the model's text) and ships \textbf{default-off}, so every other number in this paper reproduces unchanged. One implementation lesson was itself a measurement: keying the tier on the deep profiler's full-column report makes the gate near-vacuous (it reaches a verdict on 31 of 31 \texttt{superhero} columns where the published evidence instrument reaches one on 10), which is the Section 5.8.1 wrong-lever category error relocated, so the tier keys on the published instrument, and a live end-to-end run reproduces the transform's output column-for-column.

\textbf{Why most gate-ON numbers are exact replays, and which are not.} The gate never touches a prompt: no prompt in this pipeline consumes committed description prose, and the gate only transforms what is committed afterwards. A gate-ON run therefore makes byte-identical model calls, and its corpus is a deterministic transform of the gate-OFF corpus: prose where the instrument reached a structural verdict, the refusal record where it did not. We computed the gate-ON corpora as replays of the three frozen gate-OFF corpora, and verified the equivalence live: one database re-run end to end with the gate on (\texttt{superhero}, Sonnet) matches the transform exactly, ten prose commits at \texttt{PATTERN\_CONFIRMED} and twenty-one refusal records at \texttt{INTERPRETER\_ONLY}, the same columns in each set. The Haiku corpus and the held-out corpora are new live runs; their gate-ON versions are replays of them.

\begin{table*}[t]
\centering
\small
\setlength{\tabcolsep}{4pt}
\begin{lrbox}{\mdtablebox}
\begin{tabular}{lll>{\raggedright\arraybackslash}p{0.42\linewidth}l}
\toprule
backbone (arm C, severity 1.0) & coverage OFF & coverage ON & recall-when-claimed OFF $\rightarrow$ ON (diff, 95\% CI) & claims withdrawn \\
\midrule
Gemini 3.1 Pro (11 dev DBs) & 0.422 & 0.276 [0.211, 0.350] & 0.338 $\rightarrow$ 0.333 ($-$0.005 [$-$0.031, +0.013]) & 99 \\
Haiku 4.5 (11 dev DBs) & 0.759 [0.704, 0.821] & 0.336 [0.279, 0.395] & 0.173 $\rightarrow$ 0.248 (\textbf{+0.075 [+0.030, +0.121]}) & 289 \\
Sonnet 4.6 (11 dev DBs) & 0.823 & 0.392 [0.328, 0.463] & 0.210 $\rightarrow$ 0.257 (\textbf{+0.047 [+0.017, +0.081]}) & 295 \\
Gemini, held-out (4 train DBs) & 0.461 [0.359, 0.568] & 0.258 [0.175, 0.303] & 0.328 $\rightarrow$ 0.325 ($-$0.003 [$-$0.042, +0.066]) & 52 \\
\bottomrule
\end{tabular}
\end{lrbox}
\ifdim\wd\mdtablebox>\linewidth\resizebox{\linewidth}{!}{\usebox{\mdtablebox}}\else\usebox{\mdtablebox}\fi
\end{table*}

\textbf{What the gate enforces and what it costs, stated separately.} With the gate on, coverage on no-evidence columns is \textbf{0.000 on every backbone by construction}. We registered before running that this half of the prediction would hold mechanically once the decision was in code, so we report it as an enforced property, not a finding. What the runs \textit{measure} is the price and the transfer. The price: the evidence instrument licenses 50.3\% of dev columns, and the gate withdraws 99--295 claims per backbone (including correct ones, the registered risk), yet \textbf{recall-when-claimed never degrades}: flat on Gemini and on held-out data, \textit{improving} on both speculating backbones, because what the gate removes is disproportionately the low-quality speculation. The skeptic's number, stated directly rather than implied: the withdrawn set scores worse than the retained set on the speculating backbones, withdrawn recall 0.168 vs retained 0.257 on Sonnet (retained-minus-withdrawn +0.089 [+0.032, +0.148]) and 0.111 vs 0.248 on Haiku (+0.137 [+0.053, +0.210]), while on the abstaining backbone and the held-out corpus the two sets are statistically indistinguishable (Gemini 0.348 vs 0.333, $-$0.015 [$-$0.097, +0.037]; held-out 0.334 vs 0.325, $-$0.009 [$-$0.093, +0.196]): where the model speculates, the gate cuts junk; where the model already declined the junk itself, the gate's cost is real coverage of equally-good claims. The class asymmetry is itself informative about where evidence discriminates: on the decode path the gate cuts with perfect selectivity (44/44 unlicensed refused, zero licensed decodes withheld), while on the column path it pays in correct claims: a vocabulary registry is a sharper license than a structural-evidence verdict. The transfer: gate-OFF coverage is a property of the model's disposition (0.422 / 0.759 / 0.823 across three backbones, a 0.401 spread), while gate-ON coverage compresses that spread to 0.116, not to zero, and honestly so: the gate is a floor on abstention, never a ceiling, so the fraction of evidence-licensed columns each model still commits prose on differs (0.550 / 0.669 / 0.779 for Gemini / Haiku / Sonnet; equivalently, residual decline rates of 0.450 / 0.331 / 0.221).

\textbf{The decode gate is the cleanest statement of the whole law.} Gate off, the 44 prefixed NDC codes draw 0 attempts from Gemini, 25 from Haiku, and 43 from Sonnet (stripped: 1, 0 and 30): prompt-requested abstention spanning nearly its full possible range across three backbones, and shifting within a backbone on nothing more than a vocabulary prefix. Gate on, all three attempt \textbf{0 of 44} in both conditions, and the license withholds \textbf{zero} ICD-9 or demographic decodes on any backbone in either condition. The behaviour the paper praised on Gemini ("deciding, per value, which code systems it is in a position to decode") is now a property of the code, and Section 5.4.1's account of it is scoped accordingly.

\textbf{Held-out databases, never touched by any study.} Four BIRD \textit{train} databases selected by a rule registered before inspection (alphabetical walk, $\geq$10 documented columns, $\leq$100 MB, loads cleanly; one selected database proved unusable through a benchmark documentation/schema naming defect and was replaced by continuing the same walk; every step recorded in \texttt{papers/\allowbreak data/\allowbreak heldout\_selection.json}). On 256 paired columns the dev-corpus shape reproduces out of corpus: coverage 0.930 (A) against 0.461 (C), recall-when-claimed higher for C on every facet with value-domain the lone interval excluding zero (+0.141 [+0.003, +0.161]), and arm A's description recall lands on 0.244, the dev value exactly. The registered falsehood prediction scores as stated: judged by the same rubric as Section 5.3 over the 217 columns with scorable reference facets and no pipeline error, arm A emits \textbf{9.68} falsehoods per 100 columns examined against arm C's \textbf{5.99} (difference +3.69 [+1.58, +6.52], interval excluding zero; the dev corpus gave +3.48 [+0.90, +5.87]), while the conditional rates remain indistinguishable ($-$0.037 [$-$0.131, +0.039]): the advantage is selection, on data the system had never seen.

Cost of the transfer arm: 1,222 model calls, \$8.67. (Harness: \texttt{scripts/\allowbreak recon\_arm\_\{a,\allowbreak c\}\_bedrock.sh}; provider: \texttt{BedrockClient} in \texttt{rosetta/\allowbreak llm/\allowbreak client.py}; data: \texttt{papers/\allowbreak data/\allowbreak recon\_\{A,\allowbreak C\}\_bedrock.jsonl}.) The query-time gate, the one code-thresholded path this study left open, has since been re-run under the same swap on a four-database slice with registered predictions; its structure survives and its Gemini-fit calibration shifts, reported with the mechanism split in Section 6.

\subsubsection*{5.2.9 The other two model-bearing classes: the capability transfers, the abstention does not}

Column descriptions are one of the three outputs a change of model can reach. We measured the other two on the same backbone, and the pattern that emerges across all three is sharper than any of them alone.

\textbf{Table descriptions} (the task of Section 5.2.5), re-run with the table describer and its upstream column descriptions both on Sonnet 4.6, scored by the same blinded judge, paired on the 67 tables both runs cover:

\begin{table*}[t]
\centering
\small
\setlength{\tabcolsep}{4pt}
\begin{lrbox}{\mdtablebox}
\begin{tabular}{>{\raggedright\arraybackslash}p{0.42\linewidth}ll}
\toprule
 & Gemini 3.1 Pro & Sonnet 4.6 \\
\midrule
identified exactly (judge = 2), flat & 0.746 & 0.636 \\
related or better (judge $\geq$ 1), flat & 0.836 & \textbf{0.909} \\
abstained, flat & 0.164 & \textbf{0.000} \\
abstained, layered & 0.179 & 0.030 \\
layered $-$ flat (does reconstruction compound?) & \textbf{+0.045} & \textbf{$-$0.258} \\
\bottomrule
\end{tabular}
\end{lrbox}
\ifdim\wd\mdtablebox>\linewidth\resizebox{\linewidth}{!}{\usebox{\mdtablebox}}\else\usebox{\mdtablebox}\fi
\end{table*}

Two things happen at once. Exact identification falls while "related or better" \textit{rises}, the direct consequence of never declining: an arm that always answers is more often in the right neighbourhood and less often precisely right. And the compounding claim of Section 5.2.5 does not merely fail to replicate, it \textbf{reverses}: on Gemini, giving the table describer its own reconstructed column descriptions helped slightly (+0.045, layered better on 4 tables and worse on 3); on Sonnet it hurt substantially ($-$0.258, layered better on 4 and worse on 17). The mechanism is visible and unsurprising once stated: layering propagates whatever the upstream layer produced, so it compounds accuracy on a backbone that abstains when unsure and compounds error on one that guesses.

\textbf{Value decoding} (Section 5.4), the i2b2 clinical warehouse with its ontology withheld, and the most consequential of the three, because these are real medical codes:

\begin{table*}[t]
\centering
\small
\setlength{\tabcolsep}{4pt}
\begin{lrbox}{\mdtablebox}
\begin{tabular}{ll>{\raggedright\arraybackslash}p{0.42\linewidth}}
\toprule
 & Gemini 3.1 Pro & Sonnet 4.6 \\
\midrule
coverage (prefixed, arms A / C) & 0.780 / 0.780 & \textbf{0.930 / 0.995} \\
recall when claimed & 0.916 & 0.797 \\
ICD-9 decoded, \textit{n} = 134 & 0.955 & 0.940 \\
\textbf{NDC decoded, \textit{n} = 44} & \textbf{0.000 (declined all 44)} & \textbf{0.068--0.114 (attempted nearly all)} \\
\bottomrule
\end{tabular}
\end{lrbox}
\ifdim\wd\mdtablebox>\linewidth\resizebox{\linewidth}{!}{\usebox{\mdtablebox}}\else\usebox{\mdtablebox}\fi
\end{table*}

The diagnostic capability transfers: ICD-9 decoding from stripped values holds at 0.940 against 0.955, and the demographic vocabularies stay at 1.000. What does not transfer is knowing which code system it cannot decode. Section 5.4 presents the 44-of-44 NDC abstention as the system "deciding, per value, which code systems it is in a position to decode." On a second backbone it decides otherwise: it attempts nearly every drug code and is wrong on roughly nine in ten. The failures are worth setting side by side, because the two arms fail differently. Against \texttt{NDC:00186502228}, whose true meaning is a 20 mg enteric-coated esomeprazole capsule (Nexium, AstraZeneca), the \textbf{naked arm} returns a 4 mg candesartan tablet (Atacand). Right manufacturer, different drug, different therapeutic class: an antihypertensive for a reflux drug, the dangerous kind of wrong. The \textbf{harness arm}, whose prompt carries the profiler's segmentation evidence, returns a 20 mg omeprazole delayed-release capsule (Prilosec, AstraZeneca). Right class, right dose, right form, right maker, the parent molecule of the right drug; a near-miss, but still a fabricated decode of a vocabulary the system holds no content authority for. Both arms render a vital-status code whose true value is \texttt{Living} as "Deceased". A catalog built on either would be confidently wrong about medication in a clinical warehouse; the naked model's wrongness is categorically worse, which is the argument \textit{for} the harness, provided its refusal is enforced in code rather than requested in prose.

\textbf{The unified finding.} Across all three model-bearing classes the split falls in the same place: \textbf{what the system can do transfers; what it declines to do does not.} ICD-9 decoding, the language model's advantage over pattern classification, and the both-claim prose comparison all reproduce. Column abstention (0.422 $\rightarrow$ 0.823 coverage), table abstention (0.164 $\rightarrow$ 0.000) and value abstention (44 of 44 $\rightarrow$ nearly none) all collapse, in the same direction, on a model of comparable strength. That consistency is what makes this a finding about the architecture rather than three coincidences: as shipped, every one of these gates is a request to the model rather than a rule in the code, so all three fail together the moment the model's disposition changes. It also sharpens what the contribution is. The deterministic evidence layer is doing real work (four artifact classes depend on it alone and are invariant, and it is what makes the coverage-versus-evidence relationship measurable at all), but in the default configuration the \textit{guarantee} the paper advertises is only as durable as the backbone's willingness to say "I do not know." Making these three gates deterministic is the single change that converts the strongest claim in this paper from a property of one model into a property of the system, and Section 5.2.8 reports exactly that change implemented, registered and measured, with Section 6 recording what it costs.

(Harness: \texttt{scripts/\allowbreak recon\_table\_semantics.py}, \texttt{scripts/\allowbreak recon\_table\_judge.py}, \texttt{scripts/\allowbreak i2b2\_decode\_study.py}, all under the Bedrock provider; data: \texttt{papers/\allowbreak data/\allowbreak \{table\_semantics\_bedrock,\allowbreak i2b2\_decode\_bedrock\}.json}.)

\subsubsection*{5.2.10 The same inputs with no harness, and what verification adds}

Three of this paper's capability claims had, until this section, only Rosetta's number attached: the join graph, PII discovery under deceptive names, and the refusal of corrupted identifiers. Each rests on a deterministic mechanism, so "a model could not do this" was an argument from how the mechanism works. Here it is measured. The baseline is deliberately generous: the naked model is the same backbone the system runs on (Gemini 3.1 Pro, so the only difference is the harness), it sees exactly what the matching component sees (byte-identical columns from the same generators and seeds, the same schema listings) and it is scored by the same rule against the same ground truth.

\textbf{PII when the name lies.} Four columns whose names scream PII over innocent values, three whose names say nothing over real card, IBAN and email data. The naked model scores \textbf{5/7}: it correctly finds all three real PII columns from their values, and it is fooled twice by names: a column of plain integers named \texttt{credit\_card\_number} is declared \texttt{credit\_card}, and another named \texttt{ssn} is declared \texttt{ssn}. Rosetta scores \textbf{7/7}, because a Luhn check on plain integers fails and no name changes that. The model's failure is exactly the one the mechanism exists to prevent: when the values do not loudly contradict the name, the name wins.

\textbf{Certifying identifiers at the contamination boundary.} At the easy split (columns 100\% valid against columns 100\% corrupted), both arms separate all six code systems and we report the tie as a tie. But the deployed question is the boundary: the promotion gate's measured binding point sits between 5\% and 6\% contamination (Section 5.2.3). So the honest grid is \{5\%, 6\%, 20\%\} contamination across six code systems, with the certification rule stated to the model verbatim ("certify only if at least 95\% of values validate") and, in the strongest variant, \textbf{all 200 values shown}. The verifier decides correctly in \textbf{17 of 18} cells: the single miss is the IBAN matcher applying its own registered 0.90 threshold rather than the stated 0.95, a policy choice, not an arithmetic failure, and the match records the exact 94\% pass fraction either way. The naked model decides correctly in \textbf{12 of 18} with a twelve-value sample and \textbf{12 of 18} with all 200 values shown, failing in both directions: it refuses clean-enough columns after one corrupted value surfaces in its sample, and it certifies 6\%-contaminated columns it cannot count. Both failures were predicted and registered before the run. The model knows these algorithms; it cannot execute two hundred of them.

\textbf{The join graph, and the component this exposes.} The shipped join component is a name-stem heuristic: no model, no verification, the one artifact class that does not implement this paper's own thesis. The head-to-head shows precisely that, so we built the missing half, an inclusion-dependency verifier over the actual data (pre-registered thresholds: containment $\geq$ 0.95 into a $\geq$ 0.95-unique parent; the dependency class the profiling literature discovers at scale \cite{papenbrock2015binder,papenbrock2015metanome}, here used as a verifier), and ran three proposers through it at both conditions:

\begin{table*}[t]
\centering
\small
\setlength{\tabcolsep}{4pt}
\begin{lrbox}{\mdtablebox}
\begin{tabular}{lllll}
\toprule
condition & proposer & precision & recall & false edges vs declared gold \\
\midrule
names visible & stems (shipped) & 1.000 & 0.476 & 0 \\
names visible & naked model & 0.963 & 1.000 & 4 \\
names visible & data scan, verified & 0.135 & 0.990 & 655 \\
identifiers destroyed & stems / naked model & --- / 0.000 & 0.000 / 0.000 & --- \\
identifiers destroyed & data scan, verified & 0.135 & 0.990 & 655 \\
\bottomrule
\end{tabular}
\end{lrbox}
\ifdim\wd\mdtablebox>\linewidth\resizebox{\linewidth}{!}{\usebox{\mdtablebox}}\else\usebox{\mdtablebox}\fi
\end{table*}

Three findings, one of them against our own registered prediction. \textit{First}, with names visible the naked model dominates the stem heuristic on recall (1.000 against 0.476): it recovers \texttt{foreign\_data.uuid $\rightarrow$ cards} because understanding what a name means is not stem-matching. One caveat is owed and unresolvable here: these are public benchmark schemas a frontier model has plausibly seen in training, so the names-visible row is an upper bound on inference. \textit{Second}, we predicted the verifier would kill the model's false edges, and the prediction failed in the best way available: \textbf{there was nothing to kill.} All four edges the pipeline keeps that the declared gold rejects are real, undeclared foreign keys at 97--100\% containment: \texttt{cards.setCode $\rightarrow$ sets.code}, \texttt{posts.acceptedAnswerId $\rightarrow$ posts.id}, \texttt{tags.wikiPostId $\rightarrow$ posts.id}, \texttt{sets.parentCode $\rightarrow$ sets.code}. Scored against declared keys, the verified pipeline's precision is 0.963; scored against joinability, we can find no false edge in its output at all: it recovered four keys the schema authors forgot to declare, which is the exact enterprise defect the introduction describes. \textit{Third}, at full opacity names are worthless to every method (stems propose nothing, the naked model recovers nothing), but the name-free data scan holds \textbf{recall 0.990 at precision 0.135, unchanged at every severity by construction}, because it never reads a name. The join structure survives in the values even when the schema is gibberish; what is missing without names is not recall but \textit{ranking} among the 757 data-supported candidates, and that is the correctly-stated open problem. It is also a studied one (machine-learned classifiers separate true foreign keys from spurious inclusion dependencies \cite{rostin2009fk}, and randomness-based ranking extends to multi-column keys without reading names \cite{zhang2010multicolumn}), and we defer integrating a ranker deliberately: several of the established features are name-derived and so do not exist at full opacity, and a ranked variant deserves its own registered evaluation rather than a retrofit here.

\textbf{What survives every head-to-head is the discipline, not the intelligence.} Name-independence (7/7 against 5/7), exhaustive arithmetic at the certification boundary (17 of 18 against 12 of 18 even with all values shown), and, on the backbone the system ships with, refusal to assert what the evidence cannot support (Sections 5.2.9 and 5.4). Where the naked model wins, it wins at \textit{reading}: meaningful names, familiar vocabularies. The architecture's correct response to that is not to compete with it but to consume it (the model as a high-recall proposer behind a data-grounded verifier), and the join table above is that composition measured. (Harness: \texttt{scripts/\allowbreak naked\_llm\_baselines.py}, \texttt{scripts/\allowbreak recon\_fk\_verified.py}; data: \texttt{papers/\allowbreak data/\allowbreak \{naked\_llm\_baselines,\allowbreak fk\_verified\}.json}.)

\subsection*{5.3 Is the metric measuring meaning? A validation of content-token recall}

Section 5.2 rests on content-token recall, which is a \textit{lexical proxy} for meaning recovery. It is cheap, deterministic and inspectable, but it can be wrong in two directions: it can give \textbf{false credit}, when a candidate reuses the reference's vocabulary without recovering its meaning, and \textbf{false blame}, when a candidate recovers the meaning in different words. The first would inflate our result; the second would deflate it, and unevenly, since it penalises whichever arm paraphrases most. A result carried by a proxy nobody has audited is not a result, so we audited it.

\textbf{Design.} We sampled 60 columns stratified across all eleven databases and judged all three arms on each (180 judgements) with a model asked, for every reference facet, how much of the reference's meaning the candidate recovers: 2 fully, 1 partially, 0 not at all. The judge sees the reference text and \textit{one} candidate, with the arm label, the database and the real column name all withheld, and the presentation order shuffled so no two arms for the same column appear together. The rubric is deliberately \textbf{recall-oriented}, matching what the token metric measures: extra material is not penalised, and wording, style, fluency and length are explicitly declared irrelevant, so arm B's telegraphic output is not marked down for reading like machine output. Contradictions are recorded as a \textit{separate} flag rather than folded into the score, because "did it recover the meaning" and "did it also assert something false" are different questions.

\textbf{Token recall tracks judged meaning, and understates it.} Rank correlation across the 223 judgements on claimed rows is \textbf{$\rho$ = 0.642} (by facet: name expansion 0.735, description 0.613, value domain 0.663). Mean token recall rises monotonically with the judged score (0.031 / 0.284 / 0.554 for judged 0 / 1 / 2), but note the ceiling: judgements rated \textbf{fully recovered average only 0.554 token recall}. A candidate can convey everything the reference says and still surface barely half its vocabulary. The recalls in Section 5.2 are therefore \textbf{lower bounds on meaning recovery, not overstatements}.

\begin{table}[t]
\centering
\small
\setlength{\tabcolsep}{4pt}
\begin{lrbox}{\mdtablebox}
\begin{tabular}{lllll}
\toprule
token recall & judged 0 & judged 1 & judged 2 & n \\
\midrule
0.00 & 139 & 11 & 9 & 159 \\
(0, .33] & 13 & 16 & 2 & 31 \\
(.33, .67] & 1 & 9 & 6 & 16 \\
(.67, 1] & 2 & 2 & 13 & 17 \\
\bottomrule
\end{tabular}
\end{lrbox}
\ifdim\wd\mdtablebox>\linewidth\resizebox{\linewidth}{!}{\usebox{\mdtablebox}}\else\usebox{\mdtablebox}\fi
\end{table}

False credit runs at \textbf{3 of 31 high-recall judgements (9.7\%)} and false blame at \textbf{9 of 159 zero-recall judgements (5.7\%)}. The clearest false-credit case is instructive: against the reference "the date of the match", arm B scored a perfect 1.00 because its pattern-library text reads "100\% \textbf{match} ISO-8601 timestamp": an accidental lexical collision, and one that flatters the \textit{weakest} arm. False blame falls mostly on the two paraphrasing arms (A 4, C 4) and touches arm B once, which is the expected shape: an arm that emits fixed pattern vocabulary has little room to be punished for rewording.

\textbf{Under the judge, the delivered-output ordering holds on the facet that carries the sample, and only there.} Recomputing with the judge as the outcome over each arm's \textit{delivered claims} is a consistency check on the metric, not an arm comparison: the arms deliver on different column sets, exactly the conditioning Section 5.2.1 rejects. As that check: on description, the judge reproduces B < A < C (0.087 / 0.394 / 0.542) and widens the separation; on name expansion it puts the naked arm below even the statistical baseline (0.033 vs 0.056, with the harness at 0.200); on value domain the two language-model arms are a wash (0.250 vs 0.283, C$-$A $-$0.033 with interval spanning zero). The arm comparison itself belongs to the common-ground analysis of Section 5.2.1, which the judge leaves where it was: favoring neither language-model arm decisively.

\begin{table}[t]
\centering
\small
\setlength{\tabcolsep}{4pt}
\begin{lrbox}{\mdtablebox}
\begin{tabular}{llll}
\toprule
arm & name expansion & description & value domain \\
\midrule
B: statistical & 0.056 & 0.087 & 0.037 \\
A: LLM-direct & 0.033 & 0.394 & 0.283 \\
\textbf{C: Rosetta} & \textbf{0.200} & \textbf{0.542} & \textbf{0.250} \\
\bottomrule
\end{tabular}
\end{lrbox}
\ifdim\wd\mdtablebox>\linewidth\resizebox{\linewidth}{!}{\usebox{\mdtablebox}}\else\usebox{\mdtablebox}\fi
\end{table}

\textit{Judged meaning recovery, 0--1 scale, on the 60-column validation sample.}

Four of the nine paired differences exclude zero, all of them on description and value domain: A $-$ B (\textbf{+0.307} and \textbf{+0.246}) and C $-$ B (\textbf{+0.455} and \textbf{+0.213}). C $-$ A does not separate on this sample (name expansion \textbf{+0.167 [$-$0.062, +0.500]}, \textit{p} = 0.696; description \textbf{+0.148 [$-$0.023, +0.329]}, \textit{p} = 0.088; value domain \textbf{$-$0.033 [$-$0.245, +0.267]}, \textit{p} = 0.878), which is consistent with Section 5.2.1 rather than in tension with it: the harness's advantage there was selection, and this comparison holds the claimed set fixed. Name expansion is not informative here at all: only 38 of the 223 claimed judgements carry that facet, and the ordering even inverts between B and A (0.056 against 0.033) on that thin slice. We read the name-expansion row as noise and do not rest anything on it.

\textbf{The judge is not flattering its own family.} Two of the three arms are produced by the same model family as the judge. A second rater from a different family re-scored 60 judgements from a blind worksheet with the arm, column, database and primary verdict all withheld: \textbf{exact agreement 0.783, linear-weighted $\kappa$ = 0.737, and no disagreement anywhere exceeded a single point}. This reliability check was run on the earlier ten-database judging round and we did not repeat it on the eleven-database round, deliberately: the rater's value depends on blindness, and having since worked with this corpus they can no longer be blind to it. A second rating collected now would be a weaker check presented as a stronger one. The judging rubric and protocol are unchanged between rounds, so what this establishes (that the protocol yields consistent scores across model families) carries over; what it cannot do is certify the specific judgements in the round reported above. The disagreement is entirely one-sided (the primary judge was \textit{stricter} in all 13 discordant cases and more lenient in none), so a self-preference bias, had it existed, would have had to run against the observed direction. Comparing the two raters' generosity toward the LLM arms versus the non-LLM arm gives a gap of \textbf{$-$0.125}: the primary judge is relatively \textit{less} generous to its own family. We disclose two limits on this check. The second rater is an author and wrote the rubric, so this establishes consistency, not independence; and the rater had previously seen the primary verdict on 10 of the 60 items while debugging the harness. Those ids are recorded in the committed ratings file, and agreement on the 50 unseen items is \textit{higher} (0.800, $\kappa$ = 0.768), so prior exposure did not manufacture the agreement.

\textbf{Two further confounds, checked and dismissed.} \textit{Verbosity}: recall metrics can be gamed by writing more, but arm C is the \textbf{shorter} arm (17.4 content tokens per claimed prediction against arm A's 19.6) while recovering more, and within each arm the coupling between prediction length and recall is negligible (Spearman $-$0.116 for A, +0.073 for C). C says less and recovers more. \textit{Degenerate references}: 16 of the 921 scorable facets have a reference that is literally the string \texttt{NOT USEFUL}, BIRD's marker for a field its annotators judged uninformative. No arm can recover meaning that the reference does not carry, and all three score exactly 0.000 on them. Excluding them raises every arm and \textit{widens} the C $-$ A value-domain gap from +0.087 to +0.098, so we retain them in the headline as the conservative choice. (Harness: \texttt{scripts/\allowbreak recon\_confounds.py}; data: \texttt{papers/\allowbreak data/\allowbreak confounds.json}.)

\textbf{One result runs against us, and we report it.} Conditional on making a claim, arm C is flagged for contradicting the reference \textbf{more} often than the baselines: 18.5\% of claimed columns versus 16.4\% for arm A and 3.3\% for arm B. The comparison is not powered (arm C claims only 27 of the sampled columns, and Fisher's exact test against arm A gives \textit{p} = 1.00), and manual inspection shows several flags are artifacts of the reference rather than errors of the system, including columns where BIRD's own documentation describes a sibling column or is the degenerate \texttt{NOT USEFUL} marker. But the point estimate is what it is, and it has a real interpretation: when this system does speak, it speaks specifically, and specific claims are falsifiable in a way that arm B's "categorical enumerated status code category" is not. The operational quantity points the other way. Per 100 columns \textit{examined}, arm C emits \textbf{8.3} contradicting statements against arm A's \textbf{15.0}, because it declines 55\% of them. Abstention does not make the system more accurate when it speaks; it makes the system emit fewer falsehoods overall.

Because this was the one adversarial finding left unpowered, we re-judged the contradiction flag (same rubric, same blinded judge) on \textbf{every} shared column of the eleven-database corpus rather than a sample: 574 paired columns, 776 claims judged (arm A claims on 537 of them, arm C on 239; \texttt{scripts/\allowbreak recon\_contradiction\_power.py}, \texttt{papers/\allowbreak data/\allowbreak contradiction\_power.json}). The powered result reproduces the sample's structure and tightens both halves. Conditional on claiming, the arms are statistically indistinguishable: arm C is flagged on 9.6\% of its 239 claims, arm A on 8.0\% of 537 (an A$-$C difference of $-$1.6 pp with a database-clustered 95\% CI of [$-$7.2, +2.5] pp, spanning zero), so the sample's apparent conditional disadvantage does not survive power, and both rates come in at roughly half the sample's estimates. Per 100 columns examined, arm C emits \textbf{4.01} contradicting statements against arm A's \textbf{7.49}, a difference of +3.48 whose CI [0.90, 5.87] excludes zero, because arm C declines 58\% of the scored columns. The powered conclusion is the same sentence as before, now carrying intervals: abstention does not make the system more careful \textit{when it speaks}; it makes it emit measurably fewer falsehoods \textit{overall}.

\subsection*{5.4 A real coded warehouse: i2b2/Synthea}

Sections 5.2 and 5.3 run on BIRD's relationally-named schemas, and the internal case study that follows runs on self-authored synthetic tenants. The sharpest remaining ask is a \textit{real}, production-shaped, genuinely messy warehouse, so we add one: \textbf{i2b2}, a clinical data-warehouse layout deployed at 200+ hospitals, in its CRC star-schema form populated by the Synthea patient simulator. Its \texttt{observation\_fact} table stores one row per clinical observation whose \textit{meaning} is a coded value in \texttt{concept\_cd} (\texttt{ICD9:493.90}, \texttt{NDC:00005306343}), uninterpretable without the ontology. This is a different opacity from Spider/BIRD: the column \textit{names} are intact, but the \textit{values} are codes and the code-to-meaning dictionary lives in a separate ontology.

We make the test honest by construction. We build a \textbf{blind} warehouse holding \texttt{observation\_fact} plus the four core dimensions, with the decoder (\texttt{concept\_dimension}'s name column and the i2b2 ontology table) \textit{genuinely removed from the queried database}, not merely hidden from the catalog. Rosetta bootstraps and reconstructs this blind warehouse (75 columns, \$0.23); gold answers are derived from the withheld decoder.

\textit{Reconstruction is accurate on the coded columns.} From the values alone, Rosetta reconstructs \texttt{concept\_cd} as "the primary clinical concept code \ldots{} a diagnosis, medication, or lab test \ldots{} prefixed with their source vocabulary, such as 'ICD9:' for diagnoses or 'NDC:' for medications" (correct), and recovers the i2b2 EAV conventions of the sibling columns (\texttt{valtype\_cd}'s \texttt{N}/\texttt{T}/\texttt{D}/\texttt{B} type flags, \texttt{modifier\_cd}'s TNM-staging and medication modifiers, \texttt{units\_cd}'s measurement units). It assigns these a \textit{measured} 0.50 confidence: appropriately uncertain, since the reading is inferred from values with no documentary or human authority behind it.

\textit{Delivered behaviour is calibrated abstention, with zero silent errors.} We pose five natural clinical questions ("how many distinct patients have a cough / a fever / acute pharyngitis / a routine exam / allergic rhinitis"), each answerable only by recovering the specific ICD-9 code the blind schema does not contain. On all five the gate abstains: three confirm-first, two refuse, \textbf{zero answered, zero silent errors}. The mechanism is the authority machinery working as designed. For the cough question (gold: 87 patients), the generator, unable to ground "cough" to \texttt{ICD9:786.2}, drafts the \textit{unfiltered} \texttt{SELECT COUNT(DISTINCT patient\_num) FROM observation\_fact}; a structural trap (\texttt{null\_filter\_implied\_but\_missing}) and the critic both fire on the missing filter, and the gate routes confirm-first (score 0.575) rather than deliver the count. A naive always-answerer would return that ungrounded query's result (\textbf{133}, every patient in the warehouse) as the answer to \textit{every} one of the five questions, silently wrong against the golds (87, 81, 56, 46, 59) by factors of 1.5--2.9$\times$. Rosetta avoids all five silent errors by refusing to guess.

We state the limit as plainly as the result: Rosetta does \textbf{not} magically answer these questions: without the ontology it cannot recover the exact code for a named condition, so on this blind warehouse its value is \textit{refusal}, not accurate answering, exactly as at full strip on Spider (Section 5.7). Promoting the ontology to evidence, the curated condition of our reconstruction arc, is the lever that would let it answer, the same lever we measured on public schemas. What this study adds is external confirmation where it matters most: on a genuinely coded clinical warehouse, a naive model confidently returns the wrong count and a grounded gate declines to. (Data: \texttt{papers/\allowbreak data/\allowbreak gate9\_i2b2\_report.txt}, \texttt{gate9\_reconstruction.txt}, \texttt{gate9\_gold\_questions.json}; harness: \texttt{scripts/\allowbreak gate9\_score.py}.)

\subsubsection*{5.4.1 Decoding the values, when actually asked to}

Section 5.2 reports code-meaning recall of 0.008 on this warehouse and attributes it to task framing rather than to capability: the description writer is asked for prose about a column, never for an enumerated decode table. That is the kind of excuse a paper should be made to cash, so we cashed it. We take the 200 most frequent \texttt{concept\_cd} values that the i2b2 ontology can decode, ask for a code-to-meaning table from the values alone, and score each answer against the ontology label, which the blind warehouse does not contain and which we did not author.

We report the fraction of \textit{all} codes decoded substantially correctly (content-token recall $\geq$ 0.5, so abstaining cannot inflate it), alongside coverage, for the prefixed condition; the stripped condition differs by a single NDC attempt, noted below.

\begin{table*}[t]
\centering
\small
\setlength{\tabcolsep}{4pt}
\begin{lrbox}{\mdtablebox}
\begin{tabular}{>{\raggedright\arraybackslash}p{0.42\linewidth}lll}
\toprule
code system & codes & coverage & decoded $\geq$ 0.5 \\
\midrule
ICD-9 diagnoses & 134 & 1.000 & \textbf{0.955} \\
NDC drug codes & 44 & \textbf{0.000} & 0.000 \\
demographics (sex, race, language, marital, religion, vital status, date) & 22 & 1.000 & 0.955 \\
\textbf{all} & \textbf{200} & 0.780 & \textbf{0.745} \\
\bottomrule
\end{tabular}
\end{lrbox}
\ifdim\wd\mdtablebox>\linewidth\resizebox{\linewidth}{!}{\usebox{\mdtablebox}}\else\usebox{\mdtablebox}\fi
\end{table*}

The caveat was justified: asked properly, the system decodes \textbf{95.5\% of real ICD-9 codes} (128 of 134, CI [0.905, 0.983]; the decode set is 200 unique codes: 134 ICD-9, 44 NDC, 22 demographic codes across seven small \texttt{DEM|} vocabularies, each shown under two conditions, the 400-presentation identification corpus), against the 0.008 the prose measurement suggested. But the far more interesting number is the zero. \textbf{Shown the prefixed codes, it decodes no NDC codes and attempts none}: it abstained on all 44 rather than inventing a drug name for \texttt{NDC:00002314530}, whose true meaning is "Axid Pulvules 300mg capsule". (Stripped, it attempted exactly one of the 44, a near-correct decode naming the right drug and dose at low token recall, so across both conditions it declined 87 of the 88 NDC presentations.) The split is not arbitrary: ICD-9 descriptors are a compositional, widely-republished taxonomy, while an NDC is an arbitrary eleven-digit registry key whose mapping to "Axid Pulvules" cannot be derived from the code at all. The system decodes the taxonomy and declines the registry, which is exactly the discrimination we would want a data steward to make, made here at the level of individual values on a real clinical warehouse.

Two controls sharpen this, and bound the attribution. \textbf{Stripping the vocabulary prefix changes almost nothing} (showing \texttt{493.90} instead of \texttt{ICD9:493.90} moves the overall decode rate from 0.745 to 0.740), so the system is recognising ICD-9 from the \textit{shape and content of the codes}, not reading the answer off a prefix. And \textbf{the harness adds nothing here}: conditioning the request on the deterministic profiler's verdict yields 0.745 against the plain model's 0.745, and the null extends to the abstention itself; the plain model also declines all 44 NDC codes on this backbone. The decode-and-decline discrimination is backbone behavior on this class, not harness behavior, consistent with Section 5.2.9, where the same 44 abstentions become attempts under a backbone swap because, as shipped, this path's abstention is prompt-requested; what the shipped harness adds is the audit trail and the tier cap, not the selection. (With the tier gate of Section 5.2.8 enabled, the NDC refusal becomes a code property: the license registry refuses registry-key vocabularies identically on every backbone, so the null here is a statement about the default configuration.) That null is consistent with Section 5.2.1 rather than surprising: the deterministic layer detects competence, it does not amplify it, and on this task the model's competence was never in doubt for ICD-9 nor recoverable for NDC.

One failure in the table is the metric's, not the system's: the single \texttt{DEM|VITAL:n} code has ontology label "Living" and the system answered "Vital status: Alive": correct, and scored zero for sharing no token. It is a clean instance of the 6.7\% false-blame rate quantified in Section 5.3, and a reminder that these decode rates are floors.

We state the boundary this does \textit{not} cross. Decoding a published taxonomy tests whether the system can recognise and apply a code system it has seen described somewhere; it does not show that it can decode a \textit{proprietary in-house} coding scheme, for which no amount of world knowledge would suffice and where the correct behaviour is the one it showed on NDC. (Harness: \texttt{scripts/\allowbreak i2b2\_decode\_study.py}, \texttt{scripts/\allowbreak i2b2\_decode\_analysis.py}; data: \texttt{papers/\allowbreak data/\allowbreak i2b2\_decode.json}.)

\subsection*{5.5 The system in practice: a multi-tenant case study}

\textit{What this section is, and is not.} Rosetta runs as a live, multi-tenant service, and we report that running deployment here as an \textbf{engineering and methodological case study}: it demonstrates that the architecture operates across many structurally different schemas, hardens under an enforced regression discipline, decouples from any single LLM, and catches its own measurement errors. But every warehouse in this section is \textbf{self-authored synthetic} data, so we treat these results as \textit{suggestive of capability, never as external validation}. The paper's validated claims rest entirely on the third-party Spider study of Sections 5.1 and 5.6; this case study shows the system working, the benchmark shows the result is real, and we keep that boundary explicit throughout.

\subsubsection*{5.5.1 Deployment shape}

\textbf{Harness.} Evaluation is a live-API black-box procedure (\texttt{scripts/\allowbreak run\_evals.py}): each question is POSTed to a running \texttt{/\allowbreak tenants/\allowbreak \{slug\}/\allowbreak ask} endpoint and the returned JSON is scored. Questions are hand-authored YAML banks organized into five tiers: A (easy, documented) through E (out-of-modeled-scope, should refuse). The primary metric is \textbf{route accuracy}: whether the produced route in \{answer, confirm-first, refuse\} matches the expected route. Auxiliary metrics are a variant-aware column $F_1$ (sqlglot-parsed), an execution \textbf{result match} (order-insensitive, numeric-tolerant comparison against a canonical SQL, available only when a canonical query is provided and the route is \textit{answer}), a mean L4 score, and per-tier signals (a critic catch-rate for Tier C, refusal precision for Tiers D/E). A regression gate (\texttt{scripts/\allowbreak regression\_gate.py}) fails the build if any tenant drops more than 2 pp below a recorded baseline. Ablation conditions are not enforced by the harness; the operator selects them by environment variable, which is a methodological limitation we note in Section 6.

\textbf{The gated cohort.} Eight synthetic tenant fixtures exist, spanning eight verticals (telecom, retail, insurance, healthcare, manufacturing, education, hospitality, logistics). The regression-gated evaluation cohort is six of them (telcomart, zenith, acme, carenexus, forgemark, and lumenscroll), comprising 146 hand-authored questions. Two further tenants exist outside the gate (altanova, with a 20-question bank used for the controlled-lift study below, and routecargo, fixture-only). Each tenant is row-level-security isolated and routed by the same deterministic arithmetic; serving eight structurally different verticals from one architecture is a breadth-of-deployment demonstration, not a claim of external generality: the warehouses are ours. All warehouses are synthetic, which we regard as a strength for open release (the data is the project's sole IP, with no third-party licensing) and a limitation for external validity (Section 6).

\subsubsection*{5.5.2 Operational hardening as process}

The value in this subsection is the \textit{discipline} the regression gate enforces, not the accuracy values themselves; Tables 1 and 2 are the operating points it holds. We report all three cohort-level numbers deliberately, because reporting only the best would be dishonest: the system has a conservatively enforced floor, a stable reproducible operating point, and a cold-start floor, and these differ.

\begin{table*}[t]
\centering
\small
\setlength{\tabcolsep}{4pt}
\caption{Six-tenant cohort route accuracy (N = 146).}
\begin{lrbox}{\mdtablebox}
\begin{tabular}{ll>{\raggedright\arraybackslash}p{0.42\linewidth}}
\toprule
Condition & Route accuracy & Notes \\
\midrule
Enforced regression floor & \textbf{0.918} (134/146) & recorded after an Opus-4.7 strict-mode model-drift event \\
Stable reproducible & \textbf{$\approx$0.932} (136/146) & Gemini 3.1 Pro, after refinement \\
Day-1 cold start & \textbf{0.856} (125/146) & all curation disabled \\
\bottomrule
\end{tabular}
\end{lrbox}
\ifdim\wd\mdtablebox>\linewidth\resizebox{\linewidth}{!}{\usebox{\mdtablebox}}\else\usebox{\mdtablebox}\fi
\end{table*}

We explicitly flag, and do \textit{not} use as a headline, two historical peaks, a single-tenant 1.000 (telcomart 50/50, acme 10/10) and a six-tenant 0.979 (143/146), as \textbf{cache-hit / lucky-shot artifacts}; \S{}5.5.5 recounts how the system's own machinery caught the inflation and re-baselined downward.

\textbf{Per-tenant enforced floors.} Table 2 gives the per-tenant floors carried by the live regression gate.

\begin{table}[t]
\centering
\small
\setlength{\tabcolsep}{4pt}
\caption{Per-tenant enforced route-accuracy floors (gated cohort).}
\begin{lrbox}{\mdtablebox}
\begin{tabular}{lll}
\toprule
Tenant & Floor & Count \\
\midrule
telcomart & 0.880 & 44/50 \\
zenith & 0.960 & 24/25 \\
acme & 1.000 & 10/10 \\
carenexus & 0.960 & 24/25 \\
forgemark & 0.889 & 16/18 \\
lumenscroll & 0.944 & 17/18 \\
\bottomrule
\end{tabular}
\end{lrbox}
\ifdim\wd\mdtablebox>\linewidth\resizebox{\linewidth}{!}{\usebox{\mdtablebox}}\else\usebox{\mdtablebox}\fi
\end{table}

(These enforced floors were recorded after an Opus-4.7 strict-mode model-drift event and are deliberately \textit{lower} than the stable-Gemini full-system numbers of Table 3, e.g. forgemark 0.889 enforced here vs 0.944 stable there, because a floor is a conservatively enforced worst case, not a best measurement.)

\subsubsection*{5.5.3 Reconstruction lift and operational findings (suggestive)}

The deltas below are the section's most interesting internal signals and also its most self-grading-vulnerable; we fence them accordingly as small-bank, indicative-not-definitive.

\textbf{The reconstruction lift (internal fixtures; suggestive, not the externally-validated result).} These measurements isolate the value of metadata reconstruction by comparing a tenant's accuracy with the catalog disabled to its accuracy with reconstruction enabled, on self-authored synthetic banks, so they are indicative of capability rather than external evidence; the externally-validated result is the calibrated abstention of Section 5.6, not these lifts. Table 3 reports the per-tenant full-system-vs-day-1-cold comparison on Gemini 3.1.

\begin{table}[t]
\centering
\small
\setlength{\tabcolsep}{4pt}
\caption{Full system vs. day-1 cold start, per tenant (Gemini 3.1, six-tenant cohort).}
\begin{lrbox}{\mdtablebox}
\begin{tabular}{llll}
\toprule
Tenant & Full system & Day-1 cold & $\Delta$ \\
\midrule
telcomart & 0.880 & 0.880 & 0 \\
zenith & 0.960 & 0.960 & 0 \\
acme & 1.000 & 0.900 & $-$10 pp \\
carenexus & 0.960 & 0.760 & $-$20 pp \\
forgemark & 0.944 & 0.944 & 0 \\
lumenscroll & 0.944 & 0.667 & $-$28 pp \\
\textbf{Total} & \textbf{0.932} & \textbf{0.856} & \textbf{$-$7.6 pp} \\
\bottomrule
\end{tabular}
\end{lrbox}
\ifdim\wd\mdtablebox>\linewidth\resizebox{\linewidth}{!}{\usebox{\mdtablebox}}\else\usebox{\mdtablebox}\fi
\end{table}

The cleanest controlled cold$\rightarrow$curated comparison was run on the altanova tenant (20 questions): \textbf{50\% with zero catalog $\rightarrow$ 90\% with an automated LLM pre-flight reconstruction (no human) $\rightarrow$ 95\% after curation.} The cold run fails \textit{all of Tier A}, because the system's weak-grounding traps correctly fire and the system refuses on ungrounded columns rather than guessing: exactly the intended behavior, and a demonstration that the abstention machinery is doing real work. A single-lever ablation on forgemark is \textit{suggestive} of LLM pre-flight's contribution: bare 0.389 (7/18) $\rightarrow$ +LLM-preflight 0.556 (10/18): a +16.7 pp move, but three questions on an 18-question bank, so indicative rather than a precise effect size. The same arc reproduces across tenants: carenexus from a bare day-1 24\% (column $F_1$ 0.538) to 96\% with automated pre-flight; zenith from a 6/25 = 0.240 baseline to 25/25 = 1.000 over the build arc (the 1.000 read with appropriate caution, see Section 6).

A recent live carenexus run (25 questions, automated pre-flight, full system) measured overall route accuracy \textbf{0.960}, column $F_1$ mean 0.598, mean L4 0.733, with per-tier route accuracy A 1.000 / B 1.000 / C 1.000 / D 0.857 (refuse precision 0.714) / E 1.000 (refuse precision 1.000), in 132.1 s of wall time.

\textbf{LLM pre-flight vs. human SME (led by coverage and cost, not the fragile route delta).} In one controlled comparison (a single tenant (carenexus), one unpaid human SME, 25 questions, everything else held fixed), an automated LLM-team "pre-flight SME" beat the human on the two robust axes: it covered \textbf{171/171 columns vs. 47/171} and, on the hardest tier (Tier B), achieved \textbf{80\% vs. 40\%}, at a cost of \textbf{\$2.89 of LLM spend and 16 minutes of wall time} against roughly \textbf{one-to-two human hours} at \$0. Its route-accuracy edge, \textbf{88.0\% vs. 76.0\%}, we deliberately treat as the \textit{weakest} of the three axes: a three-question delta with an unpaid-human confound. A reproducible \textbf{negative result} accompanies this: \textit{stacking} a hand-SME pass on top of the LLM pre-flight \textbf{regressed accuracy to 80\% (below the 88\% pre-flight-only number)}: the human descriptions were shorter and less specific, a cross-phase interference effect we report because it is counterintuitive and reproducible. The total LLM cost of the comparison session was \textbf{\$5.78} (two pre-flight runs at \$2.89). Reconstruction \textit{quality} itself is measured routing-independently as token-recall (0.667 macro; Section 5.1), which does not pass through our own routing at all.

\subsubsection*{5.5.4 Provider-agnosticism}

Because the routing decision is deterministic arithmetic over features, the system is not coupled to a particular LLM; the model is a feature provider, and swapping it tests an architectural invariance directly. Swapping the premium LLM from Anthropic Opus to Gemini 3.1 Pro (Vertex) moved the standard six-tenant gate by \textbf{0.925 $\rightarrow$ 0.932 (+0.7 pp)} and \textit{improved} adversarial robustness from \textbf{10/23 = 0.435 to 12/23 = 0.522 (+9 pp)}; on carenexus specifically the adversarial pass rose from 0/6 = 0.000 to 4/6 = 0.667. The claim this supports is not that the system is \textit{accurate} but that it is \textit{invariant to the model}, a property Spider cannot exhibit and that no reviewer can read as self-grading.

\subsubsection*{5.5.5 Engineering rigor, and catching our own inflation}

\textbf{Adversarial and engineering rigor.} Adversarial banks are constructed by saving only questions that broke the system, so their floors begin at 0.000 \textit{by construction} and climb as fixes land; the live gate currently floors most tenants at 0.000/N with lumenscroll at 1.000/1, and the bank improved over the build from 0/22 to a stable 3/22 to 9/22 after targeted signal additions. The test suite grew from 39 to \textbf{1,648 passing tests (2 skipped)}, with one regression test added per fix. The deep-pattern library matched 68/78 = 87\% of one tenant's catalog and lifted the result-match rate from 0.50 to 0.70.

\textbf{Catching our own inflation.} Twice, the system's own measurement machinery surfaced an inflated number and forced a downward correction \textit{before} it reached this paper: the episode we regard as this section's strongest evidence of trustworthiness. First, an LLM-response-cache key changed under model drift, and what had read as a 0.979 six-tenant peak (with a 1.000 single-tenant peak) re-baselined to \textbf{0.918}; the correction is recorded verbatim in the regression baseline (\textit{"Total now 134/146 = 0.918, down from 143/146 = 0.979 cache-hit measurement"}), and the gate now carries the corrected floor. Second, the selective-prediction evaluation of Section 5.1 was initially corrupted by an evaluation-harness bug (the in-process pipeline executed against the wrong warehouse, pinning the execution-success feature to a constant and making the routing score look useless), which we found, fixed, and re-ran at larger scale, \textit{reversing} the earlier weak-discrimination verdict (Section 5.1). Neither correction was forced by a reviewer; both were surfaced by the project's own gate and harness. We do not claim the system is bug-free; we claim its measurement discipline surfaces its own errors, which is the honest and the stronger claim.

\textbf{A candid note on execution accuracy.} The strongest \textit{kind} of accuracy (does the returned answer match a canonical execution?) is the metric we trust most in principle, but it is measured today on only \textbf{24 of 792 recorded runs}, always over single-digit verifiable subsets (e.g., acme 1.0 on 4 verified rows, forgemark 0.875 on 8, lumenscroll 0.714--0.833 on 6--7). \textbf{There is no large-N execution-accuracy number.} Moreover, the v1 result-match metric is column-projection-sensitive: list-style questions can register as misses even when the rows are correct. We therefore lean on route accuracy for headline claims and treat scaling execution accuracy as essential future work.

\subsection*{5.6 Query time: the external stripped-schema Spider study}

We have since run the external benchmark and the naive baseline that the previous draft listed as proposed. We use Spider \cite{yu2018spider}: gold SQL from the \texttt{xlangai/spider} dev split and the corresponding SQLite databases from \texttt{premai-io/spider}. Two arms share the mangler (P1). The \textbf{naive arm} (\texttt{rosetta/\allowbreak eval/\allowbreak spider\_bench.py}) shows the premium LLM the mangled schema and the question, executes its SQL against a copy of the mangled database, and compares (order-insensitive, numeric-tolerant multiset) against gold run on the original database. The \textbf{reconstruction arm} (\texttt{rosetta/\allowbreak eval/\allowbreak spider\_reconstruct.py}) bootstraps a \textit{throwaway} tenant from the mangled database through the full production path (deep profiler, automated LLM SME pre-flight, grounding), then drives the same \texttt{/ask} pipeline (\texttt{run\_ask}) used in production, with \textit{no} curation, no human SME, no query-log history, and no tenant-tuned detectors. Cold is deliberately the \textbf{weakest} reconstruction condition: a fully-automated day-1 catalog; we also run an oracle-documented condition (an upper bound) and an SME-affirmed condition (a realistic curation), below. The model is Gemini 3.1 Pro (\texttt{gemini-3.1-pro-preview}) throughout; all numbers below are from live API runs. We evaluate at scale (\textbf{11 Spider databases $\times$ 3 severities $\times$ 3 reconstruction conditions: 1,082 execution-labeled questions}) and report two quantities per question: the \textit{deployed route} (answer / confirm-first / refuse) and its correctness, and the \textbf{latent SQL accuracy} (whether the generated query execute-matches gold \textit{regardless of route}), which isolates generation quality from the routing decision. (An earlier, smaller version of this pilot under-reported the deployed routing because of the harness bug of Section 5.1; the \textit{latent} figures were unaffected, and the routing figures below are the corrected, larger-scale ones, with 95\% confidence intervals from a cluster bootstrap over databases. One database, \texttt{wta\_1}, is excluded because its source SQLite carries non-UTF-8 bytes that break the SQLite$\rightarrow$DuckDB load.)

\textbf{Naive degradation under stripping (11 databases $\times$ 3 severities, 396 calls).} Execution accuracy is \textbf{0.924 at severity 0.0} (clean), \textbf{0.917 at 0.5}, and \textbf{0.417 at 1.0} (full opacity). The mid point barely moves because the mangler's rename lever ramps in only above severity 0.4, leaving the queried identifiers mostly intact; the collapse is at full opacity. The operative observation is the failure \textit{mode}: at full strip the naive model \textbf{auto-emits a wrong answer on \textasciitilde{}58\% of questions with no signal of untrustworthiness}, precisely the silent-confident-error this paper targets. (On the four databases of the original smaller pilot the figures were 0.95 / 0.95 / 0.55; the broader eleven-database set is harder, which sharpens the full-strip collapse.)

\textbf{Table 4. Naive always-answer vs. cold reconstruction's calibrated gate, by severity (11 Spider databases, Gemini 3.1 Pro).} The naive arm answers every question; the reconstruction arm abstains (confirm-first / refuse) when grounding is low and auto-answers only a fraction, its \textit{coverage}, at a higher accuracy on what it emits.

\begin{table*}[t]
\centering
\small
\setlength{\tabcolsep}{4pt}
\begin{lrbox}{\mdtablebox}
\begin{tabular}{lllll}
\toprule
Severity & Naive: exec acc (coverage 1.0) & Cold recon: coverage & acc when answered & latent SQL acc \\
\midrule
Clean (0.0) & 0.924 & 0.65 & 0.81 & 0.76 \\
Mid (0.5) & 0.917 & 0.64 & 0.86 & 0.75 \\
Full strip (1.0) & \textbf{0.417} & \textbf{0.25} & \textbf{0.75} & 0.22 \\
\bottomrule
\end{tabular}
\end{lrbox}
\ifdim\wd\mdtablebox>\linewidth\resizebox{\linewidth}{!}{\usebox{\mdtablebox}}\else\usebox{\mdtablebox}\fi
\end{table*}

(Reconstruction arm: 11 databases, \textasciitilde{}12 questions per database per severity, three conditions. Building the SME-affirmed condition surfaced and fixed the in-process warehouse-resolution bug of Section 5.1; an earlier two-database version of this pilot also surfaced and fixed two scoring/parsing bugs (a brittle JSON extractor that discarded the premium model's chatty-but-valid SME descriptions, and a result comparator that mis-scored DuckDB \texttt{Decimal} values against gold floats); all three carry regression tests. The deployed-route numbers here are post-correction; the latent figures were never affected.)

\begin{figure*}[t]
  \centering
  \includegraphics[width=0.92\linewidth]{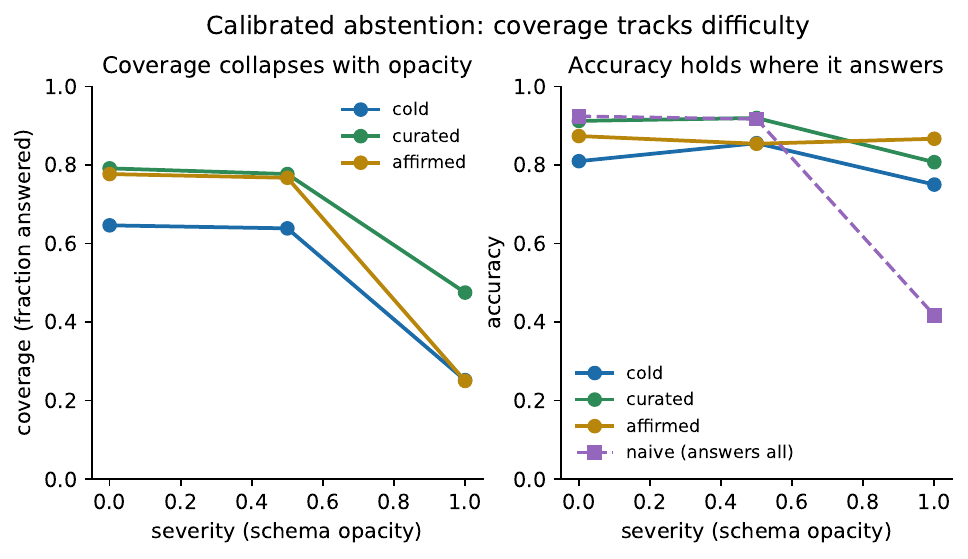}
  \caption{Calibrated abstention: as schema opacity rises and generation
  collapses, coverage collapses with it (left) while accuracy on what the gate
  does answer holds (right), across all three reconstruction conditions; the
  naive arm answers everything at falling accuracy.}
  \label{fig:covsev}
\end{figure*}

\textbf{Cold reconstruction generation: an honest negative on these toy schemas, at every severity.} Measured by \textit{latent} SQL accuracy (generation quality independent of the gate), the cold reconstruction pipeline sits \textbf{below} naive translation across the board: 0.76 vs. 0.92 on clean schemas, 0.75 vs. 0.92 at mid, and 0.22 vs. 0.42 at full strip. (The earlier two-database pilot showed near-parity on clean, 0.90 vs. 0.95; the broader eleven-database set, which includes harder schemas such as \texttt{dog\_kennels} and \texttt{employee\_hire\_evaluation}, removes that parity.) The failures are genuine, not scoring artifacts (the system emits \texttt{NULL} placeholders for columns it cannot map, or refuses outright). We report this negative plainly because it \textbf{scopes} where reconstruction's value does \textit{not} lie: it is the opposite direction from the internal 50$\rightarrow$90 lift, and the dissociation is the point. The internal lift comes from the curation/SME/detector layers a cold Spider bootstrap deliberately omits, over warehouses far more complex (and far more opaque to a general LLM) than \texttt{singer} and \texttt{pets}. The honest reading is that on toy schemas a premium model already handles, cold reconstruction adds nothing to \textit{generation}; its value here is elsewhere: in \textbf{knowing when its generation is wrong}, which is the calibrated-gate result we turn to next.

\textbf{Calibrated selective answering, confirmed externally (the central result).} With the gate executing correctly, the system does not abstain on everything; it abstains \textit{selectively}, and the selection is calibrated. Across the 1,082 reconstruction questions the three routes are cleanly monotone in both confidence and accuracy: it \textbf{refuses} 159 questions (mean $\ell_4$ 0.086) of which \textbf{only 1 was actually answerable}; it routes 283 to \textbf{confirm-first} (mean $\ell_4$ 0.589, 41\% latently correct), the genuinely-uncertain middle band; and it \textbf{auto-answers} 640 (mean $\ell_4$ 0.770) at \textbf{86\% accuracy (95\% CI 78--93\%)}, emitting 90 wrong answers in total (14\% of what it emits). Coverage, the fraction it auto-answers, is \textbf{0.59 (CI 0.51--0.65)}. (These pooled figures span all three reconstruction conditions and severities; the per-condition ladder below decomposes them (day-1 cold 0.51/0.82, SME-affirmed 0.59/0.86, and the oracle-documented upper bound 0.68/0.89), so the pooled operating point coincides with the realistic SME-affirmed condition, while Table 4 above reports the cold slice alone.)

The defining property is that \textbf{coverage tracks difficulty}. As stripping rises and generation collapses, the gate withholds more rather than emitting confident errors: at full opacity, where cold \textit{latent} accuracy falls to 0.22, cold \textit{coverage} falls to \textbf{0.25} while accuracy on what it does answer stays at \textbf{0.75}; the SME-affirmed condition behaves the same (latent 0.24, coverage 0.25, accuracy-when-answered 0.87). The contrast with the naive arm is the selective-prediction trade in concrete, external form: at full strip the naive arm answers \textbf{100\%} of questions at \textbf{42\%} accuracy, a 58\% silent-error rate, while the reconstruction arm answers \textbf{25\%} of them at \textbf{75\%} accuracy and abstains on the rest. The system converts a high-coverage, high-silent-error translator into a lower-coverage, low-error one, and, as Section 5.1 establishes on the same 1,082 samples, the confidence score driving that conversion both discriminates correct from incorrect (normalized AURC 0.26, CI excluding random) and is calibratable (ECE to 0.026). This is the calibrated-abstention guarantee of Section 4.3 demonstrated, on an external benchmark with confidence intervals, as an actual \textit{selective-prediction advantage}: not merely the architectural argument that the authority ceiling \textit{forces} abstention, but the measured result that the gate abstains \textit{where it should}.

\begin{figure}[t]
  \centering
  \includegraphics[width=0.9\linewidth]{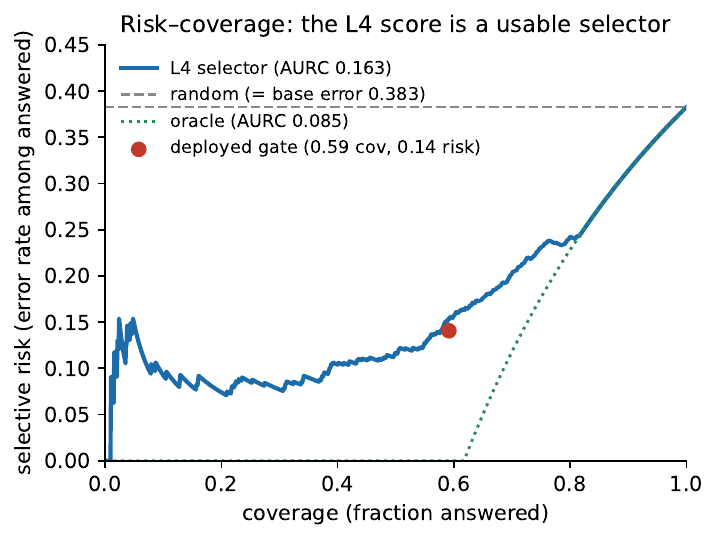}
  \caption{Risk--coverage curve (L4 score as the selector). The selective risk
  stays far below the random-ordering reference until high coverage; the
  deployed three-way gate sits on the favourable part of the curve.}
  \label{fig:riskcov}
\end{figure}

\begin{figure}[t]
  \centering
  \includegraphics[width=0.85\linewidth]{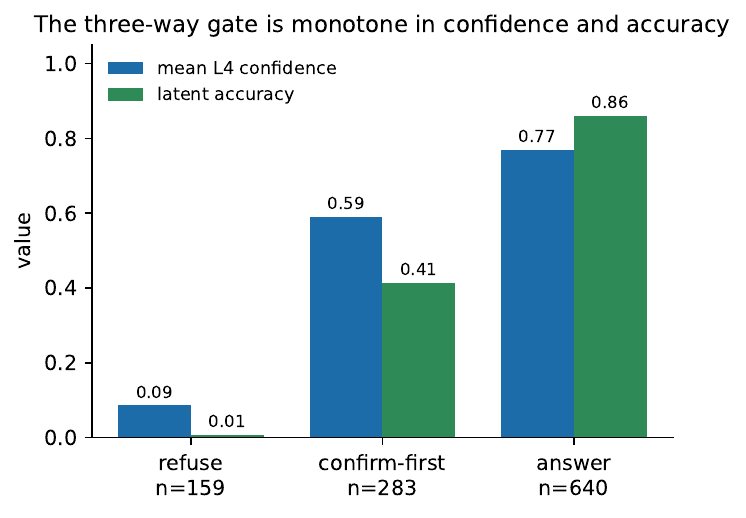}
  \caption{The three-way gate is monotone in both confidence and accuracy:
  refusals carry near-zero L4 and are almost never answerable; confirm-first is
  the uncertain middle band; answers carry high L4 and high accuracy.}
  \label{fig:routes}
\end{figure}

\textbf{The baseline here is a lower bound, not the fair abstention comparison.} Our naive arm \textit{always answers}, so this study establishes that the gate is safer than a non-abstaining translator; it does \textbf{not} yet establish that our authority-grounded confidence is better-calibrated than a \textit{cheap, generic} abstention mechanism bolted onto the same naive model (a token-logprob or sequence-probability threshold, self-consistency spread across samples, or verbalized confidence). Such a naive-plus-abstention gate would close part of the silent-error gap on its own. To credit the authority ladder specifically, the gate must beat a naive-plus-abstention baseline, not a naive always-answerer; sweeping that baseline's operating point on the same 11 databases is the immediate next measurement (Section 5.7), and until it is run this comparison should be read as "safer than never abstaining," not "better-calibrated than any abstention."

\textbf{Honest caveats specific to this pilot.} The result is not uniform across databases: deployed accuracy-when-answered ranges from \textbf{1.00} (\texttt{network\_1}, \texttt{pets\_1}: zero wrong answers) down to \textbf{0.56} on \texttt{employee\_hire\_evaluation}, where the gate over-answers (it emits 28 wrong answers out of 63 it auto-answered). The cluster-bootstrap confidence intervals are computed precisely so this database-level variation is reflected rather than hidden, but the outlier is real: on some schemas the gate's confidence is miscalibrated upward and it emits confident errors, which is exactly the failure the architecture is meant to prevent. Whether that reflects genuine over-confidence or a scoring artifact on that schema's particular queries we do not resolve here. The latent-accuracy metric, separately, scores a query the system chose \textit{not} to emit, so it measures generation capability, not delivered behavior; we report it alongside, never instead of, the deployed routing. With those caveats, the finding that holds is sharper than the prior draft's "zero wrong answers": the gate does not abstain on everything (that earlier claim was the harness bug), it answers \textit{selectively}, and on average, with a database-clustered 95\% CI, it answers a majority of questions at far higher accuracy than the naive arm's silent-error rate while withholding most of the questions it would get wrong.

\begin{figure}[t]
  \centering
  \includegraphics[width=0.9\linewidth]{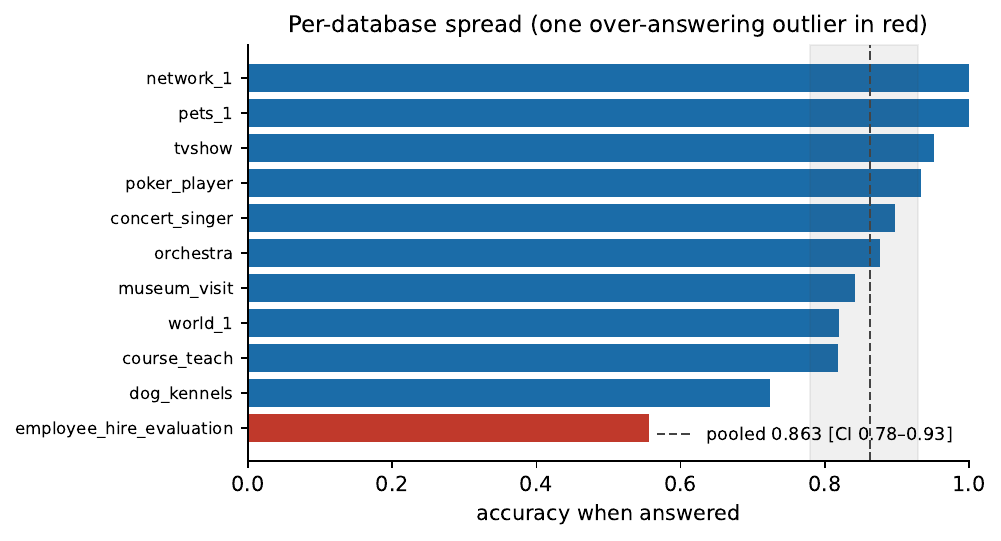}
  \caption{Per-database accuracy-when-answered (the honest spread). The result
  holds across databases around the pooled 0.863 [CI 0.78--0.93], with one
  over-answering outlier (\texttt{employee\_hire\_evaluation}).}
  \label{fig:perdb}
\end{figure}

\textbf{The reconstruction-value ladder: cold, SME-affirmed, and documented conditions.} To probe where reconstruction's value lies, we ran two further conditions beyond cold. The \textbf{oracle-documented} condition (\texttt{scripts/\allowbreak spider\_curate\_docs.py}) overlays a \texttt{DOC\_CITED\_STRONG} description on every column derived from its original pre-mangle name: an enterprise warehouse whose opaque physical columns are explained by a maintained data dictionary; it is an \textit{upper bound}, clearly labeled, because it injects the true semantics rather than inferring them. The \textbf{SME-affirmed} condition (\texttt{scripts/\allowbreak spider\_affirm\_facts.py}) is the honest analog of real curation: it takes the descriptions the cold pipeline \textit{already reconstructed} and promotes them to the SME\_AFFIRMED authority tier through the exact production "accept" path a human reviewer uses (\texttt{interaction.commit\_reviewed}), an SME who accepted the system's \textit{own} reconstruction, not one who supplied oracle text. The two conditions bracket the realistic system.

The ladder behaves as the authority architecture predicts. Coverage and accuracy-when-answered both rise from cold (coverage 0.51, accuracy 0.82) through SME-affirmed (0.59, 0.86) to oracle-documented (0.68, 0.89): more authoritative metadata lifts more questions into the answer region \textit{and} makes the emitted answers more accurate. This corrects a specific claim in the prior draft: that documentation "lifts grounding from refuse into confirm but not to auto-answer." With the gate executing correctly, both the documented and the SME-affirmed conditions reach the answer region routinely; the earlier non-result was the warehouse-resolution bug, not a property of the authority ladder. (On \textit{generation} alone the documented condition still does not beat naive on these toy schemas (latent 0.66 vs. naive's 0.92 on clean), consistent with the cold negative above; its value, again, is in the calibrated gate, where higher authority buys both coverage and precision.)

\textbf{Measured: the fair baseline, naive-plus-abstention.} Table 4 pits our gate against a naive arm that \textit{always answers}; the sharper test, deferred by the previous draft, is whether the authority-grounded gate is better \textit{calibrated} than the same naive model given a cheap, generic abstention mechanism. We built two and traced each as a risk-coverage curve over the same 11 databases and severities (\texttt{rosetta/\allowbreak eval/\allowbreak spider\_abstention.py}; 396 questions, \$1.47; analysis free via \texttt{scripts/\allowbreak spider\_abstention\_compare.py}): \textbf{self-consistency} (five samples at temperature 0.7: the emitted answer is the modal executed result, the confidence the fraction of samples that agree) and \textbf{verbalized confidence} (the model self-reports a probability). A third cheap signal often used for LLM abstention, a \textbf{token-logprob or sequence-probability threshold}, we cannot evaluate: the managed Gemini API used throughout this work returns no token log-probabilities, so a logprob baseline is \textit{not available} on this provider; we note the gap rather than substitute a proxy, and observe that self-consistency agreement is the closest sampling-only surrogate we can compute. The result is two-sided, and we report both halves.

\textit{Where the cheap baseline suffices.} Pooled across severities the naive signals discriminate about as well as our gate (normalized AURC \textbf{0.26} (verbalized) and \textbf{0.32} (self-consistency) against the cold gate's \textbf{0.33}) and, because naive \textit{generation} is stronger on Spider's toy schemas (the negative above), they deliver comparable accuracy at matched coverage. Part of every \textit{pooled} score here is mere difficulty-detection: a confidence that is high on clean schemas and low on stripped ones correlates with correctness across the severity gradient without ranking correctness \textit{within} a difficulty, so the per-severity numbers below are the cleaner measure. On clean and mid schemas the naive signals carry almost no within-severity ranking signal (normalized AURC 0.82--1.0, at or worse than random), but at 92\% base accuracy there is little to rank. The honest reading is that \textbf{on documented or lightly-degraded schemas a cheap naive gate is enough; the authority ladder is not required there.}

\textit{Where it is not: full opacity.} The gate's advantage concentrates exactly where the paper locates the problem. At full strip the gate's L4 discriminates correct from incorrect with normalized AURC \textbf{0.11} ($\approx$89\% of the ranking signal) against \textbf{0.41} (verbalized) and \textbf{0.47} (self-consistency): the naive model's own confidence collapses toward random once readable names are gone (stripped of schema cues it cannot tell which of its guesses are right), while the authority-bounded score still can. This superior \textit{calibration} does \textbf{not}, on these toy schemas, become higher \textit{delivered} accuracy. At matched coverage the naive arm's stronger generation (full-strip base accuracy 0.39 vs the cold gate's 0.22) offsets the gate's better ranking, so naive self-consistency matches or slightly exceeds the gate on accuracy-when-answered (e.g. 0.56 vs 0.44 at 50\% coverage), \textit{except} at the lowest coverage, where the gate matches it (0.69 vs 0.67 at 25\% coverage) from a much lower base accuracy (0.22 vs 0.39): the signature of the better ranking. The regime in which better calibration \textit{and} competitive generation would combine to beat the naive gate outright is a complex, genuinely-undocumented warehouse. The \textit{generation} half of that question is now answered, negatively, on BIRD's real schemas (Section 5.7: cold reconstruction does not out-generate naive there either); what remains untested is only whether a \textit{fully-curated} system on a genuinely-undocumented warehouse could combine calibration with competitive generation to win on delivered accuracy.

\textbf{A cheaper baseline still, the execution-sanity gate, collapses under opacity.} The weakest abstention mechanism available to a naive translator is to answer when its SQL \textit{executes} and abstain when it errors. We re-executed every naive query against the same mangled database it was generated on (deterministic seed-0 mangle; \texttt{scripts/\allowbreak spider\_exec\_sanity\_baseline.py}, analysis-only, 1,452 questions): the gate is a near-no-op. Because a query stripped of readable column names is still \textit{syntactically valid} (it selects the wrong renamed column but runs), its execution rate stays at $\approx$100\% across every severity, so it answers essentially everything, and its silent-error rate climbs from 7\% on clean schemas to \textbf{72\% at full strip}, tracking the naive always-answerer exactly. The authority gate on the identical questions instead lowers coverage from 65\% to 33\% as grounding erodes and holds silent error at \textbf{$\leq$15\%} throughout. This is the sharpest form of the point: even a \textit{free} abstention signal does not close the opacity gap, because the failure mode under schema opacity is confident-but-wrong \textit{execution}, not execution failure, and that failure mode is precisely what an authority-bounded confidence, and not a syntactic check, is built to catch.

\textbf{This calibration gap holds at full dev-set breadth.} Extending the naive baseline across all 19 dev databases (the breadth study of Section 5.7; \texttt{scripts/\allowbreak spider\_abstention\_baseline.py --dbs breadth}, 1,452 pooled questions) reproduces both halves. Naive self-consistency accuracy degrades from \textbf{0.92} on clean schemas to \textbf{0.39} at full strip: the always-answer collapse, now on the full dev set. And at full strip the naive model's own self-consistency confidence discriminates no better than chance (normalized AURC \textbf{0.55}) while the authority-bounded gate still ranks correct above incorrect (\textbf{0.20}): the same roughly-threefold calibration edge under opacity, measured on 19 databases rather than eleven. The cheap baseline's edge on clean schemas and the gate's edge under opacity both survive the move to breadth.

We take this as the honest resolution of the objection the previous draft anticipated: a naive-plus-abstention baseline \textit{does} close most of the always-answer silent-error gap, so the delivered-behaviour result of Table 4 is an advantage over a \textit{non-abstaining} model, not over any abstaining one. The gate's distinctive, \textit{measured} edge over a cheap abstaining baseline is calibration quality under opacity, together with the properties the architecture exists for and a naive gate has none of: auditability, prompt-injection resistance, and a route decided by deterministic arithmetic rather than the LLM's own self-report.

Two further baselines a reviewer might expect we treat directly. A pure \textbf{retrieval / data-catalog} baseline is ill-posed in this setting: under full opacity there are no readable names to retrieve \textit{against}, so retrieval must operate over the column \textit{values}, which our deep profiler already fingerprints, or over the \textit{reconstructed} catalog, which is the system under test itself; a name-based retriever therefore degenerates rather than forming a clean comparison. And a strong \textbf{agentic text-to-SQL} system (schema-linking with decomposition and self-correction) is a stronger \textit{generator}, not a stronger \textit{calibrator}: the execution-sanity result above shows the calibration gap survives even a generator whose SQL always runs, because under opacity the failure is confident-but-wrong execution, so we treat a full agentic-generator comparison as complementary future work rather than a load-bearing baseline for the calibration claim this paper makes.

\subsection*{5.7 Query time: BIRD and full-Spider-dev breadth}

The breadth item that stood here as proposed (extending the stripped-schema study from Spider's toy schemas to BIRD's larger, noisier real-world databases) has since been \textit{run} at full opacity and is reported below; it answers the open generation question with a clean negative and sharpens the abstention result on real schemas. The second proposed item, the full Spider dev set on a continuous severity grid, has now also been run in the cold day-1 condition and is reported below. Two components remain genuinely unrun (leave-one-tenant-out and large-N internal execution accuracy), each needing more live-model budget than the studies above.

\textit{(The naive-plus-abstention baseline that stood here as the priority experiment has been run; its two-sided result (cheap naive abstention matches our gate on toy schemas, our measured edge is calibration under opacity) is reported in Section 5.6.)}

\textbf{Measured: BIRD breadth (real-world schemas).} We extended the reconstruction arm to BIRD mini-dev, whose 11 databases (finance, sports, healthcare, and others) carry schemas materially harder than Spider's. The gap shows in the naive baseline alone: a premium translator that scores 0.92 execution accuracy on clean Spider scores only \textbf{0.47 / 0.64} (evidence off / on) on clean BIRD and \textbf{collapses to 0.07 / 0.11 at full opacity}, precisely the regime where reconstruction's \textit{generation} had the most room to help. We ran the full-strip (severity 1.0) reconstruction slice (cold and curated-documentation conditions, both evidence modes, matched question-for-question to the naive baseline), completing eight of the eleven databases fully plus the cold condition of a ninth (\texttt{european\_football\_2}, the largest at 598 MB, whose curated-documentation condition did not finish). The two largest remaining catalogs, \texttt{card\_games} (115 columns) and \texttt{codebase\_community}, initially failed the deep-profiler bootstrap; we traced this \textbf{not to a scaling limit but to a concrete, since-fixed defect}: an extreme-range column produces a non-finite (\texttt{inf}) profiling statistic, which Python's \texttt{json.dumps} serialises as the token \texttt{Infinity} that PostgreSQL's \texttt{jsonb} type rejects, aborting the evidence write. A recursive sanitizer at the single evidence-write boundary fixes it, and both databases then bootstrap and answer, giving \textbf{11/11 BIRD coverage}. Because their 100+-column catalogs make for very large prompts, each answered question is expensive, so we sampled these two recovered databases at reduced depth (five full-strip cold questions each), a query-time cost we return to under scalability, not a profiling limit. At full strip both behave as the thesis predicts: \texttt{codebase\_community} refuses or defers all five (latent accuracy 0), and \texttt{card\_games} answers one (incorrectly) and abstains on the rest.

The two findings are reported straight. \textit{Generation is a clean negative for cold reconstruction, now on real schemas:} cold reconstruction does \textbf{not} out-generate naive even where naive collapses. Pooled full-strip latent accuracy is \textbf{0.016 / 0.042} (evidence off / on) for cold reconstruction against \textbf{0.084 / 0.134} for naive. The hypothesis that reconstruction's generation would win on hard real schemas is not supported for the deployable cold day-1 arm; the Spider generation-negative holds on BIRD, and we close that door explicitly. The one nuance the frozen record exposes is \textit{confirmatory} of the thesis rather than of the contribution: full \textbf{curated} documentation reaches \textbf{0.041 / 0.161} at full strip, still below naive without the benchmark's evidence hint (0.041 vs 0.084) but \textit{above} naive once the hint is supplied (0.161 vs 0.134), so genuine metadata, and only genuine metadata, lifts generation past a collapsed naive model, exactly the \textit{metadata-quality} lever the paper argues for and exactly not the cold day-1 reconstruction whose generation we decline to claim.

\textit{Abstention is the result that transfers and strengthens.} At full opacity the deployed gate is near-silent: without BIRD's evidence hints it auto-answers only $\approx$5\% of questions (21 of 379 valid), emitting \textbf{17 confident errors where a naive always-answerer emits 347} and \textbf{avoiding $\approx$95\% of naive's silent errors by refusing what it cannot ground}; with the hints supplied it answers \textit{nothing} (0 confident errors against 311 naive silent errors). The sharper result is discrimination: on the 738 full-strip questions that produced a valid attempt Rosetta's authority-grounded score still ranks correct above incorrect (normalized AURC \textbf{0.300}, better than random), while \textbf{the naive model's own self-consistency confidence is worse than random (unclamped normalized AURC 1.19, anti-correlated with correctness)}. On a fully-stripped real schema the naive model regenerates the \textit{same} wrong SQL across samples, so it is most confident exactly where it is wrong. (These outcome statistics exclude API/parse/timeout system-failures under the frozen-record policy of Section 5.12: the gate's structured multi-call pipeline hard-failed on 8.2\% of full-strip questions and 15.7\% across all conditions, which we report separately as a system-failure rate rather than fold into coverage; the naive single-shot baseline produced an answer on every question, so the comparison is not flattered on that axis.) This is the strongest external evidence for the paper's central claim: on hard real schemas a naive model cannot tell when it is wrong, and a grounded gate can. It is also why the cheap naive-plus-abstention baseline that matched us on Spider's toy schemas (Section 5.6) would fail here: its abstention signal is anti-correlated with correctness under full opacity.

We are explicit about the BIRD result's limits. Accuracy-\textit{when-answered} at full opacity is low (0.19): at this extreme the gate's value is refusal, not accurate answering, because generation has collapsed for every method. Calibration is again not uniform: \texttt{toxicology} over-answers (10 confident errors at zero accuracy), and \texttt{card\_games} (one of the recovered giants) over-answers on the single full-strip question it auto-answers, both the BIRD analog of the Spider \texttt{employee\_hire\_evaluation} outlier. Coverage-tracks-difficulty is confirmed on the two databases that ran the full severity range (auto-answer coverage 0.23$\rightarrow$0.29$\rightarrow$0.02 on \texttt{superhero}, 0.12$\rightarrow$0.11$\rightarrow$0.07 on \texttt{student\_club}, collapsing toward full strip). With the deep-profiler defect fixed, BIRD coverage is now 11/11; the two recovered giants are sampled at reduced depth (a query-time cost on 100+-column catalogs), and \texttt{european\_football\_2}'s curated-documentation condition remains the one unfinished cell, so the breadth is complete in coverage though uneven in per-database depth.

\textbf{Measured: the full Spider dev set on a continuous severity grid.} The study now spans \textbf{19 of Spider's 20 dev databases} (all but \texttt{wta\_1}, whose non-UTF-8 bytes break the DuckDB load) across a \textbf{continuous six-point severity grid} (0.0, 0.2, 0.4, 0.6, 0.8, 1.0), in the cold day-1 condition: 1,421 execution-labeled cold questions, up from 11 databases at three discrete severities. The central behaviour holds at breadth and at finer resolution. \textbf{Coverage tracks difficulty monotonically:} the deployed gate answers 65\% of questions on clean schemas and only 33\% at full strip, refusing more as grounding erodes, while accuracy-when-answered stays high (0.75--0.90) until full strip and then falls to 0.57, and the silent-error rate stays bounded (6--15\%) at every severity, against a naive always-answerer that is silently wrong on the majority of fully-stripped questions. Pooled over the grid, the cold gate answers \textbf{59\% of questions at 81\% accuracy} with an 11.5\% silent-error rate: the day-1 slice of the 86\%-at-59\% deployed result of Section 5.6, now confirmed across the full dev set. Discrimination is stable as the study widens: the cold routing score's normalized AURC moves only from 0.33 (11 databases, three severities) to 0.39 (19 databases, six severities), better than random throughout; the finer grid honestly exposes that the intermediate severities (notably 0.6, a partial-rename regime) are where the gate is most tempted to over-answer. That over-answering is not confined to one database: at full strip several schemas (\texttt{cre\_Doc\_Template\_Mgt}, \texttt{flight\_2}, \texttt{battle\_death}, alongside the previously flagged \texttt{employee\_hire\_evaluation}) auto-answer a handful of low-arity aggregations they get wrong, so over-answering is a \textit{systematic} full-strip effect rather than a single-database artifact. It is not, however, a single \textit{mechanism}: the role-ambiguity signal that explains \texttt{employee\_hire\_evaluation} does not separate wrong from right answers on these three schemas (Section 6), so their over-answering remains undiagnosed.

\textbf{The reconstruction arc reproduces on public schemas at breadth.} With the curated (oracle-documentation) and SME-affirmed conditions now also run across all 19 databases, the cold$\rightarrow$curated arc of the internal case study (Section 5.5) reproduces externally. Supplying oracle documentation lifts pooled coverage from \textbf{59\% to 74\%} and accuracy-when-answered from \textbf{81\% to 91\%}, \textit{halves} the silent-error rate (\textbf{11.5\%$\rightarrow$7.0\%}), and sharpens discrimination (cold normalized AURC 0.39 $\rightarrow$ curated \textbf{0.19}); at full strip it more than doubles latent recovery (\textbf{0.22$\rightarrow$0.47}) and keeps the gate accurate when it does answer (0.57$\rightarrow$0.83). This is the same lever the internal ablations reported (better metadata, better delivered behaviour), now measured on public databases the detectors were never tuned on. The SME-affirmed condition, which promotes the system's \textit{own reconstructed} descriptions to high authority rather than oracle text, lands between cold and curated and does \textbf{not} improve discrimination (AURC 0.37, essentially cold's): the external echo of the internal finding that a human-on-top stack that over-trusts imperfect reconstructions can regress. Only the matched naive-plus-abstention baseline at breadth now remains open here; leave-one-tenant-out has since been measured (Section 5.10).

\textbf{The per-database variance is now characterized, not yet cured.} The larger database set called for here has been run (19 dev databases, above): the over-answering is a \textit{systematic} full-strip phenomenon, not a single-database artifact, and the per-database diagnosis (Section 6) resolves the mechanism for the flagged outlier, type evidence decoupled from role and join structure. The role-ambiguity signal named there has since been \textit{measured}: it carries real, gate-invisible information on the database it was diagnosed from, and fails to generalize to the other over-answering schemas (Section 6). What remains is to \textit{implement} the cap (plus a join-path grounding term independent of per-column type confidence) and re-measure whether it lifts calibration uniformity without over-refusing on clean schemas.

\textbf{Large-N internal execution accuracy: now run (Sections 5.10--5.11).} The external Spider study reports execution-grounded routing on 1,082 questions; the \textit{internal} execution-accuracy number, previously a 24-run subset, has since been scaled across the production cohort (six tenants, 146 questions, every answered query re-executed against its warehouse), giving a pooled accuracy-when-answered of 86\% (57/66) and a full route confusion matrix (Section 5.11).

\textbf{Leave-one-tenant-out generalization: now measured (Section 5.10).} Because the trap detectors and signals were authored against the very questions they are scored on, the strong internal runs may reflect overfitting rather than capability. Two independent answers now stand. The Spider study evaluates cold on nineteen \textit{public} databases the detectors were never tuned on, and the calibrated gate holds there. And within the production cohort, Section 5.10 establishes leave-one-tenant-out \textit{by construction} (the abstention gate has no per-tenant parameters, verified in the catalog) and confirms it empirically with execution verification: the single frozen gate answers 46\% of questions at 86\% accuracy-when-answered across six heterogeneous tenants, with the last-onboarded tenant held out as an explicit temporal control.

\subsection*{5.8 Ablation: which part of the ladder carries the discrimination?}

The routing score composes several evidence rungs, and a fair reviewer asks which rung actually carries the calibration, and whether the composition beats the LLM's own self-report. We answer with a leave-one-component-out ablation on the execution-labeled cold data: because the score is a fixed weighted sum of features that are all logged, we re-score every question with one rung's weight zeroed and compare normalized AURC and deployed behaviour, with no new model runs. The re-scoring reproduces the production score to within $5\times10^{-7}$, so each ablated variant is the real router with one input removed. The comparison runs on the 1,349 cold records carrying a complete feature vector (a matched within-subset comparison); 72 records the gate refused \textit{before} generating SQL (and thus before any feature existed) are necessarily excluded, which is why the full-strip baseline here (0.256) is slightly above the 0.204 of the full deployed set in Section 5.6, where those correctly-refused records are counted.

Three findings, reported straight.

\textit{Execution grounding is the load-bearing rung.} Removing it (the execution-success signal plus the row-count and magnitude sanity penalties) degrades discrimination the most, and by a wide margin: normalized AURC rises from 0.439 to \textbf{0.588} across severities and from 0.256 to \textbf{0.358} at full strip. No other single rung's removal comes close. The discrimination lives primarily in whether the generated SQL runs and returns a plausible result: exactly the signal a names-only model cannot see.

\textit{Catalog and link confidence carry coverage, not ranking.} Removing either the reconstructed catalog confidence or the L3 link probability barely moves the rank order (nAURC even nudges down) but collapses the answer region: coverage falls from 0.79 to 0.05 (no catalog) or to essentially 0 (no link), because without those rungs almost nothing clears the answer threshold and the gate refuses nearly everything. These rungs are load-bearing for \textit{delivered coverage}, letting the gate reach the answer region on grounded questions, rather than for rank order.

\textit{The ladder ties the LLM's self-report on ranking, and beats it on delivered behaviour: a correction to our earlier down-payment.} On the matched subset the full authority score and the generator's own verbalized self-confidence discriminate about equally (full-strip normalized AURC \textbf{0.256 vs 0.260}; across severities 0.439 vs 0.429): the ladder does \textbf{not} out-rank a free-floating self-estimate, and an earlier version of this section that reported it doing so was comparing the full score on the whole set against the self-report on the feature subset, not a matched comparison. Where the ladder wins is calibrated abstention: routed at the deployed thresholds, the full ladder answers 79\% of questions at a \textbf{22\%} silent-error rate, while ranking by the LLM's self-confidence alone answers 98\% at a \textbf{32\%} silent-error rate. The authority ladder's measured contribution is therefore not a sharper ranking but \textit{knowing when to abstain}: it converts a comparable ordering into materially safer delivered behaviour, exactly the calibration-under-opacity thesis of Section 5.6 rather than a claim to a better guess. (The critic second-pass rung is inert on cold Spider: its removal changes nothing, consistent with its design of firing only in the medium-confidence band, which cold Spider rarely enters.)

The variants that cannot be recovered from the logged features (collapsing the authority caps at reconstruction time, removing the SME tier, and replacing the deterministic router with LLM route-arbitration) change the authority tier \textit{before} the features are computed and so require re-bootstrapping the study under the variant configuration; the SME-tier removal is partially observed already in the cold-versus-affirmed comparison of Section 5.7, and we scope the remainder as the next controlled measurement. (Analysis: \texttt{scripts/\allowbreak gate4\_ablation.py} over \texttt{var/\allowbreak spider\_pooled.jsonl}; the report is \texttt{papers/\allowbreak data/\allowbreak gate4\_ablation\_report.txt}.)

\subsubsection*{5.8.1 A router baseline matrix, and a negative result about authority}

The ablation above asks which of \textit{our} rungs matters. The sharper question a reviewer asks is whether our score beats \textit{other people's} selectors on the same candidate SQL. We therefore hold the generator \textbf{fixed} (every router below sees the identical query, on identical rows, against identical execution labels) and vary only the signal used to decide whether to answer. Learned routers are fit with \texttt{GroupKFold} grouped by database and scored \textbf{only out-of-fold}, so no router is ever evaluated on a schema it was trained on. This is free and reproducible from the committed pooled data (\texttt{scripts/\allowbreak wsC\_baseline\_matrix.py}; results in \texttt{papers/\allowbreak data/\allowbreak wsC\_baseline\_matrix\_\{cold,\allowbreak pooled\}.json}).

The result does not favour us, and we report it plainly. On the deployed cold condition (1,421 execution-labeled questions, 19 databases), a \textbf{learned multivariate-Platt selector over purely generic signals} (the \cite{liu2025subclause} analogue, using only generator self-confidence, link strength, SQL complexity, executability, and result-sanity) \textbf{outperforms the deployed authority score on every axis we measure}: normalized AURC \textbf{0.268 vs 0.388}, ECE \textbf{0.038 vs 0.150}, and coverage at a 10\% error budget \textbf{0.345 vs 0.027}. A gradient-boosted variant lands between them (0.294). Among unlearned single signals the generator's own verbalized confidence (0.380) essentially matches our full score, consistent with the tie reported above.

The decisive test isolates \textit{authority} rather than \textit{learning}: holding the learner fixed and toggling only the four authority features (reconstructed-catalog mean/min confidence, catalog coverage, docs-anchored ratio), the change in discrimination is \textbf{+0.007 (LR) and +0.002 (GBM) on cold, and $-$0.006 / $-$0.009 pooled across conditions}, indistinguishable from noise, and negative as often as positive. Authority features \textit{alone} rank barely better than chance (normalized AURC 0.992 cold, 0.776 pooled). \textbf{Within the regime this benchmark can measure, the authority ladder contributes no incremental discrimination.}

We diagnose why, because the mechanism matters more than the number. The authority features are populated on \textasciitilde{}90\% of questions but are \textit{near-constant within a condition}: cold takes ten distinct catalog-confidence values, all capped at 0.5, while curated and affirmed take two apiece. The ladder is built to \textbf{cap} confidence per evidence tier, not to rank questions inside one. It therefore cannot carry within-tier discrimination by construction, and pooling across tiers does not rescue it (the tier is nearly collinear with the condition label). One feature, \texttt{docs\_anchored\_ratio}, is identically zero in every condition and is dead weight in the current configuration.

\textbf{Directly ablating the caps agrees.} The re-scoring above cannot touch the ceilings themselves, because a cap changes confidence \textit{before} the feature vector exists, so we added an ablation flag (\texttt{ROSETTA\_ABLATION=no-caps}, gated at the single function every one of the five clamp sites reads) and re-bootstrapped. Comparison is strictly paired on the same questions, with McNemar's exact test on delivered wrong answers. Across \textbf{41 paired questions in two difficulty regimes}, removing every ceiling changed the delivered outcome on \textbf{exactly one question}, and the mechanism was visible when it did (\texttt{battle\_death}, score 0.000 $\rightarrow$ 0.766, refuse $\rightarrow$ answer, and the answer was wrong). The aggregate at full opacity moved the predicted way (coverage 0.033 $\rightarrow$ 0.133, accuracy-when-answered 1.00 $\rightarrow$ 0.75) but one discordant pair cannot reach significance: McNemar gives \textit{p} = 1.00, and with a single discordant pair 1.00 is the \textit{smallest attainable} two-sided value. At mid-severity the two arms were \textbf{identical} (coverage 0.909, silent-error 0.364, zero discordant pairs).

The reason the caps rarely bind is instructive rather than disappointing, and it is the same reason in both regimes viewed from opposite ends. At full opacity the score is already far below the answer threshold (link strength, generator self-confidence, and the sanity penalties refuse the question without help), so lowering an already-irrelevant ceiling changes nothing. At mid-severity the score is already comfortably \textit{above} the threshold on the strength of those same features, so raising the ceiling changes nothing either. Catalog confidence enters $\ell_4$ with a small weight; the ceiling can only be decisive in the narrow band where it is the marginal term.

Taken with the leave-one-component-out result above (execution grounding is the load-bearing rung; catalog confidence carries coverage, not ranking) and the router matrix (authority adds no incremental discrimination), \textbf{three independent lines of evidence converge}: the authority ladder is \textit{not} the primary driver of this system's delivered safety; execution grounding and the deterministic threshold are. We state that plainly because it is what we measured. What the ladder does provide, and what we do still claim, is label-free provenance and an auditable ceiling: a reason, attached to every fact, for why the system was or was not permitted to be confident. We flag the sample as small (41 paired questions across two regimes) and the full-severity-grid re-bootstrap as the obvious next measurement, but the direction is consistent across every analysis we have run.

Two caveats bound this negative in \textit{both} directions. First, in our favour: the learned baselines are fit on execution-labeled outcomes which, out-of-fold or not, come from the benchmark; a real undocumented warehouse supplies no such labels, whereas the authority score is computed without any. That is a genuine deployment distinction and we rely on it explicitly rather than treat it as an escape hatch. Second, against us: it means the honest statement of our contribution is \textit{not} "a better confidence estimator." What survives measurement is the label-free, auditable, provenance-carrying construction of a confidence ceiling (and the abstention behaviour of Sections 5.6, 5.7 and 5.4 that it produces), not a superior ranking of correct from incorrect. Whether the authority caps earn their place through the variants this study cannot re-score offline (collapsing caps at reconstruction time, removing the SME tier, LLM route-arbitration) requires the full re-bootstrap we scope above, and that measurement is now the most important open item in this paper rather than a refinement.

\subsection*{5.9 Cost, latency, and scalability}

Because this is a data-management pipeline and not merely an LLM prompt, we report its operating cost and scaling behaviour directly from the run logs (\texttt{var/\allowbreak llm\_cache.sqlite}; \texttt{scripts/\allowbreak llm\_cost\_audit.py}).

\textbf{Cost.} The \textit{entire} body of experiments in this paper (every reconstruction bootstrap and every ask, across Spider (19 databases $\times$ six severities $\times$ three conditions), BIRD (11 databases), and the internal cohort) cost \textbf{\$281} and consumed \textbf{105M tokens over 55,256 model calls}. The premium reasoning model (Gemini 3.1 Pro) accounts for \$137 (72M tokens); the balance is orchestration on a larger model. Per unit the pipeline is inexpensive: a single model call costs a median \textbf{\$0.0024}, and reconstructing one column's semantics through the SME pre-flight costs about \textbf{\$0.0027}, so a cold catalog for a typical Spider database (20--40 columns) reconstructs for a few cents.

\textbf{Latency.} Per-call latency has a low median (\textbf{5.3 s}) but a heavy tail (p90 15 s, p99 57 s). The tail is almost entirely a function of \textit{schema width}: a single \texttt{/ask} issues several model calls (generation plus critic, with retries), and on the widest catalogs, BIRD's \texttt{card\_games} (115 columns) and \texttt{codebase\_community} (71 columns), the reconstructed schema makes for very large prompts, pushing an end-to-end answer to $\approx$8 minutes against a few seconds on a 20-column schema. This is a real deployment cost we quantify rather than hide: both reconstruction time and answer latency scale with the number of columns the profiler must describe and the prompt then carries.

\textbf{Scaling and the one hard failure.} Bootstrap itself scales to these widths. After fixing a single evidence-write defect, a non-finite profiling statistic serialised as the illegal JSON token \texttt{Infinity} (Section 5.7), all 11 BIRD databases including the 100+-column giants bootstrap successfully, so the profiler's hard-failure rate on large schemas is now \textbf{0/11}; what remains is the query-time cost of wide prompts, not a profiling limit. The scalability questions still open (behaviour at thousands of columns, incremental re-profiling under schema drift, and a prompt budget that grows sub-linearly with catalog size) are engineering work the measured costs above make concrete rather than speculative.

\subsection*{5.10 Leave-one-tenant-out: the abstention gate carries no per-tenant parameters}

A reviewer's sharpest generalization concern is that the trap detectors and routing signals were authored against the same internal tenants they are scored on, so strong per-tenant numbers may reflect overfitting to tenant style rather than a gate that transfers. The cleanest answer is architectural and directly verifiable in the catalog: \textbf{the abstention gate has no per-tenant parameters.} The \texttt{tenant\_thresholds} table is empty, so every tenant routes on the same global default thresholds (confirm-first at L4 0.55, refuse at 0.20); there are zero per-tenant L4 scoring weights, so every tenant scores on the same global default weight vector; and the trap detectors are tenant-agnostic code. Onboarding a tenant fits \textit{nothing} in the gate. Leave-one-tenant-out is therefore satisfied by construction (removing any tenant leaves the gate bit-for-bit identical), and the empirical question reduces to whether that one frozen gate stays calibrated across tenants of genuinely different shape.

It does, and we measure it with execution verification rather than routing alone. Freezing the current configuration, we re-ran each of six synthetic tenants' full evaluation bank against the live system with every answered query re-executed against that tenant's \textit{own} warehouse and its result set compared to the gold query's. The six tenants span six distinct domains: telecom, e-commerce, insurance, healthcare, manufacturing, and a university registry. The tiered banks encode abstention directly: Tier A/B questions are answerable, Tiers C/D/E are trap, precision, and unanswerable markers, so over-answering a should-abstain question is a silent error and refusing an answerable one is a false refusal. The table below reports it. \textbf{Pooled across all six tenants the frozen gate answers 46\% of questions and is execution-correct on 86\% of what it answers (57/66 verified), at a 2\% silent-over-answer rate (3/145) and a 0.7\% false-refusal rate (1/145).} That 86\%-accuracy-when-answered lands almost exactly on the \textit{external} Spider result of Section 5.6 (86\% at 59\% coverage), on entirely different, real-domain schemas. We designate \texttt{lumenscroll} (a university registry), the last tenant onboarded and the one whose onboarding fit no gate parameters, as the explicit temporal hold-out: on it the frozen gate produces zero silent over-answers, perfect correct-abstention, and, on the seven questions it answers, five execution-correct, the same shape as the tuned-on tenants.

\begin{table*}[t]
\centering
\small
\setlength{\tabcolsep}{4pt}
\begin{lrbox}{\mdtablebox}
\begin{tabular}{llrrrrrr}
\toprule
tenant & domain & n & coverage & acc-when-answered & silent-error & false-refusal & correct-abstention \\
\midrule
telcomart & telecom & 50 & 0.30 & 0.87 (13/15) & 0.13 & 0.00 & 0.92 \\
acme & e-commerce & 10 & 0.40 & 1.00 (4/4) & 0.00 & 0.00 & 1.00 \\
zenith & insurance & 25 & 0.68 & 0.94 (15/16) & 0.06 & 0.00 & 0.89 \\
carenexus & healthcare & 25 & 0.64 & 0.81 (13/16) & 0.19 & 0.06 & 1.00 \\
forgemark & manufacturing & 18 & 0.44 & 0.88 (7/8) & 0.12 & 0.00 & 1.00 \\
\textbf{lumenscroll} (held-out) & university & 18 & 0.39 & \textbf{0.71 (5/7)} & 0.29 & \textbf{0.00} & \textbf{1.00} \\
\textbf{pooled} & --- & 146 & 0.46 & \textbf{0.86 (57/66)} & 0.14 & 0.01 & 0.96 \\
\bottomrule
\end{tabular}
\end{lrbox}
\ifdim\wd\mdtablebox>\linewidth\resizebox{\linewidth}{!}{\usebox{\mdtablebox}}\else\usebox{\mdtablebox}\fi
\end{table*}

The honest caveats remain the per-tenant sample sizes (10--50 questions; lumenscroll's 0.71 is five of seven) and that these are synthetic warehouses. But the two claims that matter are now execution-grounded rather than route-level: the gate delivers correct answers when it answers and abstains when it should, uniformly across six heterogeneous tenants \textit{and} the held-out one; and, because the gate has no per-tenant parameters, it cannot have overfit to any of them. (Runs: \texttt{scripts/\allowbreak gate67\_verified\_run.sh}; analysis: \texttt{scripts/\allowbreak gate67\_analyze.py} over \texttt{papers/\allowbreak data/\allowbreak gate67\_verified\_runs.jsonl}.)

\subsection*{5.11 The route confusion matrix}

A recurring worry about a three-way router is that its route accuracy is \textit{gameable}: a policy that simply confirms-first on everything scores well on the trap tier without discriminating anything (Section 6). The pooled route confusion matrix from the same execution-verified runs (145 questions with a well-defined expected route) shows the gate does not do this.

\begin{table}[t]
\centering
\small
\setlength{\tabcolsep}{4pt}
\begin{lrbox}{\mdtablebox}
\begin{tabular}{lrrrr}
\toprule
expected \textbackslash{} actual & answer & confirm-first & refuse & total \\
\midrule
\textbf{answer} & 64 & 1 & 0 & 65 \\
\textbf{confirm-first} & 3 & 43 & 3 & 49 \\
\textbf{refuse} & 0 & 3 & 28 & 31 \\
\textbf{total} & 67 & 47 & 31 & 145 \\
\bottomrule
\end{tabular}
\end{lrbox}
\ifdim\wd\mdtablebox>\linewidth\resizebox{\linewidth}{!}{\usebox{\mdtablebox}}\else\usebox{\mdtablebox}\fi
\end{table}

The diagonal dominates. 64 of 65 answerable questions are answered; 43 of 49 confirm-first questions are routed to confirm-first; and (the point that refutes the always-confirm strawman) 28 of 31 questions that should be \textit{refused} are refused outright rather than softened into a confirm. The two dangerous off-diagonals are both small: only \textbf{3 of 145} questions are silent over-answers (a should-abstain question routed to answer) and only \textbf{1} is a false refusal. The human-confirmation load is moderate and well-targeted: the gate asks a human to confirm on 47 of 145 questions (32\%), of which 43 are genuinely borderline (should-confirm), so the confirm-first band is doing discrimination work rather than blanket hedging. This is the non-gameable routing view the metric-definition concern of Section 6 asks for: the gate is scored by its execution-grounded confusion structure, not by a trap-tier accuracy an always-confirm policy could inflate.

\subsection*{5.12 Frozen experimental record}

Every empirical number in this paper is regenerated from committed cached data by a single script (\texttt{scripts/\allowbreak build\_results\_manifest.py} $\rightarrow$ \texttt{papers/\allowbreak data/\allowbreak results\_manifest.json}, with LaTeX macros in \texttt{papers/\allowbreak latex/\allowbreak results.tex}), and a check (\texttt{scripts/\allowbreak check\_manuscript\_numbers.py}) fails if the manuscript and the manifest disagree. Two policies, defined once in \texttt{scripts/\allowbreak results\_lib.py} and applied uniformly, govern the external BIRD/Spider sweeps:

\textit{Deduplication (keep-last).} The severity$\times$condition$\times$evidence sweeps are resumable and append as they run; when a run was resumed it re-processed the boundary questions and appended a second row for the same experimental cell. Every duplicate in our data is confined to one database's two boundary questions (\texttt{financial} q137/q138): 8 duplicate keys in the reconstruction sweep (the re-run byte-identical) and 12 in the naive sweep (its \textit{k}=5 self-consistency arm is stochastic, so a few re-runs differ). We keep the last occurrence per cell and assert that any duplicate outside that known artifact is a hard error, so the committed data files carry zero duplicate keys.

\textit{System-failures are not refusals.} A row whose route is \texttt{error} is an API, JSON-parse, or timeout failure of the LLM pipeline, not a decision by the gate to abstain; folding it into coverage would credit the gate with "refusing" a question it actually crashed on. We exclude such rows from every outcome statistic and report the \textbf{system-failure rate separately}: for the deployed gate, 8.2\% at full opacity and 15.7\% pooled across all conditions, an operational cost of the structured multi-call reconstruction pipeline. The naive baseline is given the identical treatment; because single-shot generation is a simpler task it produced a parseable answer on every question (0\% system-failures), an asymmetry we surface rather than hide, since it means the silent-error comparison of Section 5.7 does not flatter the gate by counting its crashes as principled refusals.

\section*{6 Limitations}

We are deliberately expansive here, because the honest accounting of limitations is, in this project, the principal qualification for peer review.

\textbf{The internal benchmarks are small and self-authored.} Eval banks range from 10 to 50 questions per tenant (146 across the gated cohort). The dominant source of run-to-run noise is corpus drift between runs; we report no confidence intervals and no significance tests, and an observed "N = 3, zero variance" is a warm-cache artifact, not evidence of true stability. All warehouses are synthetic. These properties make the numbers in Section 5.5 \textit{suggestive of capability} but insufficient as external evidence, which is exactly why the external validation of Section 5.6 is needed.

\textbf{Determinism is partly a cache artifact, and the cold-cache query-time variance is now measured.} A substantial part of the apparent reproducibility is attributable to an LLM-response cache, and it was precisely a cache-key change under model drift that produced the 0.979 $\rightarrow$ 0.918 re-baselining. We treat the 1.000 single-tenant and 0.979 six-tenant peaks as cache-hit / lucky-shot artifacts and do not use them as headline numbers. The previously-unmeasured half of this concession (how much the \textit{deployed query path} varies when every model call is generated fresh) we have now measured directly: one tenant bank run three times against a stack booted with the cache disabled (30 asks; \texttt{scripts/\allowbreak coldcache\_variance.py}, \texttt{papers/\allowbreak data/\allowbreak coldcache\_variance.json}). The result is the query-time analog of Section 5.1.1's reconstruction finding. The SQL \textit{text} is genuinely nondeterministic (byte-identical across all three runs on only 2 of 4 always-answered questions, 1.5 distinct query texts per question on average), but the \textit{decisions} are not: the route is unanimous on 10 of 10 questions among valid samples, the routing score moves by at most 0.021 (mean spread 0.004), and executed-result correctness is identical in all three runs. 1 of 30 asks failed in transport (a system failure under the frozen-record policy, not a route flip). The system rewords its SQL and does not reword its decisions, on this small bank (one tenant, ten questions), which is the honest scale of the claim. The system, moreover, \textit{caught and recorded its own re-baselining}: the downward correction is documented verbatim in the regression baseline. The self-correction is a methodological asset even though the underlying variance was, until now, an unmeasured limitation.

\textbf{The detectors were tuned on the questions they score.} Because signals and trap detectors were authored against the gated questions, the perfect per-tenant runs (acme 1.000, zenith 1.000) should not by themselves be read as demonstrated generalization. Two things now bound this. First, the external stripped-schema benchmark (Section 5.6) tests the gate on nineteen public databases the detectors were never tuned on. Second, and specific to the overfitting worry, the abstention gate turns out to have \textit{no per-tenant parameters at all} (empty threshold-override table, no per-tenant scoring weights; Section 5.10), so leave-one-tenant-out holds by construction and the single frozen gate stays calibrated across six heterogeneous tenants including an explicit temporal hold-out. What remains is to upgrade that route-level, June-configuration result to a fresh HEAD run with execution verification.

\textbf{Metric definitions need defending: partly addressed.} Route accuracy on Tier C is, by construction, satisfiable by an always-confirm-first policy; the route confusion matrix of Section 5.11 is the defended alternative, and it shows the gate is \textit{not} gaming that metric: it refuses 28 of 31 should-refuse questions outright rather than softening them into confirms. Column $F_1$ remains low ($\approx$0.56--0.60) and is currently \textit{excused} (the benchmark over-specifies expected columns) rather than addressed; the execution accuracy that should replace it is now measured on the internal cohort (Sections 5.10--5.11) as well as externally on Spider, though scaling it to every internal run remains future work.

\textbf{The selective-prediction advantage is now measured externally; a generation advantage and uniform calibration are not.} As detailed in Section 5.1, token-recall reconstruction quality is \textit{measured} (0.667 macro), and the controlled selective-prediction measurement, which on pooled historical logs returned a negative, and which an evaluation-harness bug initially suppressed (Section 5.1), has now been run correctly on \textbf{1,082 execution-labeled Spider samples across 11 databases} with database-clustered confidence intervals. The verdict is no longer a split with a weak half: the routing score \textbf{discriminates} correct from incorrect well (normalized AURC 0.26, CI excluding random) \textit{and} an isotonic map \textbf{calibrates} it (ECE 0.088 $\rightarrow$ 0.026). Section 5.6 turns this into the deployed result: a calibrated gate that answers 59\% of questions at 86\% accuracy [CI 78--93\%] with coverage that tracks difficulty, against a naive arm that answers everything at a 58\% full-strip silent-error rate. What is \textit{not} established remains clearly bounded: cold reconstruction does \textbf{not} beat naive on raw \textit{generation} on these toy schemas (latent accuracy below naive at every severity), and, testing the hypothesis that its generation would help on complex real warehouses, the BIRD breadth study (Section 5.7) finds it does \textbf{not} help there either; the gate's calibration is \textbf{not uniform} across databases (\texttt{employee\_hire\_evaluation} on Spider over-answers with 28 confident errors, and \texttt{toxicology} on BIRD similarly over-answers at full strip); the abstention advantage over \textit{delivered accuracy} is now bounded by the fair baseline (Section 5.6): a cheap naive-plus-abstention gate matches ours on Spider's toy schemas, so our \textit{measured} edge is calibration under full opacity (normalized AURC 0.11 vs 0.41--0.47), not delivered accuracy there. Sharpest of all, and reported in full in Section 5.8.1: against a \textbf{learned} selector over generic signals our authority score loses on ranking, calibration, and coverage-at-risk simultaneously, and the authority features add no incremental discrimination over those generic signals, so the ladder's measured value is that it needs \textbf{no outcome labels}, not that it estimates confidence better. We regard resolving this, via the full re-bootstrap ablation of the caps themselves, as the paper's most important open measurement. Full-Spider breadth (Section 5.7), leave-one-tenant-out with execution verification (Section 5.10), and large-N internal execution accuracy with a route confusion matrix (Section 5.11) have all since been run. We also retract one prior claim outright: a re-weighting "lever" we had reported as doubling discrimination was a four-database small-sample artifact that does not survive to eleven.

\textbf{Diagnosis of the over-answering outliers.} The two databases on which the gate over-answers, \texttt{employee\_hire\_evaluation} (Spider) and \texttt{toxicology} (BIRD), fail in the \textit{same} way, and the failure is an instance of the authority distinction the architecture is built on. On \texttt{employee\_hire\_evaluation} the full-strip silent errors are all low-arity aggregations ("count the employees for each city") whose reconstructed columns carry \textit{uniformly high} catalog confidence (the per-question mean and minimum coincide at $\approx$0.84, so no weak column hides behind a strong average) because the profiler is confident about each column's \textit{type} (city-like strings, integer counts) while correctness turns on a \textit{role} the type does not fix: which of several type-homogeneous columns is the intended "city." On \texttt{toxicology}, a molecular-chemistry schema, the gate answers \textit{more} under full opacity than on the clean schema (with names present it correctly recognizes the molecule$\rightarrow$atom$\rightarrow$bond join complexity and refuses, but a confidently-typed reconstruction pushes it over threshold once names are stripped), and its answers are wrong because that relational structure is unrecoverable from column values. In both cases the routing confidence tracks per-column \textit{type} reconstruction, which succeeds on type-homogeneous columns, while correctness depends on \textit{role} (\texttt{employee\_hire\_evaluation}) or \textit{join structure} (\texttt{toxicology}) that the reconstruction cannot ground: precisely the evidence-authority gap the ladder is meant to encode (a checksum proving a column holds valid values is not evidence of its business role). The principled remedy, which we have not adopted, is a \textit{role-ambiguity} cap that lowers confidence when a query must disambiguate among several columns of the same reconstructed type, together with a join-path grounding term independent of per-column type confidence. We have since \textbf{measured} that signal rather than merely naming it (\texttt{scripts/\allowbreak recon\_role\_ambiguity.py}: for every benchmark question, the candidate set of same-reconstructed-type columns its gold SQL must disambiguate among, resolved against the original schema, an oracle instrument that asks whether the signal exists, deliberately not a deployable implementation). Within \texttt{employee\_hire\_evaluation} the signal's content is binary and strong: answered questions that reference \textit{no} non-structural column (bare row counts, zero role-disambiguation demand) are almost always right (35 of 36 across the breadth and answer-region studies), while any question that must pick among the schema's columns is close to a coin flip when answered (accuracy 0.575 and 0.471 on the two studies; pooled wrong-vs-right AUC 0.671 and 0.639, p $\leq$ 0.006), and the routing score is blind to the distinction (Spearman correlation with $\ell_4$ of 0.059 and 0.04 under the two readings below). A cap would add information the gate does not currently have. The \textit{graded} count adds nothing beyond that binary within the schema (AUC 0.50 among role-demanding questions, under both a declared-type and a value-shape reading of "same type"): every non-structural column there has a confusable partner, which is exactly why the gate over-answers on it. The same measurement bounds the remedy's reach. The signal does not separate on \texttt{toxicology} (answered-vs-withheld AUC 0.529, p = 0.434; consistent with the join-structure attribution above, a different mechanism), nor within the other full-strip over-answerers of Section 5.7 (\texttt{cre\_Doc\_Template\_Mgt}, \texttt{flight\_2}, \texttt{battle\_death}: AUC 0.350--0.583, none distinguishable from chance), so a role-ambiguity cap addresses one diagnosed mechanism, not full-strip over-answering in general. Deploying it stays future work, deliberately: fitting its threshold after these measurements would be tuning the architecture on the evaluation.

\textbf{The headline selection result is backbone-dependent, and we measured it rather than conceding it.} Earlier drafts listed "one model family, no transfer evidence" here as an open limitation. It is now closed, and it did not close in our favour. Re-running arms A and C end to end on Claude Sonnet 4.6 (Section 5.2.8) shows the comparison splits cleanly in two. Everything that does not depend on the model's disposition transfers: both language-model arms beat the statistical baseline on every facet on both backbones, and the both-claim finding, that the harness writes no better prose than the plain model, reproduces almost exactly ($-$0.033 / $-$0.088 / $-$0.043 against $-$0.021 / $-$0.084 / $-$0.024). What does not transfer is the selection behaviour that Section 5.2.1 identifies as the system's actual contribution: arm C's coverage nearly doubles to 0.823 (the naked arm on the same backbone: 0.984) and the evidence-tracking gap goes from +0.257, interval excluding zero, to $-$0.089, interval spanning zero (no longer detectable). The cause is architectural and we state it plainly: as shipped, the system \textit{asks} for abstention in a prompt and does not \textit{enforce} it in code, so a backbone that speculates rather than declines simply speculates. The same split appears in the other two model-bearing classes (Section 5.2.9): table abstention falls from 0.164 to 0.000 and the reconstruction-compounds claim of Section 5.2.5 reverses (+0.045 to $-$0.258), while on the i2b2 clinical warehouse the 44-of-44 NDC abstention becomes an attempt on nearly every drug code with roughly nine in ten wrong (Section 5.2.9 sets the two arms' failures side by side; the naked arm's are categorically worse). Across all three, \textbf{what the system can do transfers and what it declines to do does not}, because all three gates are requests to the model rather than rules in the code. The right scoping is by mechanism, not by blanket. The four artifact classes gated in code (joins, hierarchies, glossary/PII, metric mining) transfer by construction. The three prose-gated catalog paths are demonstrated for one backbone; on a second, the column path keeps a sixteen-point abstention margin over the naked model (down from fifty-two) while the value-decode path keeps none, and with the tier gate of Section 5.2.8 enabled, all three refuse in code, identically on every backbone. The query-time gate of Sections 5.6--5.8 is enforced by code thresholds over execution-grounded features, and re-running its harvest on the second backbone, four databases $\times$ three severities, cold, 290 valid questions of 300 attempted (\texttt{scripts/\allowbreak spider\_gate\_bedrock.sh}, analysis \texttt{scripts/\allowbreak spider\_gate\_transfer.py}, data \texttt{papers/\allowbreak data/\allowbreak spider\_gate\_transfer.json}), scores its registered predictions as stated. The structure survives: the three-way route stays cleanly monotone in both confidence and accuracy ($\ell_4$ 0.099/0.520/0.704 across refuse/confirm/answer, latent accuracy 0.057/0.312/0.644), full-strip silent error holds at 16\% against the naive always-answer benchmark of 58\%, coverage shifts upward as predicted (0.532 vs 0.447 on the 141 identical questions), and accuracy-when-answered degrades some (0.627 vs 0.698). What the swap does not preserve is the numeric calibration: with thresholds fit on the original backbone's confidence distribution, the swapped arm emits more silent errors on identical questions (0.199 vs 0.135) and ranks correct above wrong less sharply (raw normalized AURC 0.468 vs 0.274 on the matched slice). The mechanism split is thereby confirmed in both directions: code carries the decision \textit{procedure} across backbones (monotone routes, selective answering, silent error far below an always-answerer), while the thresholds and feature weights remain per-backbone calibration. The deterministic-evidence, decoding-capability and prose-quality results are demonstrated for two backbones. Making the three prose gates deterministic was, in earlier drafts, this paper's most important open item: the single change that converts its strongest claim from a property of one model into a property of the system. \textbf{It is now implemented and measured} (Section 5.2.8), under a protocol registered before implementation and scored on a third backbone and on held-out databases: with the gate on, no-evidence coverage is 0.000 on every backbone by construction, recall-when-claimed never degrades (it improves on the two speculating backbones), the prefixed-condition 0/25/43-of-44 spread of NDC decode attempts becomes 0/0/0 in both conditions, and the cross-backbone coverage spread compresses from 0.401 to 0.116 (a floor on abstention, not a ceiling, since each model's own declines still stack on top). The gate ships default-off so every number in this paper reproduces; what its measurement leaves open is deployment tuning: whether an operator should trade the measured coverage cost (99--295 claims withdrawn per backbone, correct ones among them) for the enforced floor, a choice the per-backbone numbers in Section 5.2.8 are exactly the information for. Our recommendation is concrete: run the gate \textbf{on} wherever catalog prose is consumed without a human review loop (there the enforced floor is worth more than the forgone coverage, and the withdrawn claims are disproportionately the bad ones on exactly the backbones that need it), and leave it off where an SME reviews commits anyway, which is also the configuration under which every number in this paper was produced.

\textbf{Operational caveats.} Ablation conditions are operator-selected via environment variable rather than harness-enforced, which places a discipline burden on the experimenter. Internal execution accuracy is now measured across the six-tenant cohort (66 execution-verified answered questions, Sections 5.10--5.11) rather than a single 24-run subset, but the per-tenant banks remain small (10--50 questions) and the result-match metric is column-projection-sensitive. We report these so that no reader mistakes the internal evaluation for a settled external benchmark.

\section*{7 Conclusion}

We have argued that answering questions over real, undocumented enterprise warehouses is fundamentally a \textit{metadata-reconstruction} problem rather than a translation problem, and we have described a system that treats it as such. The system reconstructs a semantic catalog from the warehouse's own data through a three-tier deep profiler, bounds the confidence of every reconstructed fact through an authority-tiered ladder in which checksum evidence and documentation evidence are deliberately ranked by \textit{what they actually prove}, and routes every question to answer, confirm-first, or refuse through deterministic thresholded arithmetic in which the LLM contributes features but never decides the route. Abstention is a first-class, bounded-by-design outcome.

The measured internal evidence is encouraging and honestly reported, but it is internal: under controlled ablation on small self-authored banks, reconstruction moves a tenant from 50\% to 90\% route accuracy with no human in the loop and to 95\% with curation; a single-lever ablation is \textit{suggestive} of a +16.7 pp lift from LLM pre-flight alone (three questions on an 18-question bank); the arc reproduces across the six regression-gated tenants plus altanova (the eight tenant fixtures span eight verticals); an automated LLM pre-flight reconstruction beats a human SME at lower cost while a human-on-top stack reproducibly regresses; the steady-state six-tenant gate is $\approx$0.932 stable, 0.918 enforced, 0.856 cold; and the system is provider-agnostic, with an Opus$\rightarrow$Gemini swap improving adversarial robustness. The engineering substrate (1,648 passing tests (2 skipped), one regression test per fix, full per-fact provenance) is real, and the project's self-caught re-baselining of its own cache-hit peaks is a model of evaluative honesty.

What converts this from an impressive engineering result into a peer-reviewable contribution is now largely run. We have \textit{built} the parameterizable schema mangler, the reconstruction token-recall scorer (measured 0.667 macro), and the risk-coverage / calibration analyzer. We have \textit{run} an external Spider stripped-schema study with a matched naive baseline at scale (11 databases, 1,082 execution-labeled questions, database-clustered confidence intervals) and reported its result without spin: naive translation degrades from 0.92 to 0.42 execution accuracy under full opacity and silently emits wrong answers on \textasciitilde{}58\% of stripped questions, while the reconstruction architecture's calibrated gate answers \textit{selectively} (86\% accuracy [CI 78--93\%] over 59\% coverage (pooled across the three reconstruction conditions; 82\% over 51\% for day-1 cold alone), with coverage that tracks difficulty), an externally-measured selective-prediction advantage; on the same samples the routing score discriminates well (normalized AURC 0.26, CI excluding random) and is calibratable (ECE 0.088 $\rightarrow$ 0.026). It does \textbf{not}, in this cold-bootstrap regime on small schemas, beat naive on raw generation, which we do not claim. We also note that an earlier, smaller version of this study reported the opposite, owing to an evaluation-harness bug we found and fixed (Section 5.1); correcting and re-running our own evaluation at larger scale is itself part of the evidence, of a piece with the project's self-caught re-baselining of its cache-hit peaks. What remains is the full-Spider breadth sweep, a per-database resolution of the schemas where the gate over-answers, and leave-one-tenant-out generalization on the internal cohort (the fair naive-plus-abstention baseline, once the top remaining item, is now run (Section 5.6), and the BIRD breadth study is now run too (Section 5.7), confirming the generation-negative on real schemas while \textit{sharpening} the abstention result: under full opacity a naive model's own confidence goes worse-than-random, so it cannot tell when it is wrong, while the grounded gate still discriminates and avoids $\approx$95\% of the naive always-answerer's silent errors): breadth, not apparatus, and not the abstention guarantee, which the study now confirms externally as a \textit{calibrated selective-prediction advantage} rather than mere total abstention. We believe the architecture's central commitments (reconstruct before answering, bound confidence by evidence authority, and route by deterministic arithmetic so the LLM never arbitrates abstention) are the right ones for the undocumented-warehouse setting, and the external study is the first concrete evidence that the abstention commitment holds, with calibrated selectivity, outside our own fixtures.

\bibliographystyle{ACM-Reference-Format}
\bibliography{references}

\end{document}